\documentclass[draft,11pt]{article}
\usepackage[utf8]{inputenc}

\usepackage[final]{microtype}

\usepackage[paper = letterpaper,
centering,
lmargin = 1in,
rmargin = 1in,
tmargin = 1in,
bmargin = 1in]{geometry}

\usepackage{amsmath,
	amssymb,
	mathtools,
	amsthm,
	physics,
	bm,
	physics,
	mathrsfs,
	stmaryrd,
	upgreek,
	textcomp,
	soul%
}
\usepackage[np,autolanguage]{numprint}

\usepackage[final]{graphicx}
\DeclareGraphicsExtensions{.pdf, .jpg, .png}
\usepackage{color}
\usepackage{array,
	multirow,
	booktabs,
	float,
	rotating,
	subfigure,
	caption,
	afterpage,
	url%
}

\usepackage[symbol,perpage]{footmisc}
\usepackage[square,numbers,sort&compress]{natbib}

\usepackage{enumerate}
\usepackage{nth}

\usepackage{ifthen}
\usepackage{calc}
\usepackage{xspace}

\usepackage{hyperref}
\hypersetup{final}

\usepackage{comment}
\excludecomment{FCSkip}

\usepackage[colorinlistoftodos]{todonotes}
\usepackage{acronym}
\acrodef{ALE}[ALE]{Arbitrary Lagrangian--Eulerian}
\acrodef{FSI}[FSI]{fluid-structure interaction}
\acrodef{FEM}[FEM]{finite element method}
\acrodef{BVP}[BVP]{boundary value problem}
\acrodef{DAE}[DAE]{differential-algebraic equations}
\acrodef{BDF}[BDF]{backward differentiation formulas}
\acrodef{TENG}[TENG]{Tissue Engineered Nerve Guide}
\acrodef{CUPE}[CUPE]{cross-linked urethane-doped polyester elastomers}
\acrodef{IO}[IO]{interpolation order}
\acrodef{PDE}[PDE]{partial differential equation}

\newcommand{\scdot}{\ensuremath{\!\cdot\!}}
\newcommand{\Grad}[2][nonsense]{%
	\ifthenelse{\equal{#1}{nonsense}}%
	{\nabla#2}%
	{\nabla_{#1}#2}%
}
\newcommand{\Div}[2][nonsense]{%
	\ifthenelse{\equal{#1}{nonsense}}%
	{\nabla\scdot#2}%
	{\nabla_{#1}\scdot#2}%
}
\newcommand{\Lin}[2][nonsense]{%
	\ifthenelse{\equal{#1}{nonsense}}%
	{\ensuremath{\mathscr{L}(#1,#1)}}%
	{\ensuremath{\mathscr{L}(#1,#2)}}%
}
\newcommand{\bv}[1]{\boldsymbol{\mathbf{#1}}}
\newcommand{\tensor}[1]{\mathsf{#1}}
\newcommand{\transpose}[1]{#1^{\scriptscriptstyle\mathrm{T}}}
\newcommand{\inversetranspose}[1]{\ensuremath{#1^{\scriptscriptstyle-\mathrm{T}}}}
\newcommand{\ord}[3][nonsense]{%
	\ifthenelse{\equal{#1}{nonsense}}%
	{\ensuremath{#2^{\scriptscriptstyle(#3)}}}%
	{\ensuremath{#2^{\scriptscriptstyle#1(#3)}}}%
}
\newcommand{\colondot}{\mathbin{:}}

\title{A mixture theory-based finite element formulation for the study of biodegradation of poroelastic scaffolds\footnote{Submitted for publication in \textit{Journal of Fluids and Structures}}}
\author{Priyanka Patki$^{b,c}$\footnote{Corresponding author}\\
W-321 Millenium Science Complex\\
The Pennsylvania State University,\\
University Park, PA 16802, USA\\
pap185@psu.edu
\and
Francesco Costanzo$^{a,b,c}$\\
W-315 Millenium Science Complex\\
The Pennsylvania State University,\\
University Park, PA 16802, USA\\
fmcostanzo@psu.edu
\and
$^{a}$Department of Engineering Science and Mechanics\\
$^{b}$Department of Mechanical Engineering\\
$^{c}$Center for Neural Engineering\\
The Pennsylvania State University,\\
University Park, PA 16802, USA
}	
\date{}

\begin{document}
	\maketitle
	\allowdisplaybreaks{                             
		\begin{abstract}
			We derive a mixture theory-based mathematical model of the degradation of a poroelastic solid immersed in a fluid bath. The evolution of the solid's mechanical and transport properties are also modeled.  The inspiration for the model is the study of the temporal evolution of biodegradable \acp{TENG}, which are surgical implants supporting the alignment and re-growth of damaged nerves. The model comprises of the degrading solid, the degradation reaction products, and the fluid in which the solid is immersed. The weak formulation of the \acp{PDE} so derived is numerically implemented using a \ac{FEM}. The numerical model is studied for stability and convergence rates using the Method of Manufactured Solutions.
			\vspace{5pt}
		\end{abstract}

\section{Introduction}
\label{sec:Intro}
Peripheral nerves carry sensory-motor impulses from the brain to different parts of the body and vice versa. Injuries to peripheral nerves can have dire consequences, such as loss of sensitivity, motion, or both, in the injured organ. The inherent mechanism of the body to mend the damaged or broken nerves can have retargeting errors when rejoining broken nerve ends with large gaps. External support for realignment and mechanical strength of such nerve injuries involving large gaps is often provided in the form of surgical nerve autografts \cite{pns1-dodla20191223, pns2-bellamkonda2006peripheral} (nerve segments obtained from a healthy donor site in the patient). Unfortunately, this method has several shortcomings, such as scarcity of viable donor cites, possibility of neuroma at the donor location and mismatch between the size of the graft and injured nerve. \acfp{TENG} are an attractive alternative to nerve autographs. \acp{TENG} are sleeve-like porous scaffolds, preferably biodegradable, that are sutured to the severed nerve ends \cite{Nguyen2015Tissue-Engineer-0, pns1-dodla20191223}. \acp{TENG} are expected to degrade and be bioresorped as the nerve regrows. Ideally, the rate of regeneration of the nerve matches the \ac{TENG}'s degradation rate. Moreover, \acp{TENG} are expected to provide mechanical support and alignment to the broken nerve ends and are usually designed to be porous in order to enable effective supply of drugs and nutrients to the recovering nerve \cite{pns1-dodla20191223}. Thus, \ac{TENG} design requires careful study of their degradation chemistry and the corresponding evolution of their transport and mechanical properties. In this study, we have derived a formulation capable of simulating the complex chemistry, mechanics, and transport aspects of a typical \ac{TENG}. An example of a typical \ac{TENG} material is \ac{CUPE} \cite{CUPE1, CUPE2}, and our eventual goal is to develop a model which can accurately estimate the degradation and mechanical response of this material when inserted at a target cite.

Studies of degradation of biopolymers (e.g.\ \cite{Hydrolysis_chemistry, hydrolysis_chemistry2}) show that the rate of hydrolytic degradation of elastomers is governed by the concentration of the polymer as well as the  degradation reaction products. When a poroelastic polymer scaffold is inserted in a dynamic environment such as in a limb, it will likely be subject to stresses along with diffusive momentum transfer between the bodily fluids and the solid. This necessitates the study of the mass and momentum balance equations of the polymer, the degradation products and the fluid in which the scaffold is inserted. Mixture theory provides an excellent framework for the study of all components in the mixture. Thus, we resorted to this continuum framework for the derivation of our governing equations.

Reactive mixture theoretic formulations for biomechanical applications have been limited mostly to aneurysms (c.f. e.g \cite{baek2006theoretical}) and growth and remodeling of cartilage, bone and soft tissues (cf., e.g., \cite{taber1995biomechanics,cartilageGrowth1,klisch2005cartilage,ateshian2007theory,humphrey2002constrained,ambrosi2010insight,GARIKIPATI20041595}). Most of these works are based on volumetric or surface growth, through decomposition, without directly involving chemical kinetics. Moreover, the growth or remodeling is assumed to occur mostly through mechanical loading. Considerable focus of these studies is mainly on the evaluation of residual forces, tissue swelling or other mechanical phenomena. 
Constitutive models for the growth in biological tissues by applying the Clausius-Duhem inequality have been derived in \cite{GARIKIPATI20041595}, in which the authors derive coupled mass transport-mechanics based formulation for growth of tissues, by considering the viscous fluid and solid to be a single vicoselastic entity. Balance equations are then obtained for this viscoelastic tissue. They obtain the formulation in terms of entities such as the mixing entropy, which may not be a readily available input data. Moreover, as mentioned earlier, a component-wise modeling of our problem is essential, as the chemical kinematics of the degradation of the poro-elastic solid may depend on the local concentrations of various components in the mixture. Thus, from a modeling point of view, degradation cannot be viewed simply as a reversal of tissue growth. 

Chemo-mechanical coupling has been modeled for non-biomechanical applications as well. For example, in \cite{karra2012model} and \cite{Roy2008} a model for the oxidative degradation of polyimides at high temperatures has been derived. The evolution of properties in a chemically reacting mixture is determined mainly by two processes - diffusion and chemistry. Oxidative degradation of polyimides and growth of biological tissues are examples of reaction-dominated and diffusion-dominated thermochemical processes respectively. An excellent derivation of these two asymptotic extremes in a diffusive-reactive system consisting of a visco-elastic fluid diffusing through a thermoelastic body have been derived in \cite{HallRajagopalDiffusionChemistry}. Diffusion-dominated processes lead to occurrence of reactions all over the body, whereas reaction-dominated processes often lead to advancing reaction fronts in the body, thus the modeling approach for these two cases is different.

A more general modeling approach is essential when the diffusion-reaction rates are comparable, or when a more generalized formulation is sought. Hence, in our formulation, the chemistry as well as mechanical and transport properties of different components in the mixture are derived as a function of the spatial location of material points in the mixture domain. Any heterogeneity in the mixture is naturally taken care of in this approach, which also allows for more flexibility in determining the local behavior of the mixture, based on the responses of its components. We choose a classical continuum mechanics approach for mixtures (c.f. e.g. \cite{bowenbook,bowenContinuumBook}) for the derivation of constitutive models for flows through poro-elastic degrading solids. Balance equations are derived for each component in the mixture and constitutive forms are then derived for stresses in the solid as well as forces of interaction between several components of the mixture. The mixture theoretic equations governing the motion of fluid flow through poro-elastic solids have been derived in \cite{bowen1980incompressible} and implemented using an \ac{FEM} framework in \cite{costanzo2017arbitrary}. In this work, we extend this theory to include bio-degradation of the poro-elastic polymer. This constitutes a reactive \ac{FSI} problem. We extend our previous non-reactive \ac{FSI} formulation by including the relevant terms in the mass and momentum balance equations of all the mixture constituents and re-deriving the formulation. In this preliminary effort, we assume that a single product of the degradation reaction is produced, which remains within the system. Thus, we model our thermochemical system as a three component mixture, comprising of the poro-elastic solid, the degradation product and the base fluid in which the solid is inserted, assuming all the components to be incompressible. The constitutive equations are derived by invoking entropy maximization for the entire mixture (second law of thermodynamics for mixtures), which is a standard technique in mixture theory, \cite{bowenbook, gurtinbook}.

Notations used in the paper, along with kinematics and general mixture theoretic balance laws are given in Section~\hyperref[sec:basic math]{\ref*{sec:basic math}}. Constitutive equations for the specific case of a degrading poroelastic solid immersed in a fluid are derived in Section~\hyperref[sec:Const]{\ref*{sec:Const}}.  Here, we also derive the Eulerian strong form of the governing equations, as well as their final and simplified quasi-static \ac{ALE} strong form. Corresponding weak formulation is presented in Section~\hyperref[sec:FEM]{\ref*{sec:FEM}}, along with the introduction of functional spaces for the trial and weighting functional spaces. The results pertaining to the numerical implementation of the \ac{FEM} scheme are discussed in Section~\hyperref[sec:Results]{\ref*{sec:Results}}, along with a discussion on the convergence and stability of the numerical scheme. We conclude the paper with a summary and discussion section.

\section{Kinematics and balance laws for mixtures}
\label{sec:basic math}

\subsection{Notation and kinematics}
\label{subsect: notation and kinematics}
For a mixture of $N$ constituents, we posit that each of these has its own reference configuration. Figure~\ref{fig:configuratons}
\begin{figure}[htb]
\centering
{\includegraphics[width=5in]{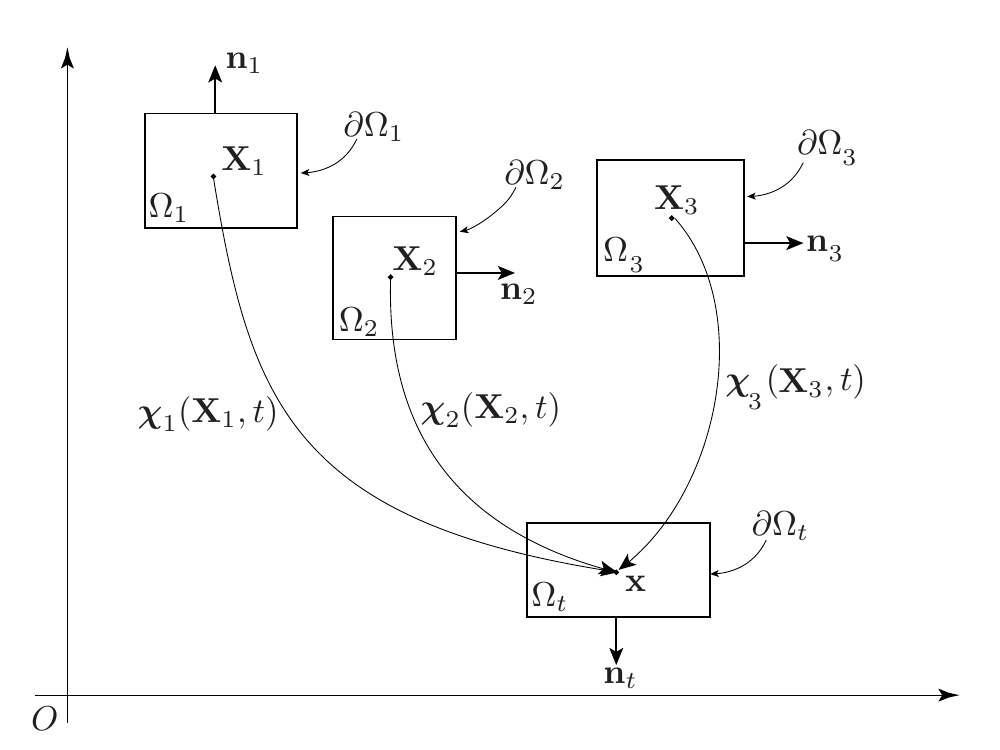}}
\caption{Configurations: $\Omega_1,~\Omega_2,~\Omega_3$ are the reference configurations of species 1, 2 and 3 respectively, and $\bv{X}_1,~\bv{X}_2,~\bv{X}_3$ are labels given to particles of the respective constituents. $\Omega_t$ denotes the current configuration, and $\bv{x}$ is the common image of the diffeomorphisms $\bv{\chi}_1$, $\bv{\chi}_2$, $\bv{\chi}_3$. $\bv{n}_a$'s denote the unit normals to the boundaries $\partial\Omega_a$'s. Section~\ref{sec:Const} onwards, the subscript~1 will be replaced by $s$ and will denote the solid component in a mixture consisting of $N-1$ fluids and a degrading poroelastic solid. Numerical subscripts will denote various fluids in the mixture.}
\label{fig:configuratons}
\end{figure}
shows a 3-component mixture, where $\Omega_a$ and $\bv{X}_a$ ($a = 1, \ldots, 3$) denote the reference configuration and the referential particle position of component $a$, respectively. For our analysis, we assume that the reference configurations for all components coincide with the initial configuration, and hence, $\bv{X}_a$ also denotes the initial location of the particle. $\Omega_t$ denotes the current configuration. $\Omega_a$ and $\Omega_t$ are subsets of $\mathscr{E}^{n_d}$, the $n_d$-dimensional Euclidean point space, $d = 2, 3$. The translation vector space of $\mathscr{E}^{n_d}$ is denoted by $\mathscr{T}^{n_d}$. The particle at $\bv{X}_a$ is mapped to a spatial point $\bv{x}$ in the current configuration via the diffeomorphism $\bv{\chi}_a:\Omega_a \cross [0,T] \rightarrow \Omega_t$ given by,
\begin{equation}
\bv{x}=\bv{\chi}_a(\bv{X}_a,t).
\label{eq:diffeomorphism}
\end{equation}
$\bv{X}_a$ is also used to denote the inverse map of the above mapping, that is,
\begin{equation}
\bv{X}_a = \bv{\chi}_a^{-1}(\bv{x},t),
\label{eq:inverse_diffeomorphism}
\end{equation}
where $\bv{\chi}_a^{-1}:\Omega_t \rightarrow \Omega_s$.

The properties of the particle are similarly mapped and hence evolve with time. As an example, if a constituent is generated in the mixture as a result of a chemical reaction, then the mass density of a `particle' of this constituent in the reference frame may increase with time. 
Also, using Eqs.~\eqref{eq:diffeomorphism} and~\eqref{eq:inverse_diffeomorphism}, for any property $\bar{\alpha}(\bv{X}_a,t)$ in the reference frame (or, following the particle) currently at $\bv{x}$, given by $\breve{\alpha}(\bv{x},t)$ in the current frame, we can write,
\begin{equation}
\alpha = \bar{\alpha}(\bv{X}_a,t) = \breve{\alpha}(\bv{x},t) = \bar{\alpha}\circ\bv{\chi}_a^{-1}(\bv{x},t) = \breve{\alpha}\circ\bv{\chi}_a(\bv{X}_a,t).
\label{eq:Change_of_frame}
\end{equation}
Henceforth, any property will simply be denoted by $\alpha$, and its domain as a function will be decided based on the context.

As is common in mixture theory, all components are assumed to co-exist at any given point $\bv{x}$ in the current configuration.  Each constituent is characterized by a mass density distribution $\rho_a=\rho_a(\bv{x},t)$, which represents the mass of component $a$ per unit mixture volume.  Then, the mixture mass density is defined as
\begin{equation}
\label{eq: mixture density definition}
\rho=\rho(\bv{x},t)=\sum\limits_{a=1}^{N}\rho_a(\bv{x},t).
\end{equation}
We assume that each constituent is incompressible. In this case, it is meaningful to talk about the intrinsic or true mass density of the constituent, which we denote by $\gamma_{a}$, and which is the mass of constituent $a$ when said constituent is in pure form.  The incompressibility assumption is embodied by the statement that $\gamma_{a}$ ($a=1,\ldots,N$) is a (known) constant. Relating $\rho_{a}$ to $\gamma_{a}$ is done by introducing the notion of \emph{volume fraction} of constituent $a$, which we denoted by $\phi_{a}$ and define through the following relation:
\begin{equation}
\rho_a=\gamma_a\phi_a.
\label{eq:phi_gamma_rho_relation}
\end{equation}
We assume that the constituents of the mixture occupy the volume filled by the mixture.  This assumption is expressed by the following relation, referred to as the saturation condition:
\begin{equation}
\sum\limits_{a=1}^N \phi_a = 1.
\label{eq:saturation}
\end{equation}

Spatial gradients with respect to $\bv{x}$ and $\bv{X}_a$, respectively, are denoted as,
\begin{equation}
\Grad[\bv{x}] = \pdv{\bv{x}}
\quad\text{and}\quad
\Grad[\bv{X}_a]= \pdv{\bv{X}_a}.
\end{equation}

For $a=1,\ldots,N$,
We define the displacement, $\bv{u}_a$, deformation gradient, $\tensor{F}_a$ and its Jacobian $J_a$ as
\begin{align}
\bv{u}_a(\bv{X}_a,t) &= \bv{\chi}_a(\bv{X}_a,t) - \bv{X}_a \text{~or~} \bv{u}_a(\bv{x},t) = \bv{x} - \bv{\chi}^{-1}_a(\bv{x},t),
\label{eq:u_a} \\
\tensor{F}_a(\bv{X}_a,t) &= \Grad[\bv{X}_a]{\bv{\chi}_a(\bv{X}_a,t)},
\label{eq:F_a}\\ 
J_a(\bv{X}_a,t) &= \det\bigr(\tensor{F}_a(\bv{X}_a,t)\bigl). \label{eq:basics_Ja}
\end{align}
Also, the right Cauchy-Green tensor is given by $\tensor{C}_a = \transpose{\tensor{F}}_a\tensor{F}_a$.
The Eulerian description of the material velocity of constituent $a$ is given by
\begin{equation}
\bv{v}_a(\bv{x}_a,t) = \pdv{\bv{\chi}_a(\bv{X}_a,t)}{t}\bigg|_{\bv{X}_a=\bv{\chi}_a^{-1}(\bv{x},t)} = \pdv{\bv{u}_a}{t}\bigg|_{\bv{X}_a=\bv{\chi}_a^{-1}(\bv{x},t)}.
\label{eq:Displacement_velocity_relationship}
\end{equation}
Let $\alpha_{a}(\bv{x},t)$ be the Eulerian description of a quantity pertaining substance $a$. Then the \emph{material} time derivative of $\alpha_{a}$ is denoted by $\grave{\alpha}_{a}$ and is defined by,
\begin{equation}
\grave{\alpha}_a = \pdv{\alpha_{a}(\bv{\chi}_a(\bv{X}_a,t),t)}{t} =
\pdv{\alpha_a(\bv{x},t)}{t} + \Grad[\bv{x}]{\alpha_a(\bv{x},t)}\cdot\bv{v}_a,
\end{equation}
where, for generic vectors $\bv{u}$ and $\bv{v}$, $\bv{u}\cdot\bv{v}$ denotes the inner product of $\bv{u}$ and $\bv{v}$.  The barycentric velocity of the mixture is defined by
\begin{equation}
\bv{v} = \frac{1}{\rho}\sum\limits_{a=1}^N\rho_a\bv{v}_a.
\end{equation}
The \emph{time derivative of a quantity $\alpha(\bv{x},t)$ following the barycentric velocity} is denoted by $\dot{\alpha}$ and is given by,
\begin{equation}
\dot{\alpha} = \pdv{\alpha(\bv{x},t)}{t} + \Grad[\bv{x}]{\alpha}\cdot\bv{v}.
\end{equation}
We will denote by $\tilde{\bv{v}}_a$ the diffusion velocity of substance $a$, which is given by:
\begin{equation}
\tilde{\bv{v}}_a = \bv{v}_a-\bv{v},
\label{eq:diffusion velocity}
\end{equation}
Another important concept for our analysis is that of filtration velocity of any fluid $b$, which we define as,
\begin{equation}
\bv{v}_{\text{fil}_b} = \phi_b(\bv{v}_b-\bv{v}_s),
\label{eq:filtration_vel_definition}
\end{equation} 
which is the relative velocity of the fluid with respect to the solid, scaled by the fluid volume fraction. Details of filtration velocity can be found in \cite{bowen1980incompressible}, and further discussion from the numerical \ac{FSI} point of view can be found in \cite{costanzo2017arbitrary}. In the text, we derive constitutive equations for a mixture consisting of one incompressible poroelastic solid and $N-1$ incompressible fluids, and later specialize the formulation for 2 fluids.

\subsection{Balance of mass}
\label{subsec:Mass balance}
The balance of mass for species $a$ in an Eulerian framework is given by,
\begin{equation}
\pdv{\rho_a}{t}\bigg|_{\bv{x}} + \Div[\bv{x}]{(\rho_a\bv{v}_a)} = \hat{c}_a,
\label{eq:mass_balance_component_Eulerian}
\end{equation}
where $\hat{c}_a$ is the rate of production of species $a$ per unit volume of the mixture.
Dividing by $\gamma_{a}$, the true mass density of the species $a$, and recalling that $\gamma_{a}$ has been assumed to be a constant, we get
\begin{equation}
\pdv{\phi_a}{t}\bigg|_{\bv{x}} + \Div[\bv{x}]{(\phi_a\bv{v}_a)} = \frac{\hat{c}_a}{\gamma_a} = \hat{\gamma}_a,
\label{eq:intermediate_mass}
\end{equation}
where the term $\hat{\gamma}_a$ is the rate of production of specific volume (volume of a species per unit volume of the mixture) of constituent $a$. Rewriting the equation in terms of the filtration velocity, we get,
\begin{equation}
\pdv{\phi_a}{t}\bigg|_{\bv{x}} - \Div[\bv{x}]{(\phi_a\bv{v}_s)} + \Div[\bv{x}]{\bv{v}_{\text{fil}_a}}  - \hat{\gamma}_a = 0.
\label{eq:Eulerian_mass}
\end{equation}
Pulling back Eq.~\eqref{eq:intermediate_mass} to the reference configuration of the solid, $\Omega_s$, we have,
\begin{equation}
\pdv{\phi_a}{t}\bigg|_{\bv{X}_s} - \Grad[\bv{x}]{\phi_a}\cdot\bv{v}_s + \Div[\bv{x}]{(\phi_a\bv{v}_a)}  - \hat{\gamma}_a = 0.
\end{equation}
This remapping operation can be viewed as adopting an \ac{ALE} framework (cf.\ \cite{CardiovascularMathematics-book-2009}) in which the \ac{ALE} map is the motion of the solid phase.
Rewriting the \ac{ALE} equation in terms of the filtration velocity, we obtain
\begin{equation}
\pdv{\phi_a}{t}\bigg|_{\bv{X}_s} + \phi_a\inversetranspose{\tensor{F}_s}\colondot\Grad[\bv{X}_s]{\bv{v}_s} + \inversetranspose{\tensor{F}_s}\colondot\Grad[\bv{X}_s]{\bv{v}_{\text{fil}_a}}-\hat{\gamma}_a = 0,
\label{eq:ALE_mass_simple}
\end{equation}
where, for any two (second order) tensors $\tensor{A}$ and $\tensor{B}$, $\tensor{A}\colondot\tensor{B}$ denotes the (tensor) inner-product of $\tensor{A}$ and $\tensor{B}$.
Since $\bv{v}_{\text{fil}_s}=\bv{0}$, in the \ac{ALE} framework, the mass balance equation for the solid is
\begin{equation}
\pdv{\phi_s}{t}\bigg|_{\bv{X}_s} + \phi_s\inversetranspose{\tensor{F}_s}\colondot\Grad[\bv{X}_s]{\bv{v}_s} = \hat{\gamma}_s. \label{eq:ALE_mass_solid_Js}
\end{equation}
The mass balance equation for the mixture is postulated to be the result of adding all of the constituent-wise equations with the stipulation that: 
\begin{equation}
\sum\limits_{a=1}^{N}\hat{c}_a = 0.
\label{eq:c_a_Constraint}
\end{equation}
Equation~\eqref{eq:c_a_Constraint} is a constraint relation stating that no net mass is produced or consumed in the mixture.  This said, it is important to observe that, in general, $\sum\limits_{a=1}^{N}\hat{\gamma}_a \ne 0$. 
Summing each contribution provided by Eq.~\eqref{eq:mass_balance_component_Eulerian}, and separating the contribution of the solid phase from the other contributions, 
we get the mass balance equation for the mixture in the Eulerian framework, given by,
\begin{equation}
\Div[\bv{x}]{\left(\bv{v}_s+\sum\limits_{a=2}^N\bv{v}_{\text{fil}_a}\right)} = \sum\limits_{a=1}^N\hat{\gamma}_a.
\label{eq:Eulerian_mass_mixture}
\end{equation}
Similarly, summing Eqs.~\eqref{eq:ALE_mass_simple} for $a=2, 3,\ldots, N$ along with Eq.~\eqref{eq:ALE_mass_solid_Js}, enforcing the constraint in Eq.~\eqref{eq:saturation}, the mass balance of the mixture in the \ac{ALE} framework can be given the following form
\begin{equation}
\inversetranspose{\tensor{F}_s}\colondot\Grad[\bv{X}_s]{\left(\sum\limits_{a=2}^N\bv{v}_{\text{fil}_a}+\bv{v}_s\right)} = \sum\limits_{a=1}^N\hat{\gamma}_a.
\label{eq:ALE_mass_mixture}
\end{equation}

\subsection{Balance of momentum}
\label{subsec:Momentum balance}
For any component $a$ in the mixture, the Eulerian form of momentum balance equation is given by
\begin{equation}
\underbrace{\rho_a\left( \pdv{\bv{v}_a}{t}\bigg|_{\bv{x}} + \bv{v}_a\scdot\Grad[\bv{x}]{\bv{v}_a} \right)}_{\rho_a\grave{\bv{v}}_a}
=  \Div[\bv{x}]\tensor{T}_a+\rho_a\bv{b}_a+\hat{\bv{p}}_a,
\label{eq:momentum_original_a}
\end{equation}
where, for generic vector fields $\bv{u}$ and $\bv{v}$, the notation $\bv{u}\scdot\Grad[\bv{x}]{\bv{v}}$ denotes the action of the gradient of $\bv{v}$ onto $\bv{u}$,\footnote{This notation is typical in the fluid mechanics literature (cf.\ \cite{Chorin2000A-Mathematical--0}).  In the continuum mechanics literature, the operation in question is typically denoted by $(\Grad[\bv{x}]{\bv{v}})\bv{u}$ (cf.\ \cite{gurtinbook,bowenbook}).} and where $\tensor{T}_a$ and $\bv{b}_a$ are the partial Cauchy stress tensor and external body force density, respectively. The term $\hat{\bv{p}}_a$ is the momentum supply for constituent $a$.  Sometimes, this quantity is also referred to as the local or internal force acting on constituent $a$ due to its interaction with the other components in the mixture \cite{bowenbook}. The constitutive relations for $\hat{\bv{p}}_a$ and $\tensor{T}_a$ depend upon the type of mixture involved. 

Following \cite{bowenbook}, we define the stress tensor for the mixture as
\begin{equation}
\label{eq: Mixture stress tensor def}
\tensor{T} = \sum\limits_{a=1}^{N}(\tensor{T}_a - \rho_a\tilde{\bv{v}}_a\otimes\tilde{\bv{v}}_a).
\end{equation}
In addition, the momentum balance postulate for the mixture is taken to be
\begin{equation}
\rho\left( \pdv{\bv{v}}{t}\bigg|_{\bv{x}} + \bv{v}\scdot\Grad[\bv{x}]{\bv{v}} \right) = \Div[\bv{x}]{\tensor{T}} + \rho\bv{b},
\label{eq:mixture_momentum_postulate}
\end{equation}
where $\bv{b} = \frac{1}{\rho}\sum\limits_{a=1}^{N}\rho_a\bv{b}_a$.  Equations~\eqref{eq:mixture_momentum_postulate} and~\eqref{eq: Mixture stress tensor def}, along with the balance of momentum equations for each constituent, yields the following constraint relation:
\begin{equation}
\sum\limits_{a=1}^N(\hat{\bv{p}}_a+\hat{c}_a\tilde{\bv{v}}_a) = \bv{0}
\quad\Leftrightarrow\quad
\sum\limits_{a=1}^N(\hat{\bv{p}}_a+\hat{c}_a\bv{v}_a) = \bv{0}.
\label{eq:mixture_pa_ca_constraint}
\end{equation}
Another equivalent statement of the balance of momentum for the mixture is obtained by summing Eq.~\eqref{eq:momentum_original_a} over all components in the mixture, and applying the constraints given by  Eqs.~\eqref{eq:c_a_Constraint} and~\eqref{eq:mixture_pa_ca_constraint}.  So doing yields
\begin{equation}
\sum\limits_{a=1}^N\rho_a\left( \pdv{\bv{v}_a}{t}\bigg|_{\bv{x}}+\Grad[\bv{x}]{\bv{v}_a}\cdot\bv{v}_a \right) = \Div[\bv{x}]{\tensor{T}_I} + \rho\bv{b} - \sum\limits_{a=1}^N(\hat{c}_a\bv{v}_a),
\label{eq:Momentum_balance_mixture_useful}
\end{equation}
where $\tensor{T}_I$ is referred to as the internal Cauchy stress tensor of the mixture and is given by
\begin{equation}
\tensor{T}_I = \sum\limits_{a=1}^{N}\tensor{T_a}.
\end{equation}

\subsection{Energy balance and second axiom of thermodynamics}
\label{subsec:Energy 2nd law}
Following \cite{bowenbook}, we choose the energy balance equation for the mixture as
\begin{equation}
\rho\dot{\epsilon} = \tr(\tensor{T}\tensor{L}) - \Div[\bv{x}]{\bv{q}} + \rho r + \sum\limits_{a=1}^N\rho_a\tilde{\bv{v}}_a\cdot\bv{b}_a,
\end{equation}
along with the constraint,
\begin{equation}
0 = \sum\limits_{a=1}^N[\hat{\epsilon}_a+\tilde{\bv{v}}_a\cdot\hat{\bv{p}}_a+\hat{c}_a(\epsilon_a+\tfrac{1}{2}\tilde{\bv{v}}_a^2)],
\end{equation}
where $\epsilon_a$ and $\hat{\epsilon}_a$ represent the internal energy and energy supply per unit volume of the component $a$, respectively. The quantities
\begin{equation*}
\epsilon = \sum\limits_{a=1}^N(\epsilon_a+\frac{1}{2\rho}\rho_a\tilde{v}_a^2),\quad \bv{q},\quad\text{and}\quad r
\end{equation*}
denote the internal energy density, heat flux vector and external heat supply density for the mixture, respectively. The energy equation contributes to the governing equations of the system only when we are considering non-isothermal systems, in which temperature of the system varies. With this in mind, the energy equation also contributes to a framework for the discussion of the constitutive equations based on the entropy production inequality.  We follow the treatment of this inequality in \cite{bowenbook}.  As such, we assume that all constituents of the mixture are at the same temperature $\theta$.  Furthermore, we denote by $\psi_{a}$ the Helmholtz free energy of species $a$ per unit mass of said species.  Then, we also define the vector
\begin{equation}
\label{eq: h def}
\bv{h} = \bv{q} - \sum\limits_{a=1}^N\rho_a(\tensor{K}_a+\tfrac{1}{2}\tilde{v}_a^2\tensor{I})\tilde{\bv{v}}_a
\end{equation}
to represent the mixture energy flux, where $\tensor{K}_a$ is the chemical potential tensor defined by
\begin{equation}
\tensor{K}_a = \psi_a\tensor{I} - \tensor{T}_a/\rho_a.
\label{eq:chemical_potential}
\end{equation}
In addition to $\bv{h}$ and $\tensor{K}_{a}$, we also introduce $\eta_{a}$ as the specific entropy of constituent $a$, as well as the following quantities:
\begin{equation}
\label{eq: more entropy defs}
\eta = \frac{1}{\rho}\sum\limits_{a=1}^N\rho_a\eta_a
\quad\text{and}\quad
\Psi_a = \rho_a\psi_a = \rho_a(\epsilon_a - \theta\eta_a).
\end{equation}
The quantities $\eta$ and $\Psi_{a}$ denote the specific entropy of the mixture and the Helmholtz free energy of species $a$ per unit volume of the mixture.  With the above definitions, we then have the general form of the second axiom of thermodynamics as \cite{bowenbook, bowen1980incompressible}:
\begin{equation}
-\sum\limits_{a=1}^N\grave{\Psi}_a - \rho\eta\dot{\theta} - \sum\limits_{a=1}^N\trace{(\rho_a\tensor{K}_a\tensor{L}_a)} - \frac{\bv{h}\cdot\Grad[\bv{x}]{\theta}}{\theta} - \sum\limits_{a=1}^N\left(\tilde{\bv{v}}_a\cdot\hat{\bv{p}}_a+\tfrac{1}{2}\hat{c}_a\tilde{\bv{v}}_a^2\right) \geq 0,
\label{eq:second_law_general}
\end{equation}
where $\tensor{L}_a$ is the spatial gradient of $\bv{v}_{a}$, i.e.,
\begin{equation}
\tensor{L}_a = \Grad[\bv{x}]{\bv{v}_a}.
\label{eq:L_a_definition}
\end{equation}
Using the definition of $\tilde{\bv{v}}_a$, the last term of inequality~\eqref{eq:second_law_general} can be adjusted so that the inequality can be written as,
\begin{equation}
-\sum\limits_{a=1}^N\grave{\Psi}_a - \rho\eta\dot{\theta} - \sum\limits_{a=1}^N\trace{(\rho_a\tensor{K}_a\tensor{L}_a)} - \frac{\bv{h}\cdot\Grad[\bv{x}]{\theta}}{\theta} - \sum\limits_{a=1}^N(\hat{\bv{p}}_a+\tfrac{1}{2}\hat{c}_a(\bv{v}_a-\bv{v}_s))\cdot(\bv{v}_a-\bv{v}_s) \geq 0.
\label{eq:second_law_final}
\end{equation}

\paragraph*{Mass balance and saturation constraint}
If we add all the component-wise balance of mass equations (given by Eq.~\eqref{eq:mass_balance_component_Eulerian}), and use the saturation constraint given by Eq.~\eqref{eq:saturation}, we get,
\begin{equation}
\sum\limits_{a=1}^N\left[\bv{v}_a\cdot\Grad[\bv{x}]{\phi_a} + \phi_a\Div[\bv{x}]\bv{v}_a - {\hat{\gamma}_a} \right] = 0.
\label{eq:sat_constraint_primary}
\end{equation}
In terms of $\tensor{L}_a$, Eq.~\eqref{eq:sat_constraint_primary} can be re-written as,
\begin{equation}
\sum\limits_{a=1}^N\left[ (\bv{v}_a - \bv{v}_s)\cdot\Grad[\bv{x}]{\phi_a} + \phi_a\tensor{I}\colondot\tensor{L}_a -\hat{\gamma}_a \right] = 0.
\label{eq:constraint}
\end{equation}

\section{Constitutive relations}
\label{sec:Const}
We derive general constitutive relations for a mixture comprising of an elastic solid and $N-1$ fluids, and all the constituents are assumed to be incompressible. We assign mass production rates $(\hat{c}_a)$ to all species. The value of $\hat{c}_a$ is 0 in the case of species which are not chemcially reactive. We consider a mixture which allows for combined effects of elasticity of the solid, heat conduction, diffusion and buoyancy (or hydrodynamic diffusive forces resulting from density gradients). The variables for which we would like to obtain constitutive forms are $\Psi_a,~ \eta,~ \hat{\bv{p}}_a,~ \tensor{T}_a,~ \bv{h}$ and $\hat{\gamma}_a$. We will first derive the form of $\tensor{K}_a$, and then, with the help of Eq.~\eqref{eq:chemical_potential}, we will derive the form for $\tensor{T}_a$.

We define a thermochemical process for a single temperature $\theta$ as a set of $7N+4$ functions whose values are given by
\begin{equation}
\begin{alignedat}{4}
\bv{x} &= \bv{\chi_a}(\bv{X}_a,t), \quad &\bv{b}_a &= \bv{b}_a(\bv{X}_a,t), \quad &\hat{\bv{p}}_a &= \hat{\bv{p}}_a(\bv{X}_a,t), \quad &\Psi_a &= \Psi_a(\bv{X}_a,t),\\
\tensor{T}_a &= \tensor{T}_a(\bv{X}_a,t), &\hat{c}_a &= \hat{c}_a(\bv{X}_a,t), &\phi_a &= \phi_a(\bv{X}_a,t), \quad &\bv{q} &= \bv{q}(\bv{x},t),\\
r &= r(\bv{x},t), &\eta &= \eta(\bv{x},t), &\theta &= \theta(\bv{x},t).
\label{eq:thermochemical_process}
\end{alignedat}
\end{equation}
In writing Eqs.~\eqref{eq:thermochemical_process} we have made use of the relations in Eq.~\eqref{eq:Change_of_frame}.

We require that the constitutive equations be consistent with the second law of thermodynamics, along with relevant balance laws and
the saturation constraint in Eq.~\eqref{eq:constraint}. Enforcing this constraint along with the entropy inequality, inequality~\eqref{eq:second_law_final}, we obtain a variational inequality written in terms of a Lagrange multiplier, $\lambda$,
\begin{multline}
-\sum\limits_{a=1}^N\grave{\Psi}_a - \rho\eta\dot{\theta} - \sum\limits_{a=1}^N\trace{((\rho_a\tensor{K}_a-\lambda\phi_a\tensor{I})\tensor{L}_a)} - \frac{\bv{h}\cdot\Grad[\bv{x}]{\theta}}{\theta} \\
-\sum\limits_{a=1}^N \left[ \left((\hat{\bv{p}}_a-\lambda\Grad[\bv{x}]{\phi_a}) +\tfrac{1}{2}\hat{c}_a(\bv{v}_a-\bv{v}_s)\right)\cdot(\bv{v}_a-\bv{v}_s) + \lambda\hat{\gamma}_a\right] \geq 0,
\label{eq:final_variational_inequality}
\end{multline}
for all admissible thermochemical processes.


Due to the indeterminacy introduced by the saturation constraint, constitutive equations need to be defined for $(\rho_a\tensor{K}_a-\lambda\phi_a\tensor{I})$ and $(\hat{\bv{p}}_a-\lambda\Grad{\phi_a})$ instead of $\tensor{K}_a$ and $\hat{\bv{p}}_a$ respectively. Thus, for the assumptions for the mixture as stated in the beginning of this section, we assume general constitutive forms of several state variables as follows,
\begin{equation}
(\Psi_a, \eta,(\hat{\bv{p}}_a-\lambda\Grad[\bv{x}]{\phi_a}), (\rho_a\tensor{K}_a-\phi_a\lambda\tensor{I}), \bv{h}, \hat{\gamma}_a) =
\bar{f}(\theta,\Grad[\bv{x}]{\theta},\tensor{F}_s,\Grad[\bv{x}]\tensor{F}_s,\phi_b,\Grad[\bv{x}]\bv{\phi}_b,\bv{v}_a).
\label{eq:assumptions1}
\end{equation}
Here, b = 2, 3,\ldots, $N$, the $N-1$ fluid components in the system.  Equation~\eqref{eq:assumptions1} states that we view the quantities on the left of the equal sign to be functions of the kinematic variables listed as arguments of the function $\bar{f}$. In reality, it is well known that, as stated, the dependence implied by the left-hand side of Eq.~\eqref{eq:assumptions1} does not guarantee compatibility with the axiom of material frame-indifference (cf.\ \cite{bowen1969thermochemistry, bowenbook, bowenContinuumBook, bowen1969diffusion}).  We therefore specialize Eq.~\eqref{eq:assumptions1} as follows:
\begin{equation}
\bar{f} = \left(\Psi_a, \eta, \tensor{F}_s\hat{\bv{p}}_a^*, \tensor{F}_s\tensor{K}_a^*\transpose{\tensor{F}_s}, \tensor{F}_s\bv{h}^*, \hat{\gamma}_a  \right) = f^*(\theta,\bv{g},\tensor{C}_s,\tensor{G}_s,\phi_b,\bv{d}_b,\bv{w}_b),
\label{eq:assumptions}
\end{equation}
where, $\bv{g}=\Grad[\bv{X}_s]{\theta} = \transpose{\tensor{F}_s}\Grad[\bv{x}]{\theta}$, $\tensor{G}_s = \Grad[\bv{X}_s]{\tensor{C}_s} = \transpose{\tensor{F}_s}\Grad[\bv{x}]{\tensor{C}_s}$, $\bv{d}_b=\Grad[\bv{X}_s]{\phi_b} = \transpose{\tensor{F}_s}\Grad[\bv{x}]{\phi_b}$ and $\bv{w}_b=\transpose{\tensor{F}_s}(\bv{v}_b-\bv{v}_s)$. The constitutive equations have exactly the same form as in the incompressible and non-reactive case in \cite{bowen1980incompressible}, except that we assume the dependence of $\Psi_a$ on the exact same parameters as the other state variables.
In \cite{ambrosi2010insight} (a mixture theoretic model of soft tissue growth and remodeling), the authors present arguments for
a relatively simple constitutive form for $\Psi_a$, given by $\Psi_a = \Psi_a(\tensor{F}_s,\rho_a)$, $a = 1,2,\ldots,N$.
We rather derive the most general constitutive forms using the assumptions in Eqs.~\eqref{eq:assumptions}, and introduce specializations later.
Also, we define the internal free energy of the mixture as,
\begin{equation}
\Psi_I = \Psi_I(\theta,\bv{g},\tensor{C}_s,\tensor{G}_s,\phi_b,\bv{d}_b,\bv{w}_a) = \sum\limits_{a=1}^{N}\Psi_a.
\end{equation}
Then, following the constitutive assumptions in Eqs.~\eqref{eq:assumptions1}, we have,
\begin{equation}
\begin{split}
\grave{\Psi}_a &= \pdv{\Psi_a}{\theta}\left( \pdv{\theta}{t}+\Grad[\bv{x}]\theta\cdot\bv{v}_a \right) + \pdv{\Psi_a}{\tensor{C}_s}\colondot\left( \pdv{\tensor{C}_s}{t}+\Grad[\bv{x}]\tensor{C}_s\cdot\bv{v}_a \right) \\
&+ \pdv{\Psi_a}{\bv{g}}\cdot\left( \pdv{\bv{g}}{t} + \Grad[\bv{x}]{\bv{g}}\cdot\bv{v}_a \right)
+ C\left[\pdv{\Psi_a}{\tensor{G}_s}\otimes\left( \pdv{\tensor{G}_s}{t}+\Grad[\bv{x}]\tensor{G}_s\cdot\bv{v}_a \right)\right] \\
&+ \sum\limits_{b=2}^N \left[
\pdv{\Psi_a}{\bv{d}_b}\cdot\left( \pdv{\bv{d}_b}{t}+\Grad[\bv{x}]\bv{d}_b\cdot\bv{v}_a \right) + \pdv{\Psi_a}{\phi_b}\left( \pdv{\phi_b}{t}+\Grad[\bv{x}]\phi_b\cdot\bv{v}_a \right) + \pdv{\Psi_a}{\bv{w}_b}\cdot\left( \pdv{\bv{w}_b}{t} + \Grad[\bv{x}]{\bv{w}_b}\cdot{\bv{v}_a} \right) \right]
\label{eq:psi}
\end{split}
\end{equation} 
Based on the assumptions in Eqs.~\eqref{eq:assumptions}, inequality~\eqref{eq:final_variational_inequality} becomes,
\begin{equation}
\begin{split}
&-\pdv{\Psi_I}{\bv{g}}\cdot\dot{\bv{g}} - C\pdv{\Psi_I}{\tensor{G}_s}\otimes\dot{\tensor{G}}_s - \sum\limits_{b=2}^N\pdv{\Psi_I}{\bv{w}_b}\cdot\dot{\bv{w}}_b - \left(\pdv{\Psi_I}{\theta} +\rho\eta \right)\dot{\theta} + \sum_{b=2}^{N}\phi_b\pdv{\Psi_I}{\bv{d}_b}\cdot\Grad[\bv{X}_s]{\tr(\tensor{L}_b)} \\ 
&-\sum\limits_{a=1}^{N}\left(\pdv{\Psi_a}{\bv{g}}\cdot\Grad[\bv{x}]{\bv{g}}+C\sum\limits_{a=1}^{N}\pdv{\Psi_a}{\tensor{G}_s}\otimes\Grad[\bv{x}]{\tensor{G}_s}\right) \cdot\tilde{\bv{v}}_a -\trace{\left[ \left(\tensor{F}_s\left( \tensor{K}_s^*+2\pdv{\Psi_I}{\tensor{C}_s} \right) -\sum\limits_{a=1}^{N}\tilde{\bv{v}}_a\otimes\pdv{\Psi_a}{\bv{w}_b} \right) \transpose{\tensor{F}}_s\tensor{L}_s \right]} \\
&-\sum\limits_{b=2}^{N}\trace{ \left[\left( \tensor{F}_s\tensor{K}_b^*\transpose{\tensor{F}}_s-\phi_b\pdv{\Psi_I}{\phi_b}\tensor{I} - \left(\pdv{\Psi_I}{\bv{d}_b} \cdot\bv{d}_b\right)\tensor{I}- \left(\pdv{\Psi_I}{\bv{d}_b} \otimes\inversetranspose{\tensor{F}_s}\bv{d}_b\right)\transpose{\tensor{F}}_s + \sum\limits_{a=1}^{N}\tilde{\bv{v}}_a\otimes\pdv{\Psi_a}{\bv{w}_b}\transpose{\tensor{F}_s} \right)\tensor{L}_b \right]} \\
&-\frac{\bv{m}\cdot\bv{g}}{\theta} -\sum\limits_{b=2}^N\left[ \hat{\gamma}_b\left( \lambda + \pdv{\Psi_I}{\phi_b} +\frac{\gamma_b}{2}(v_b-v_s)^2 \right)+\pdv{\Psi_I}{\bv{d}_b}\cdot\Grad[\bv{X}_s]{\hat{\gamma}_b}\right] - \lambda{\hat{\gamma}_s} \\
&-\sum\limits_{a=1}^N\left(\hat{\bv{f}}_a - \left(\pdv{\Psi_I}{\bv{d}_a} - \sum\limits_{b=2}^N\pdv{\Psi_a}{\bv{d}_b}\right)\cdot\Grad[\bv{x}]{\bv{d}_a}\right)\cdot(\bv{v}_a-\bv{v}_s) \geq 0,
\label{eq:Var_inequality_final}
\end{split}
\end{equation}
where $C$ denotes the contraction operator as defined in, for example, \cite{bowen1969diffusion} and we have defined $\bv{m}$ and $\hat{\bv{f}}_a$ as,
\begin{equation}
\begin{split}
\bv{m} &= \bv{h} + \theta\sum\limits_{a=1}^N\pdv{\Psi_a}{\theta}\tilde{\bv{v}}_a, \\
\hat{\bv{f}}_a &= \tensor{F}_s\hat{\bv{p}}_a^* - \pdv{\Psi_I}{\phi_a}\Grad[\bv{x}]\phi_a + \pdv{\Psi_a}{\tensor{C}_s}\Grad[\bv{x}]{\tensor{C}_s} + \sum\limits_{b=2}^N\pdv{\Psi_a}{\phi_b}\Grad[\bv{x}]\phi_b,
\end{split}
\label{eq:m_fa}
\end{equation}
and, the following relations hold (as is evident from Eqs.~\eqref{eq:m_fa} and~\eqref{eq:assumptions}),
\begin{equation}
\begin{split}
\bv{m} = \tensor{F}_s\bv{m}^*(\theta,\bv{g},\tensor{C}_s,\tensor{G}_s,\phi_b,\bv{d}_b,\bv{w}_a),\\
\hat{\bv{f}}_a = \tensor{F}_s\hat{\bv{f}}_a^*(\theta,\bv{g},\tensor{C}_s,\tensor{G}_s,\phi_b,\bv{d}_b,\bv{w}_a).
\end{split}
\end{equation}
The inequality~\eqref{eq:Var_inequality_final} needs to hold for all admissible thermochemical processes, for all values of  $\dot{\bv{g}},~\dv{\tensor{G}_s}{t},~\dv{\bv{v}_b}{t},~\tensor{L}_a,~\Grad[\bv{X}_s]{\tr(\tensor{L}_a)},~\dot{\theta}$ and $\Grad[\bv{x}]{\bv{d}_a}$, holding values of all other independent variables fixed. Hence, $\forall~b = 2,\ldots,N-1$, we have
\begin{subequations}
\begin{gather}
\label{eq:Variable_elimination_1standlast}
\pdv{\Psi_I}{\bv{g}} = \bv{0},\quad
\pdv{\Psi_I}{\tensor{G}_s} = \tensor{0},\quad
\pdv{\Psi_I}{\bv{v}_b} = \bv{0}, \quad
\pdv{\Psi_I}{\bv{d}_b}\phi_b = \bv{0}, \quad 
\pdv{\Psi_I}{\theta} = -\rho\eta,\quad
\pdv{\Psi_I}{\bv{d}_a} - \sum\limits_{b=2}^N\pdv{\Psi_a}{\bv{d}_b} = \bv{0},
\\
\sum\limits_{a=1}^{N}\left(\pdv{\Psi_a}{\bv{g}}\cdot\Grad[\bv{x}]{\bv{g}}+C\sum\limits_{a=1}^{N}\pdv{\Psi_a}{\tensor{G}_s}\otimes\Grad[\bv{x}]{\tensor{G}_s}\right)\cdot\tilde{\bv{v}}_a =0, ~\forall~\Grad[\bv{x}]{\bv{g}},~\forall~\Grad[\bv{x}]{\tensor{G}_s},
\\
\tensor{K}_s^*
=-2\pdv{\Psi_I}{\tensor{C}_s} + \tensor{F}_s^{-1}\sum\limits_{a=1}^{N}\tilde{\bv{v}}_a\otimes\pdv{\Psi_a}{\bv{w}_b}, \label{eq:Ks_general_relation}
\\
\tensor{F}_s\tensor{K}_b^*\transpose{\tensor{F}_s} = \phi_b\pdv{\Psi_I}{\phi_b}\tensor{I} 
+ \left(\pdv{\Psi_I}{\bv{d}_b} \cdot\bv{d}_b\right)\tensor{I} + \pdv{\Psi_I}{\bv{d}_b} \otimes\inversetranspose{\tensor{F}_s}\bv{d}_b -\sum\limits_{a=1}^{N}\tilde{\bv{v}}_a\otimes\pdv{\Psi_a}{\bv{w}_b}\transpose{\tensor{F}_s}.
\label{eq:Kb_general relation}
\end{gather}
\end{subequations}
Substituting these results in inequality~\eqref{eq:Var_inequality_final}, we have
\begin{equation}
-\frac{\bv{m}\cdot\Grad[\bv{x}]{\theta}}{\theta} - \sum\limits_{a=1}^N\hat{\bv{f}}_a\cdot(\bv{v}_a-\bv{v}_s) - \sum\limits_{b=2}^N\left[\hat{\gamma}_b\left( \lambda + \pdv{\Psi_I}{\phi_b} +\frac{\gamma_b}{2}(v_b-v_s)^2 \right) + \pdv{\Psi_I}{\bv{d}_b}\cdot\Grad[\bv{X}_s]{\hat{\gamma}_b}\right] - \lambda{\hat{\gamma}_s} \geq 0.
\label{eq:Variational_Inequality_ReducedMax}
\end{equation}
The first four of Eqs.~\eqref{eq:Variable_elimination_1standlast}
also show that
\begin{equation}
\Psi_I = \Psi_I(\theta, \tensor{C}_s, \phi_b, \bv{d}_b),
\end{equation}
so the dependence of $\Psi_I$ on other independent variables is eliminated. Another important observation, stemming from inequality~\eqref{eq:Variational_Inequality_ReducedMax}, is that the presence of volume fraction (or density) gradients can cause production of entropy, unlike the non-reactive case presented in \cite{bowen1980incompressible}. This phenomenon is also observed in the case of reactive mixture of elastic materials, \cite{bowen1969thermochemistry}.


From Eqs.~\eqref{eq:Ks_general_relation}, \eqref{eq:Kb_general relation}, and~\eqref{eq:chemical_potential}, it is easy to see that the partial stress tensors for the solid and fluids, $\tensor{T}_s$ and $\tensor{T}_a$, $a = 2,\ldots,N$, and the total internal stress tensor, $\tensor{T}_I$ are given by,
\begin{subequations}
\begin{align}
\tensor{T}_s &= (\Psi_s-\phi_s\lambda)\tensor{I} + 2\tensor{F}_s\pdv{\Psi_I}{\tensor{C}_s}\transpose{\tensor{F}_s}, \label{eq:T_s_derived}\\
\tensor{T}_b &= \left[\Psi_b - \phi_b\left(\lambda+\pdv{\Psi_I}{\phi_b}\right) -
\pdv{\Psi_I}{\bv{d}_b}\cdot\bv{d}_b\right]\tensor{I} - \pdv{\Psi_I}{\bv{d}_b}\otimes\inversetranspose{\tensor{F}_s}\bv{d}_b\cdot\transpose{\tensor{F}_s} + \sum\limits_{a=1}^{N}\tilde{\bv{v}}_a\otimes\pdv{\Psi_a}{\bv{w}_b}\transpose{\tensor{F}_s}, \label{eq:T_a_derived_1}\\
\tensor{T}_I &= (\Psi_I-\lambda)\tensor{I} + 2\tensor{F}_s\pdv{\Psi_I}{\tensor{C}_s}\transpose{\tensor{F}_s} - \sum\limits_{b=2}^{N}\left[\left(\phi_b\pdv{\Psi_I}{\phi_b}-\pdv{\Psi_I}{\bv{d}_b}\cdot\bv{d}_b\right)\tensor{I} + \left(\pdv{\Psi_I}{\bv{d}_b}\otimes\inversetranspose{\tensor{F}_s}\bv{d}_a\right)\transpose{\tensor{F}_s}\right], \label{eq:T_I_derived_1}
\end{align}
\end{subequations}
where we have used the result: $\sum\limits_{b=2}^N\sum\limits_{a=1}^{N}\tilde{\bv{v}}_a\otimes\pdv{\Psi_a}{\bv{w}_b}\transpose{\tensor{F}_s} = \tensor{0}$, the proof of which can be found in, for example, \cite{bowen1969diffusion}.

\subsection{Simplified modelling assumptions}
\label{subsec:Simplified Model}
The constitutive relations presented so far are the most general form of constitutive equations that can be derived under the assumptions in Eqs.~\eqref{eq:assumptions} for flow through porous, chemically degrading media. In this study, we focus on a simplified form of constitutive equations, and the finite element based numerical implementation of the resultant model. As a first step of simplification, we ignore the dependence of the partial free energy on gradients and higher order derivatives of the primary variables, and on the component velocities. That is, we adopt the simplification proposed by \cite{bowen1980incompressible}, so that we assume $\Psi_a = \Psi_a(\theta, \tensor{C}_s, \phi_b)$ $(b = 2, 3,\ldots, N)$. Under these assumptions we see that Eqs.~\eqref{eq:T_a_derived_1} and~\eqref{eq:T_I_derived_1} simplify to,
\begin{align}
\tensor{T}_b &= \left[\Psi_b - \phi_b\left(\lambda+\pdv{\Psi_I}{\phi_b}\right)\right]\tensor{I}, \label{eq:T_a_simplified}\\
\tensor{T}_I &= (\Psi_I-\lambda)\tensor{I} + 2\tensor{F}_s\pdv{\Psi_I}{\tensor{C}_s}\transpose{\tensor{F}_s} - \sum\limits_{b=2}^{N}\phi_b\pdv{\Psi_I}{\phi_b}, \label{eq:T_I_simplified}
\end{align}
while the constitutive form for the partial stress tensor for the solid is still given by Eq.~\eqref{eq:T_s_derived}. Thus, the constitutive forms of stress tensors in the reactive and non-reactive case are exactly the same. Following \cite{bowen1980incompressible}, we see that the internal stress tensor can be re-written in terms of the free energy per unit volume of the undeformed solid, $W_I(\theta,\tensor{C}_s,J_s\phi_b) = J_s\Psi_I$, as
\begin{equation}
\tensor{T}_I = -\lambda\tensor{I} + \tensor{T}^e,
\label{eq:T_I_Ref_Conf}
\end{equation}
where the tensor $\tensor{T}^e$ is given by,
\begin{equation}
\tensor{T}^e = \frac{2}{J_s}\tensor{F}_s\pdv{W_I}{\tensor{C}_s}\transpose{\tensor{F}_s}.
\end{equation}
Hence, inequality~\eqref{eq:Variational_Inequality_ReducedMax} reduces to
\begin{equation}
-\frac{\bv{m}\cdot\Grad[\bv{x}]{\theta}}{\theta} - \sum\limits_{a=1}^N\hat{\bv{f}}_a\cdot(\bv{v}_a-\bv{v}_s) - \sum\limits_{b=2}^N\hat{\gamma}_b\left( \lambda + \pdv{\Psi_I}{\phi_b} +\frac{\gamma_b}{2}(v_b-v_s)^2 \right) - \lambda{\hat{\gamma}_s} \geq 0. \label{eq:Variational_Inequality_simplified}
\end{equation}

\paragraph*{Pore pressure:}
Stress tensors in fluids moving through porous media are typically written so as to include a component called \emph{pore pressure}. 
A mixture theoretic perspective of pore pressure can be found in \cite{bowen1980incompressible} and \cite{liu2014solid}. In the current context, the pore pressure for fluid $b$ can be defined as,
\begin{equation}
p_b = \lambda + \pdv{\Psi_I}{\phi_b}.
\label{eq:Pore_pressure}
\end{equation} 
This pressure consists of two components, a hydrostatic or base pressure, given by $p_{h} = \lambda$ and a component-specific pressure, given by $p_{b_c} = \pdv{\Psi_I}{\phi_b}$, so that for all fluids $b$,
\begin{equation}
p_b = p_h + p_{b_c}.
\end{equation}
Also, only the hydrostatic component of pressure acts on the solid, so that for the solid pressure, $p_s$ we can write,
\begin{equation}
p_s = \lambda = p_h.
\end{equation}
Thus, we see that the hydrostatic component appears in the stress tensor of all the components in the mixture, including the solid, whereas $p_{b_c}$ appears only in the stresses in fluid components.

In our current model, we will be assuming that all the fluids are equally wetting and that the pores are perfectly saturated with the liquids. 
Hence, we assume that capillary effects can be ignored, so that $p_{b_c} \approx 0$, which gives, 
\begin{equation}
p_b \approx p = p_s,
\label{eq:p_Simplified}
\end{equation}
for $b = 2, 3,\ldots, N$. In other words, we assume that the contribution of hydrostatic pressure (which acts on the solid as well as all the fluids) in the pore pressure is more significant than the contribution of the component-specific part. This condition is also achieved when we are considering a mixture of miscible fluids, for which interfacial fluid-fluid forces are absent.

Substituting Eqs.~\eqref{eq:m_fa}, \eqref{eq:T_a_simplified}, \eqref{eq:Pore_pressure}, and~\eqref{eq:p_Simplified} into Eq.~\eqref{eq:momentum_original_a}, we get,
\begin{equation}
\rho_b\left( \pdv{\bv{v}_b}{t}+\Grad[\bv{x}]{\bv{v}_b}\cdot\bv{v}_b \right) =  -\phi_b\Grad[\bv{x}]{p} + \pdv{\Psi_b}{\theta}\Grad[\bv{x}]{\theta} + \rho_b\bv{b}_b + \hat{\bv{f}}_b,
\end{equation}
for all fluid components, $b = 2, 3,\ldots, N$. Similarly, the balance of momentum equation for the solid and the mixture can be obtained from Eqs.~\eqref{eq:momentum_original_a}, \eqref{eq:Momentum_balance_mixture_useful}, \eqref{eq:T_s_derived}, \eqref{eq:T_I_simplified}, and~\eqref{eq:p_Simplified}.

\paragraph*{Equilibrium and linearization assumptions:}
With the definition of pore pressure (Eq.~\eqref{eq:Pore_pressure}), the inequality~\eqref{eq:Variational_Inequality_simplified} shows that the state given by 
\begin{equation}
\begin{split}
\Grad[\bv{x}]{\theta} &= \bv{0},\\
\bv{w}_b &= \bv{0}, \quad b = 2, 3,\ldots, N 
\end{split}
\label{eq:constraint for equilibrium 1}
\end{equation}
and
\begin{equation}
\begin{rcases}
&p_b = 0, \quad b = 2, 3,\ldots, N \\
&p_s =0,
\end{rcases}
\text{(with assumption in Eq.~\eqref{eq:p_Simplified}: $p=0$),}
\label{eq:constraint for equilibrium 2}
\end{equation}
corresponds to the state of thermodynamic equilibrium. In the absence of chemical reactions $(\hat{\gamma_b}\equiv 0)$, this system reduces to that described by Bowen, \cite{bowen1980incompressible}. A state of equilibrium for a non-reactive porous medium flow exists simply when the conditions in Eqs.~\eqref{eq:constraint for equilibrium 1} are satisfied, and the additional constraint in Eq.~\eqref{eq:constraint for equilibrium 2} is essential only when chemical reactions are involved. Thus, the conditions in Eqs.~\eqref{eq:constraint for equilibrium 1} correspond to the state of thermo-mechanical equilibrium, and the addition of the constraint in Eq.~\eqref{eq:constraint for equilibrium 2} corresponds to the state of chemo-thermo-mechanical, or thermodynamic equilibrium of the system.

At the state of thermodynamic equilibrium, we observe that the following conditions are satisfied,
\begin{align}
\bv{m}^*(\theta,\bv{0},\tensor{C}_s,\tensor{G}_s,\phi_b,\bv{d}_b,\bv{0}) &= \bv{0}, \label{eq:equilibrium1}\\
\hat{\bv{f}}_a^*(\theta,\bv{0},\tensor{C}_s,\tensor{G}_s,\phi_b,\bv{d}_b,\bv{0}) &= \bv{0}.
\label{eq:equilibrium2}
\end{align}
We assume that departures from the equilibrium state are small, so that the linearized assumptions for $\bv{m}$ and $\hat{\bv{f}}_a$ presented in \cite{bowen1980incompressible} hold. We observe that the balance of momentum equations in the reactive and non-reactive cases are exactly the same, except for the addition of the term $-\sum\limits_{a=1}^N\hat{c}_a\bv{v}_a$ to the momentum balance equation for the mixture, as seen from Eq.~\eqref{eq:Momentum_balance_mixture_useful}. Also, since the constitutive forms of $\hat{\bv{f}}_a$, and in turn of $\hat{\bv{p}}_a$ as given by Eq.~\eqref{eq:m_fa} have been evaluated only for $a = 2,\ldots, N$, the interaction force for the solid, $\hat{\bv{p}}_s$ will be evaluated using the last of Eqs.~\eqref{eq:mixture_pa_ca_constraint}.

We now introduce our final modelling assumption, that all the processes (chemical reactions, deformations etc.) in our mixture are isothermal. Moreover, we also assume that the mixture is in thermal equilibrium, which means that all the components in our mixture are at a constant, uniform temperature $\theta$ at all times. Thus, we can disregard the energy equation and the constitutive restrictions imposed on $\bv{m}$.

\subsection{Balance of momentum equations}
Under the assumptions stated so far, the Eulerian form of momentum balance equations for the fluids, solid and mixture are given, respectively by,
\begin{subequations}
\begin{align}
\rho_b\left( \pdv{\bv{v}_b}{t}+\Grad[\bv{x}]{\bv{v}_b}\cdot\bv{v}_b \right) &= -\phi_b\Grad[\bv{x}]{p} + \rho_b\bv{b}_b - \phi_b^2\frac{\mu_b}{\kappa_s}(\bv{v}_b-\bv{v}_s), \\ 
\rho_s\left( \pdv{\bv{v}_s}{t}+\Grad[\bv{x}]{\bv{v}_s}\cdot\bv{v}_s \right) &= -\phi_s\Grad[\bv{x}]{p} + \rho_s\bv{b}_s + \sum\limits_{b=2}^N\phi_b^2\frac{\mu_b}{\kappa_s}(\bv{v}_b-\bv{v}_s) + \Div[\bv{x}]\tensor{T^e} - \sum\limits_{a=1}^N\hat{c}_a\bv{v}_a  
\\
\sum\limits_{a=1}^N\rho_a\left( \pdv{\bv{v}_a}{t}+\Grad[\bv{x}]{\bv{v}_a}\cdot\bv{v}_a \right) &= -\Grad[\bv{x}]{p} + \rho\bv{b} + \Div[\bv{x}]\tensor{T}^e - \sum\limits_{a=1}^N\hat{c}_a\bv{v}_a.
\end{align}
\end{subequations}

\paragraph*{Qausi-static approximation:} 
We are interested in applications wherein the accelerations and thus, inertia effects are negligible - which we call quasi-static processes. Hence, for such applications, the momentum conservation equations reduce to,
\begin{subequations}
\begin{align}
\bv{0} &= -\phi_b\Grad[\bv{x}]{p} + \gamma_b\phi_b\bv{b}_b - \phi_b\frac{\mu_b}{\kappa_s}\bv{v}_{\text{fil}_b},
\label{eq:Eulerian_Momentum_fluid_quasistatic}
\\
\bv{0} &= -\phi_s\Grad[\bv{x}]{p} + \gamma_s\phi_s\bv{b}_s + \sum\limits_{b=2}^N\phi_b\frac{\mu_b}{\kappa_s}\bv{v}_{\text{fil}_b} + \Div[\bv{x}]\tensor{T}^e - \sum\limits_{a=1}^N\hat{c}_a\bv{v}_a  
\label{eq:Eulerian_Momentum_solid_quasistatic}
\\
\bv{0} &= -\Grad[\bv{x}]{p} + \Div[\bv{x}]\tensor{T}^e + \sum\limits_{a=1}^N(\gamma_a\phi_a\bv{b}_a - \hat{c}_a\bv{v}_a).
\label{eq:Eulerian_Momentum_mixture_quasistatic}
\end{align}
\end{subequations}
We have used Eqs.~\eqref{eq:phi_gamma_rho_relation} and~\eqref{eq:filtration_vel_definition} in writing Eqs.~\eqref{eq:Eulerian_Momentum_fluid_quasistatic} and~\eqref{eq:Eulerian_Momentum_solid_quasistatic}. Equations~\eqref{eq:Eulerian_Momentum_fluid_quasistatic}--\eqref{eq:Eulerian_Momentum_mixture_quasistatic} can be written in the \ac{ALE} framework as,
\begin{subequations}
\begin{align}
\bv{0} &= J_s\left(-\phi_b\inversetranspose{\tensor{F}_s}\Grad[\bv{X}_s]{p} + \gamma_b\phi_b\bv{b}_b - \phi_b\frac{\mu_b}{\kappa_s}\bv{v}_{\text{fil}_b}\right),
\label{eq:ALE_Momentum_fluid_quasistatic}
\\
\bv{0} &= -J_s\phi_s\inversetranspose{\tensor{F}_s}\Grad[\bv{X}_s]{p} + J_s\gamma_s\phi_s\bv{b}_s + J_s\sum\limits_{b=2}^N\phi_b\frac{\mu_b}{\kappa_s}\bv{v}_{\text{fil}_b} + \Div[\bv{X}_s]\tensor{P}^e - J_s\sum\limits_{a=1}^N\frac{\hat{c}_a}{\phi_a}\bv{v}_{\text{fil}_a}  
\label{eq:ALE_Momentum_solid_quasistatic}
\\
\bv{0} &= -J_s\inversetranspose{\tensor{F}_s}\Grad[\bv{X}_s]{p} + \Div[\bv{X}_s]\tensor{P}^e + J_s\sum\limits_{a=1}^N\left(\gamma_a\phi_a\bv{b}_a - \frac{\hat{c}_a}{\phi_a}\bv{v}_{\text{fil}_a}\right),
\label{eq:ALE_Momentum_mixture_quasistatic}
\end{align}
\end{subequations}
where $\tensor{P}^e = J_s\tensor{T}^e\inversetranspose{\tensor{F}}_s$, the first Piola-Kirchhoff elastic stress tensor. For our analysis, we assume the free energy of the mixture to be given by,
\begin{equation}
\Psi_I = \phi_s W_s(\tensor{C}_s,t),
\end{equation}
where $W_s$ is the free energy of the pure solid in the current configuration. Assuming the solid to be hyper-elastic, $\tensor{T}^e$ and $\tensor{P}^e$ are then given by,
\begin{equation}
\tensor{T}^e = 2\phi_s\tensor{F}_s\pdv{W_s}{\tensor{C}_s}\transpose{\tensor{F_s}},
\label{eq:Constitutive form_Solid stress}
\end{equation}

\begin{equation}
\tensor{P}^e = 2J_s\phi_s\tensor{F}_s\pdv{W_s}{\tensor{C}_s}.
\label{eq:Piola_constitutive form}
\end{equation}
We model our problem based on the degradation of biopolymers. In particular, we follow the hydrolytic degradation of polyesters as described in \cite{Hydrolysis_chemistry} to model our problem. The reduced hydrolytic degradation reaction as described in \cite{Hydrolysis_chemistry} involves three constituents: the polymeric solid, the product of degradation (monomer) and water, which is present in abundance in the system and acts as a base fluid. Thus, we model our numerical problem as a three-component system comprising of these three components. We assume the degradation product to have a fluid-like behavior, and will be referred to as the `monomeric fluid' in the rest of the paper. 
Following the notation used so far, the properties pertaining to the solid, monomeric and base fluid will be denoted by the subscripts $s$, 2 and 3 respectively.

\section{Finite Element Implementation}
\label{sec:FEM}
In this section we present a \ac{FEM} based numerical framework, based on the reduced mixture theoretic model derived in the previous \hyperref[subsec:Simplified Model]{section}. Also, due to the saturation condition, Eq.~\eqref{eq:saturation}, the volume fraction of one of the components in the mixture can be evaluated if the volume fractions of all other components in the system are known. Hence, we eliminate the base fluid volume fraction, $\phi_3$, from the set of primary variables and for each time step, evaluate $\phi_3$ from Eq.~\eqref{eq:saturation} and solutions for $\phi_s$ and $\phi_2$. As is customary and for the ease of application of boundary conditions, we resort to the \ac{ALE} framework.

\subsection{Intial and boundary conditions and \ac{ALE} strong form}
\subsubsection{Initial conditions.}
\label{para:initial}
The initial conditions for the density and velocity fields are set to
\begin{equation}
\phi_a(\bv{x},0) = \phi_a^0(\bv{x}), \quad \bv{u}_s(\bv{x},0) = \bv{u}_s^0(\bv{x}), \quad \bv{v}_{\text{fil}_b}(\bv{x},0) = \bv{v}^{0}_{\text{fil}_b}(\bv{x}),
\label{eq:initial_conditions}
\end{equation}
for $a = s,~2$ and $b = 2,~3$, $\forall~\bv{x}\in \Omega_t$. Also, $\bv{v}_s^0 = \dv{\bv{u}^g_s}{t}\big|_{\bv{x}~\text{fixed}}$.
\subsubsection{Boundary conditions.}
\label{para:boundary}
Referring to Fig. \ref{fig:configuratons}, the boundary $\partial\Omega_t$ is partitioned into subsets $\left(\partial\Omega_t\right)^D_d$ and $\left(\partial\Omega_t\right)^N_d$, such that $\left(\partial\Omega_t\right)^D_d \cap \left(\partial\Omega_t\right)^N_d = \emptyset$ and $\left(\partial\Omega_t\right)^D_d \cup \left(\partial\Omega_t\right)^N_d = \partial\Omega_t$, where $d = \phi_s$, $\phi_2$, and $\bv{u}_s$. The superscripts $D$ and $N$ stand for `Dirichlet' and `Neumann,' respectively. 
For all $t>0$, the following boundary conditions are then admitted for our problem,
\begin{alignat}{2}
\phi_d(\bv{x},t) &= \phi_d^g(\bv{x},t),
&&\quad\text{for $\bv{x} \in \left(\partial\Omega_t\right)_{\phi_d}^D$}, \label{eq:Dirichlet_volume fraction}
\\
\bv{u}_s(\bv{x},t) &= \bv{u}_s^g(\bv{x},t), &&\quad\text{for $\bv{x} \in \left(\partial\Omega_t\right)_{\bv{u}_s}^D$},
\label{eq:Dirichlet_us}
\\
\tensor{T}_s(\bv{x},t)\bv{n}_s(\bv{x},t) &= \bv{s}^g(\bv{x},t), &&\quad\text{for $\bv{x} \in \left(\partial\Omega_t\right)_{\bv{u}_s}^N$}, \label{eq:Neumann_us}
\end{alignat}
where
$d = s, 2$ 
and where
$\phi_d^g$, $\bv{u}_s^g$, and $\bv{s}^g$ are the prescribed volume fractions, solid displacement, and traction distribution for the solid.  The boundary conditions involving the filtration velocity will be discussed in the next paragraph.  Finally, we note that
\begin{equation}
\label{eq: total traction definition}
\tensor{T}_s \coloneqq -p \tensor{I} + \tensor{T}^{e}
\end{equation}
represents the total stress in the mixture.
As mentioned earlier, our final formulation is written in terms of an \ac{ALE} framework, for which it is essential to pull back all equations and boundary conditions to the solid reference configuration. In this context, we also partition the boundary $\partial\Omega_s$ into $\left(\partial\Omega_s\right)^D_d$ and $\left(\partial\Omega_s\right)^N_d$, such that these partitions are simply obtained by pulling back the corresponding partitions $\left(\partial\Omega_t\right)^D_d$ and $\left(\partial\Omega_t\right)^N_d$ to the solid reference frame. The corresponding boundary values can then be assigned onto these boundary partitions on the basis of Eqs.~\eqref{eq:inverse_diffeomorphism} and~\eqref{eq:Change_of_frame}.

\paragraph*{Boundary conditions for the filtration velocity:}
\label{para:impermeability boundary condition}
As is well-known (cf., e.g., \cite{Masud_darcy_stabilization,Masud2007A-Stabilized-Mixed-0}), equations like those governing Darcy flow do not support the specification of boundary values for the filtration velocity.  The only control on boundary values of the filtration velocity pertains to its normal component.  Hence, the boundary condition involving the filtration velocity is as follows: for all $t>0$
\begin{equation}
\bv{v}_{\text{fil}_b}\cdot\bv{n}_s = {v}_{\text{fil}_{b}}^n(\bv{x},t) \quad \text{for $\bv{x} \in (\partial\Omega_s)_{\bv{v}_{\text{fil}_b}}^{D_n}$},
\label{eq:impermeability}
\end{equation}
where $(\partial\Omega_s)_{\bv{v}_{\text{fil}_b}}^{D_n} \cup (\partial\Omega_s)_{\bv{v}_{\text{fil}_b}}^{N} = \partial\Omega_s$, and where ${v}_{\text{fil}_{b}}^n$ is a prescribed scalar function.  If the boundary is impermeable, then ${v}_{\text{fil}_{b}}^n = 0$.  With a slight abuse of terminology, we will still categorize the boundary condition in Eq.~\eqref{eq:impermeability} as being a Dirichlet boundary condition limited to the normal component of the velocity.  The weak implementation of Eq.~\eqref{eq:impermeability} will be discussed in the following section.  Finally, when it comes to the admissible traction boundary conditions pertaining to the filtration velocity equations, said traction can only act along the normal to the boundary.

\paragraph*{Pressure constraint:}
\label{para:pressure boundary condition and constraint}
Similar to incompressible flow problems (cf.\ \cite{Masud_darcy_stabilization,Masud2007A-Stabilized-Mixed-0,costanzo2017arbitrary}),
under pure Dirichlet velocity boundary conditions the pressure field must be constrained in order to obtain a unique solution (the pressure solution is unique up to an additive constant). We therefore adopt the following (standard) constraint on the pressure:
\begin{equation}
\int\limits_{\Omega_s}p\dd{\Omega_s} = 0.
\label{eq:pressure global constraint}
\end{equation}

Thus, two types of boundary conditions are investigated in our work:
\begin{enumerate}
\item Dirichlet boundary condition for $\bv{u}_s$ on all boundaries along with the above pressure constraint.

\item Neumann boundary condition for the mixture in which the boundary traction shown in Eq.~\eqref{eq:Neumann_us} is prescribed on a part of the domain boundary and a Dirichlet boundary condition is prescribed on the remaining boundary.
\end{enumerate}
For both the boundary conditions, Eq.~\eqref{eq:impermeability} for the filtration velocity is enforced on the entire boundary (i.e. $(\partial\Omega_s)_{\bv{v}_{\text{fil}_b}}^{D_n} \equiv \partial\Omega_s$).

\subsubsection{Strong form, Eulerian framework}
\label{subsubsection:Eulerian Strong form}
In an Eulerian framework, the strong form of the problem we consider is: Given the 
\begin{itemize}
\item body force fields, $\bv{b}_s, \bv{b}_2, \bv{b}_3 : \Omega_s \cross [0,T] \rightarrow \mathscr{T}^d$, pure component densities, $\gamma_s$, $\gamma_2$, $\gamma_3 \in \mathbb{R}^+$;
\item prescribed boundary values, $\phi_s^g : \left(\partial\Omega_t\right)_{\phi_a}^{D} \cross (0,T] \rightarrow [0,1]$, 
$\phi_2^g : \left(\partial\Omega_t\right)_{\phi_2}^{D} \cross (0,T] \rightarrow [0,1]$,
$\bv{u}_s^g : \left(\partial\Omega_t\right)_{\bv{v}_s}^{D} \cross (0,T] \rightarrow \mathscr{T}^d$,
$\bv{s}^g : \left(\partial\Omega_t\right)_{\bv{v}_s}^{N} \cross (0,T] \rightarrow \mathscr{T}^d$,
$v_{\text{fil}_2}^n : (\partial\Omega_t)_{\bv{v}_{\text{fil}_2}}^{D_n} \cross (0,T] \rightarrow \mathbb{R}$,
$v_{\text{fil}_3}^n : (\partial\Omega_t)_{\bv{v}_{\text{fil}_3}}^{D_n} \cross (0,T] \rightarrow \mathbb{R}$,
\item prescribed initial conditions, $\phi_s^0(\bv{x}) : \Omega_0 \rightarrow [0,1]$, $\phi_2^0(\bv{x}) : \Omega_0  \rightarrow [0,1]$, $\bv{u}_s^0(\bv{x}) : \Omega_0 \rightarrow \mathscr{T}^d$, $\bv{v}_{\text{fil}_2}(\bv{x}) : \Omega_0 (\bv{x}) \rightarrow  \mathscr{T}^d$,
$\bv{v}_{\text{fil}_3}(\bv{x}) : \Omega_0 \rightarrow  \mathscr{T}^d$,
\item and constitutive equation for $\tensor{T}_e$, Eq.~\eqref{eq:Constitutive form_Solid stress},
\end{itemize}
find $\phi_s : \Omega_t \cross (0,T] \rightarrow [0,1]$, $\phi_2 : \Omega_t \cross (0,T] \rightarrow (0,T]$, $\bv{u}_s^0 : \Omega_t \cross (0,T] \rightarrow \mathscr{T}^d$, $\bv{v}_s^0 : \Omega_t \cross (0,T] \rightarrow \mathscr{T}^d$, $\bv{v}_{\text{fil}_1} : \Omega_t \cross [0,T] \rightarrow \mathscr{T}^d$, $\bv{v}_{\text{fil}_2} : \Omega_t \cross (0,T] \rightarrow \mathscr{T}^d$  and $p: \Omega_t  \cross (0,T] \rightarrow \mathbb{R}$ (where $\Omega_0 \equiv \Omega_t|_{t=0} \equiv \Omega_t$), \\
such that, $\forall~\bv{X}_s \in \Omega_t$ and $\forall ~t \in [0,T]$, Eqs.~\eqref{eq:Eulerian_mass} and~\eqref{eq:Eulerian_Momentum_fluid_quasistatic} are satisfied for $a = s, 2$ and $b = 2, 3$ along with Eqs.~\eqref{eq:Eulerian_mass_mixture},  \eqref{eq:Eulerian_Momentum_mixture_quasistatic}, \eqref{eq:saturation}, and the initial conditions (Eqs.~\eqref{eq:initial_conditions}) as well as the prescribed boundary conditions: Eq.~\eqref{eq:Dirichlet_volume fraction} and Eq.~\eqref{eq:impermeability}, with either
\begin{itemize}
\item Eqs.~\eqref{eq:Neumann_us} and~\eqref{eq:Dirichlet_us}, or, 
\item Eq.~\eqref{eq:Dirichlet_us} and the constraint in Eq.~\eqref{eq:pressure global constraint}.
\end{itemize}

\subsubsection{Strong form, ALE framework.}
\label{subsubsection:ALE strong form}
Referring to Eqs.~\eqref{eq:diffeomorphism}, \eqref{eq:inverse_diffeomorphism}, and~\eqref{eq:Change_of_frame}, we define the map $\bv{\chi}_s: \Omega_s \rightarrow \Omega_t$, so that we have $\bv{x} = \bv{\chi}_s(\bv{X}_s,t)$, then the \ac{ALE} strong form of the problem reads - 
given the
\begin{itemize}
\item body force fields, $\bv{b}_s, \bv{b}_2, \bv{b}_3 : \Omega_s \cross [0,T] \rightarrow \mathscr{T}^d$, pure component densities, $\gamma_s$, $\gamma_2$, $\gamma_3 \in \mathbb{R}^+$;
\item prescribed boundary values, $\phi_s^g : \left(\partial\Omega_t\right)_{\phi_a}^{D} \cross (0,T] \rightarrow [0,1]$, 
$\phi_2^g : \left(\partial\Omega_t\right)_{\phi_2}^{D} \cross (0,T] \rightarrow [0,1]$,
$\bv{u}_s^g : \left(\partial\Omega_t\right)_{\bv{v}_s}^{D} \cross (0,T] \rightarrow \mathscr{T}^d$,
$\bv{s}^g : \left(\partial\Omega_t\right)_{\bv{v}_s}^{N} \cross (0,T] \rightarrow \mathscr{T}^d$,
$v_{\text{fil}_2}^n : (\partial\Omega_t)_{\bv{v}_{\text{fil}_2}}^{D_n} \cross (0,T] \rightarrow \mathbb{R}$,
$v_{\text{fil}_3}^n : (\partial\Omega_t)_{\bv{v}_{\text{fil}_3}}^{D_n} \cross (0,T] \rightarrow \mathbb{R}$,
\item prescribed initial conditions, $\phi_s^0(\bv{x}) : \Omega_0 \rightarrow [0,1]$, $\phi_2^0(\bv{x}) : \Omega_0  \rightarrow [0,1]$, $\bv{u}_s^0(\bv{x}) : \Omega_0 \rightarrow \mathscr{T}^d$, $\bv{v}_{\text{fil}_2}(\bv{x}) : \Omega_0 (\bv{x}) \rightarrow  \mathscr{T}^d$,
$\bv{v}_{\text{fil}_3}(\bv{x}) : \Omega_0 \rightarrow  \mathscr{T}^d$,
\item and constitutive equation for $\tensor{P}_e$, Eq.~\eqref{eq:Piola_constitutive form},
\end{itemize}
find $\phi_s : \Omega_s \cross (0,T] \rightarrow [0,1]$, $\phi_2 : \Omega_s \cross (0,T] \rightarrow (0,T]$, $\bv{u}_s^0 : \Omega_s \cross (0,T] \rightarrow \mathscr{T}^d$, $\bv{v}_s^0 : \Omega_s \cross (0,T] \rightarrow \mathscr{T}^d$, $\bv{v}_{\text{fil}_1} : \Omega_s \cross [0,T] \rightarrow \mathscr{T}^d$, $\bv{v}_{\text{fil}_2} : \Omega_s \cross (0,T] \rightarrow \mathscr{T}^d$  and $p: \Omega_s  \cross (0,T] \rightarrow \mathbb{R}$ (where $\Omega_0 \equiv \Omega_t|_{t=0} \equiv \Omega_s$), \\
such that, $\forall~\bv{X}_s \in \Omega_s$ and $\forall ~t \in [0,T]$, Eqs.~\eqref{eq:ALE_mass_simple} and~\eqref{eq:ALE_Momentum_fluid_quasistatic} are satisfied for $a = s, 2$ and $b = 2, 3$ along with
Eqs.~\eqref{eq:ALE_mass_mixture},  \eqref{eq:ALE_Momentum_mixture_quasistatic}, \eqref{eq:saturation}, and the initial conditions (Eqs.~\eqref{eq:initial_conditions}) as well as the prescribed boundary conditions in Eqs.~\eqref{eq:Dirichlet_volume fraction} and~\eqref{eq:impermeability}, with either
\begin{itemize}
\item Eqs.~\eqref{eq:Neumann_us} and~\eqref{eq:Dirichlet_us}, or,
\item Eq.~\eqref{eq:Dirichlet_us} and the constraint in Eq.~\eqref{eq:pressure global constraint}.
\end{itemize}

\subsection{Functional settings and weak form}
We define the time dependent spaces for the solid displacement, velocities, pressure and volume fractions as,
\begin{equation}
\begin{split}
&\mathcal{V}^{\bv{u}_s} = \{ \bv{u}_s \big| \bv{u}_s
\in L^2(\Omega_t)^{n_d}, \Grad[\bv{x}]{\bv{u}_s} \in L^\infty(\Omega_t)^{n_d\cross n_d}, \bv{u}_s = \bv{u}_s^g \text{~on~} (\partial\Omega_t)_{\bv{v}_s}^D \}, \\
&\mathcal{V}^{\bv{v}_{\text{fil}_b}} = \{ \bv{v}_{\text{fil}_b} \big| \bv{v}_{\text{fil}_b} \in L^2(\Omega_t)^{n_d}, \Div[\bv{x}]{\bv{v}_{\text{fil}_b}} \in L^2(\Omega_t)^{n_d\cross n_d} \}, \\
&\mathcal{V}^p = L^2(\Omega_t), \\
&\mathcal{V}^{\phi_a} = \{ \phi_a \big| \phi_a \in (0,1] \cap L^\infty (\Omega_t), \Grad[\bv{x}]{\phi_a} \in L^2(\Omega_t), \phi_a = \phi_a^g \text{~on~} (\partial\Omega_t)^D_{\phi_a} \},
\label{eq:Functional spaces - trial}
\end{split}
\end{equation}
for $b=2, 3$, $a = s, 1$ and $n_d = 2, 3$ is the dimension of the Euclidean vector space. The corresponding Galerkin-weighting spaces for the variables are then given by,
\begin{equation}
\begin{split}
&\mathcal{V}^{\bv{u}_s}_w = \{ \bv{u}_s \big| \bv{u}_s
\in L^2(\Omega_t)^{n_d}, \Grad[\bv{x}]{\bv{u}_s} \in L^\infty(\Omega_t)^{n_d\cross n_d}, \bv{u}_s = \bv{0} \text{~on~} (\partial\Omega_t)_{\bv{v}_s}^D \}, \\
&\mathcal{V}^{\bv{v}_{\text{fil}_b}}_w = \mathcal{V}^{\bv{v}_{\text{fil}_b}}, \\
&\mathcal{V}^p_w = \mathcal{V}^p, \\
&\mathcal{V}^{\phi_a}_w = \{ \phi_a \big| \phi_a \in (0,1] \cap L^\infty (\Omega_t), \Grad[\bv{x}]{\phi_a} \in L^2(\Omega_t), \phi_a = 0 \text{~on~} (\partial\Omega_t)^D_{\phi_a} \},
\label{eq:Functional spaces - weight}
\end{split}
\end{equation}

\subsubsection*{Weak implementation of the boundary conditions for the filtration velocity.
}
\label{para:Lagrange multipliers for impermeability}
The boundary condition in Eq.~\eqref{eq:impermeability} was implemented using a modified version of the method of Lagrange multipliers, as outlined by Babu\v{s}ka in \cite{Babuska1973}. We introduce an auxiliary boundary variable, $\lambda_b \in H^{\frac{1}{2}}(\partial\Omega_t)$ and its associated test function, $\tilde{\lambda}_b \in H^{\frac{1}{2}}(\partial\Omega_t)$. The impermeability boundary condition for $\bv{v}_{\text{fil}_b}$ using the method of Lagrange multipliers is then given by,
\begin{equation}
\int\limits_{\partial\Omega_t}\left[ \tilde{\lambda}_b(\bv{v}_{\text{fil}_b}\cdot\bv{n} - v^n_{\text{fil}_b}) + \lambda_b(\tilde{\bv{v}}_{\text{fil}_b} \cdot\bv{n})\right]\dd{\Gamma} = 0,
\label{eq:Eulerian_Normal_constraint}
\end{equation}
where $\bv{n}$ is the unit outward normal to the boundary $\partial\Omega_t$. In the \ac{ALE} framework the condition is written as,
\begin{equation}
\int\limits_{\partial\Omega_s}\left[ J_s \tilde{\lambda}_b(\tensor{F}_s^{-1}\bv{v}_{\text{fil}_b}\cdot\bv{n}_s - v^{n_s}_{\text{fil}_b}) + \lambda_b(\tensor{F}_s^{-1}\tilde{\bv{v}}_{\text{fil}_b} \cdot\bv{n})\right]\dd{\Gamma_s} = 0,
\label{eq:ALE_Normal_constraint}
\end{equation}
where $\bv{n}_s$ is the unit outward normal to the boundary $\partial\Omega_s$.
The expected value of the boundary field $\lambda_{b}$ will be discussed after the weak formulation is presented.


\subsubsection{Weak form in the Eulerian framework.}
\label{subsubsection:weak form, Eulerian}
Given the same data as in the strong formulation,
find $\bv{u}_s \in \mathcal{V}^{\bv{u}_s}$, $\bv{v}_{\text{fil}_1} \in \mathcal{V}^{\bv{v}_{\text{fil}_1}}$,
$\bv{v}_{\text{fil}_2} \in \mathcal{V}^{\bv{v}_{\text{fil}_2}}$,
$p \in \mathcal{V}^p$, $\phi_2 \in \mathcal{V}^{\phi_2}$ and $\phi_s \in \mathcal{V}^{\phi_s}$ such that $\forall$ $\tilde{\bv{u}}_s \in \mathcal{V}_w^{\bv{u}_s}$,  $\tilde{\bv{v}}_{\text{fil}_2} \in \mathcal{V}_w^{\bv{v}_{\text{fil}_2}}$,
$\tilde{\bv{v}}_{\text{fil}_3} \in \mathcal{V}_w^{\bv{v}_{\text{fil}_3}}$,
$\tilde{p} \in \mathcal{V}_w^p$, $\tilde{\phi}_2 \in \mathcal{V}_w^{\phi_2}$ and $\tilde{\phi}_s \in \mathcal{V}_w^{\phi_s}$,
\begin{gather}
\label{eq:Weak form_Eulerian - concentrations}
\int\limits_{\Omega_t}
\Bigl[
\tilde{\phi}_2\bigl(
\partial_{t}\phi_2|_{\bv{x}}
+ \Div[\bv{x}]{\bv{v}_{\text{fil}_2}} + \phi_2\Div[\bv{x}]{\bv{v}_s} - \hat{\gamma}_2
\bigr)
+
\tilde{\phi}_s\bigl(
\partial_{t}\phi_s|_{\bv{x}}
+ \phi_s\Div[\bv{x}]{\bv{v}_s} - \hat{\gamma}_s
\bigr)
\Bigr]\dd{\Omega_t} = 0
\\
\label{eq:Weak form_Eulerian - displacement}
\int\limits_{\Omega_t}
\biggl[
\tilde{\bv{u}}_s\cdot
\biggl(
\frac{\hat{c}_2}{\phi_2}
\bv{v}_{\text{fil}_2}
-\sum\limits_{a=1}^N\phi_a\bv{b}_a
\biggr) 
-
(\Div[\bv{x}]{\tilde{\bv{u}}_s}) p
+
\Grad[\bv{x}]{\tilde{\bv{u}}_s} \colondot
\tensor{T}^e
\biggr]\,\dd{\Omega_t}
+
\int\limits_{\left(\partial\Omega_t\right)_{\bv{u}_s}^N}
\tilde{\bv{u}}_s \cdot \bigl(p \bv{n} - \bv{s}^g\bigr)\,\dd{\Gamma} = 0,
\\
\label{eq:Weak form_Eulerian - pressure}
-\int\limits_{\Omega_t}\tilde{p}\left(\Div[\bv{x}]{(\bv{v}_s+\bv{v}_{\text{fil}_2} + \bv{v}_{\text{fil}_3})} - \sum\limits_{a=1}^{N}\hat{\gamma}_a \right)\dd{\Omega_t} = 0,
\\
\label{eq:Weak form_Eulerian - filtration 2}
\begin{multlined}[b]
\int\limits_{\Omega_t}\biggr[\tilde{\bv{v}}_{\text{fil}_2}\cdot
\biggl(\frac{\mu_2}{\kappa_s}\bv{v}_{\text{fil}_2} - \bv{b}_2 \biggr) - (\Div[\bv{x}]{\tilde{\bv{v}}_{\text{fil}_2}}) p \biggr] \dd{\Omega_t}
\\
\qquad\qquad+
\int\limits_{\partial\Omega_t}
\bigl[(\tilde{\bv{v}}_{\text{fil}_2}\cdot\bv{n})(p + \lambda_{2})
+
\tilde{\lambda}_2(\bv{v}_{\text{fil}_2}\cdot\bv{n} - v^n_{\text{fil}_2})
\bigr]\,\dd{\Gamma} = 0,
\end{multlined}
\\
\label{eq:Weak form_Eulerian - filtration 3}
\begin{multlined}[b]
\int\limits_{\Omega_t}\biggl[\tilde{\bv{v}}_{\text{fil}_3}\cdot\biggl(\frac{\mu_3}{\kappa_s}\bv{v}_{\text{fil}_3} - \bv{b}_3 \biggr) - (\Div[\bv{x}]{\tilde{\bv{v}}_{\text{fil}_3}}) p \biggr] \dd{\Omega_t}
\\
\qquad\qquad+
\int\limits_{\partial\Omega_t}
\bigl[
(\tilde{\bv{v}}_{\text{fil}_3}\cdot\bv{n})(p+\lambda_{3})
+
\tilde{\lambda}_3(\bv{v}_{\text{fil}_3}\cdot\bv{n} - v^n_{\text{fil}_3})
\bigr]\,\dd{\Gamma} = 0.
\end{multlined}
\end{gather}
As can be seen in Eqs.~\eqref{eq:Weak form_Eulerian - filtration 3} and~\eqref{eq:Weak form_Eulerian - filtration 3}, the Lagrange multipliers $\lambda_{2}$ and $\lambda_{3}$ enforcing the boundary conditions for the filtration velocities have dimensions of pressure.  This is not surprising as the pressure is the work conjugate of the divergence of said velocities. Following the arguments presented in \cite{Babuska1973}, the expected solution for the $\lambda_{2}$ and $\lambda_{3}$ is the trivial solution everywhere on $\partial\Omega_{t}$.

\subsubsection{Weak form in the ALE framework.}
\label{subsubsection:weak form, ALE} 
Given the same data as in the \hyperref[subsubsection:ALE strong form]{ALE strong problem}, find $\bv{u}_s \in \mathcal{V}^{\bv{u}_s}$, $\bv{v}_{\text{fil}_1} \in \mathcal{V}^{\bv{v}_{\text{fil}_1}}$,
$\bv{v}_{\text{fil}_2} \in \mathcal{V}^{\bv{v}_{\text{fil}_2}}$,
$p \in \mathcal{V}^p$, $\phi_2 \in \mathcal{V}^{\phi_2}$ and $\phi_s \in \mathcal{V}^{\phi_s}$ such that $\forall$ $\tilde{\bv{u}}_s \in \mathcal{V}_w^{\bv{u}_s}$,  $\tilde{\bv{v}}_{\text{fil}_2} \in \mathcal{V}_w^{\bv{v}_{\text{fil}_2}}$,
$\tilde{\bv{v}}_{\text{fil}_3} \in \mathcal{V}_w^{\bv{v}_{\text{fil}_3}}$,
$\tilde{p} \in \mathcal{V}_w^p$, $\tilde{\phi}_2 \in \mathcal{V}_w^{\phi_2}$ and $\tilde{\phi}_s \in \mathcal{V}_w^{\phi_s}$,
\begin{gather}
\label{eq:Weak form_ALE - concentrations}
\begin{multlined}[b]
\int\limits_{\Omega_s}J_s\left[\tilde{\phi}_2\left(\partial_t{\phi_2}|_{\bv{X}_s} +  \inversetranspose{\tensor{F}_s}\colondot\Grad[\bv{X}_s]{\bv{v}_{\text{fil}_2}} + \phi_2\inversetranspose{\tensor{F}_s}\colondot\Grad[\bv{X}_s]{\bv{v}_s} - \hat{\gamma}_2 \right) \right. \\
\qquad\qquad\qquad\qquad\qquad\qquad \left. + \tilde{\phi}_s\left( \partial_t{\phi_s}|_{\bv{X}_s} + \phi_s\inversetranspose{\tensor{F}_s}\colondot \Grad[\bv{X}_s]{\bv{v}_s} - \hat{\gamma}_s
\right)\right] \dd{\Omega_s} = 0,
\end{multlined} \\
\label{eq:Weak form_ALE - solid displacement}
\begin{multlined}[b]
+ \int\limits_{\Omega_s} \left[J_s \tilde{\bv{u}}_s\cdot\left( -\sum\limits_{a=1}^N\phi_a\bv{b}_a + \frac{\hat{c}_2}{\phi_2}\bv{v}_{\text{fil}_2}\right) - J_s p \Grad[\bv{X}_s]{\tilde{\bv{u}}_s}\colondot\inversetranspose{\tensor{F}_s} + \Grad[\bv{X}_s]{\tilde{\bv{u}}_s}\colondot\tensor{P}^e \right]\dd{\Omega_s} \\ 
\qquad\qquad + \int\limits_{\left(\partial\Omega_s\right)_{\bv{v}_s}^N}\left[ J_s p \tensor{F}_s^{-1}\tilde{\bv{u}}_s\cdot\bv{n}_s - \bv{s}_s^g \right]\dd{\Gamma_s} = 0,
\end{multlined}\\
\label{eq:Weak form_ALE - mass balance}
-\int\limits_{\Omega_s}J_s\tilde{p}\left( \inversetranspose{\tensor{F}_s}\colondot\Grad[\bv{X}_s]{(\bv{v}_s+\bv{v}_{\text{fil}_2} + \bv{v}_{\text{fil}_3})} - \sum\limits_{a=1}^{N}\hat{\gamma}_a \right)\dd{\Omega_s} =0,\\
\begin{multlined}
\label{eq:Weak form_ALE - filtration velocity 2}
+\int\limits_{\Omega_s}J_s\left[\tilde{\bv{v}}_{\text{fil}_2}\cdot\left(\frac{\mu_2}{\kappa_s}\bv{v}_{\text{fil}_2} - \bv{b}_2 \right) - \Grad[\bv{X}_s]{\tilde{\bv{v}}_{\text{fil}_2}}\colondot  \inversetranspose{\tensor{F}_s}p \right]\dd{\Omega_s} \\ 
\qquad + \int\limits_{\partial\Omega_s} \left[ \tilde{\lambda}_3\left(J_s \tensor{F}_s^{-1}\bv{v}_{\text{fil}_2}\cdot\bv{n}_s - v^{n_s}_{\text{fil}_2})\right) + J_s \tilde{\bv{v}}_{\text{fil}_2}\cdot (p+\lambda_2) \inversetranspose{\tensor{F}_s}\bv{n}_s \right] \dd{\Gamma_s} = 0,
\end{multlined}\\
\label{eq:Weak form_ALE - filtration velocity 3}
\begin{multlined}
+ \int\limits_{\Omega_s}J_s\left[\tilde{\bv{v}}_{\text{fil}_3}\cdot\left(\frac{\mu_3}{\kappa_s}\bv{v}_{\text{fil}_3} - \bv{b}_3 \right) - \Grad[\bv{X}_s]{\tilde{\bv{v}}_{\text{fil}_3}}\colondot \inversetranspose{\tensor{F}_s}p \right] \dd{\Omega_s} \\
\qquad + \int\limits_{\partial\Omega_s}\left[ \tilde{\lambda}_3\left(J_s \tensor{F}_s^{-1}\bv{v}_{\text{fil}_2}\cdot\bv{n}_s - v^{n_s}_{\text{fil}_3})\right) + J_s \tilde{\bv{v}}_{\text{fil}_3}\cdot (p+\lambda_3) \inversetranspose{\tensor{F}_s}\bv{n}_s \right]\dd{\Gamma_s} = 0.
\end{multlined}
\end{gather}
Here, $\bv{s}_s^g$ is the prescribed traction field on $(\partial\Omega)_s$, which is related to $\bv{s}$ as follows
\begin{equation}
\bv{s}_s = J_s||\inversetranspose{\tensor{F}_s}\bv{n}_s||\bv{s}.
\end{equation}

\begin{figure}
\centering     
\subfigure[$\|\bv{v}_{\text{fil}_2} \|$]{\includegraphics[width=0.3\textwidth]{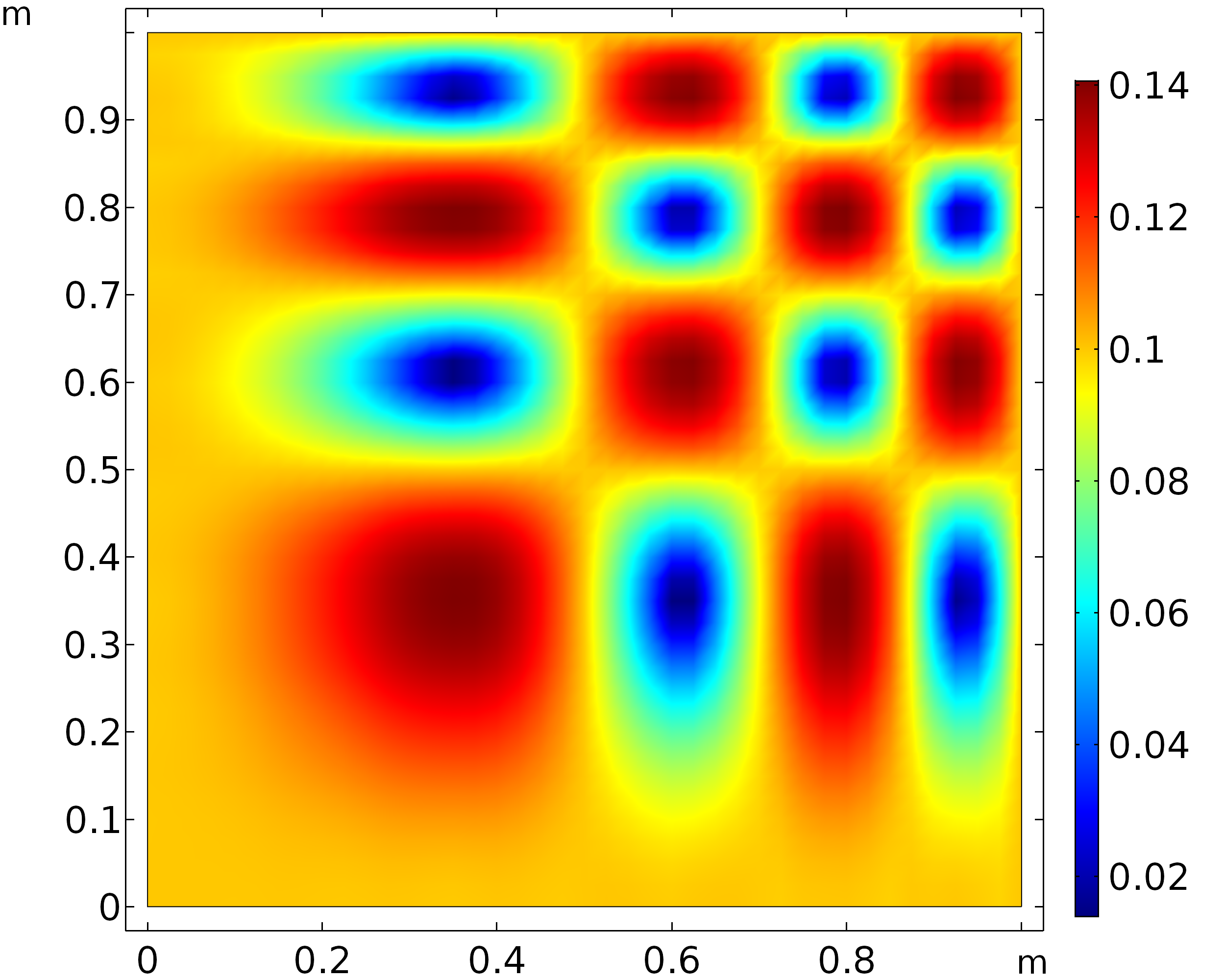}}
\subfigure[$\|\bv{v}_{\text{fil}_3}\|$]{\includegraphics[width=0.3\textwidth]{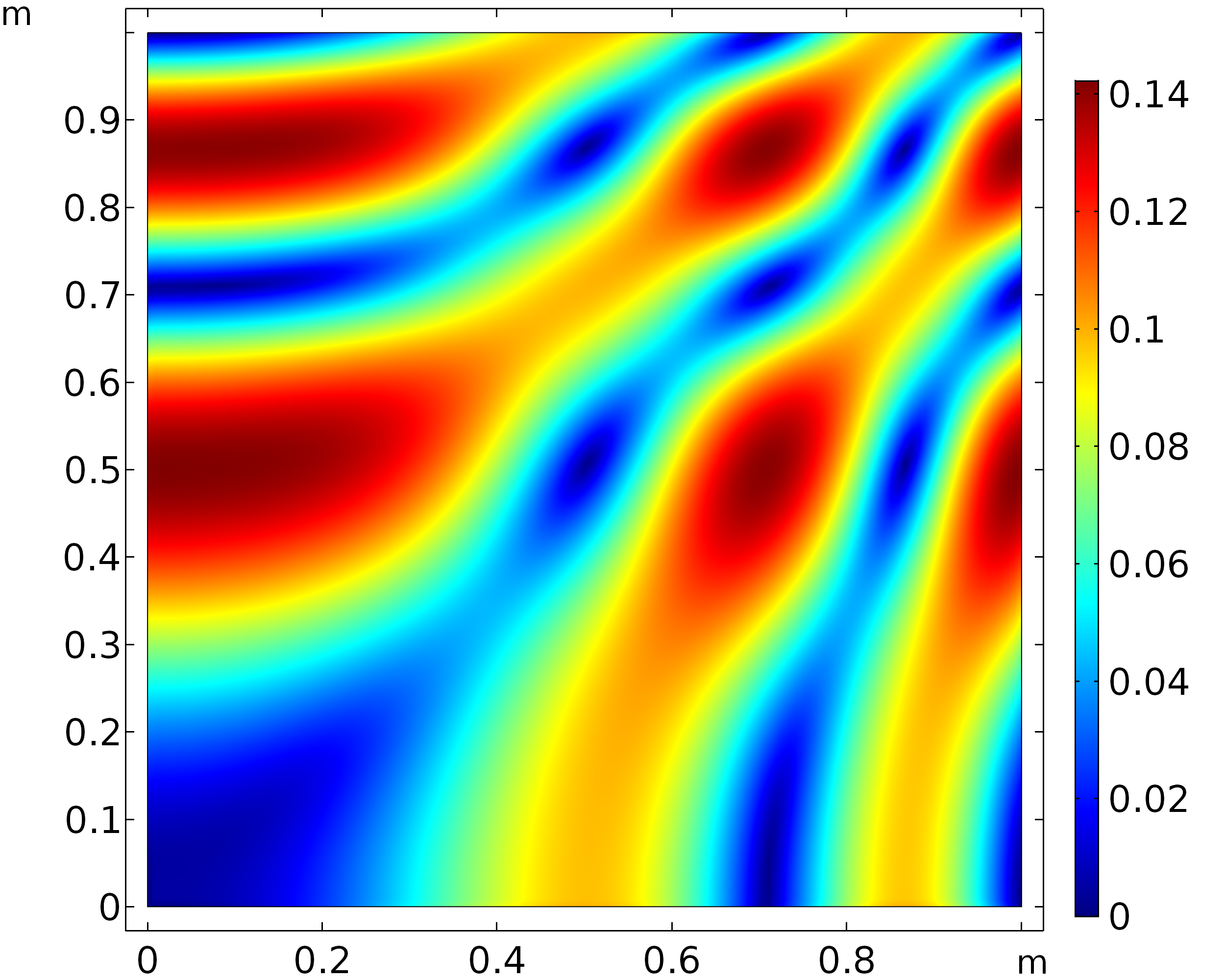}}
\subfigure[$\| p \|$]{\includegraphics[width=0.3\textwidth]{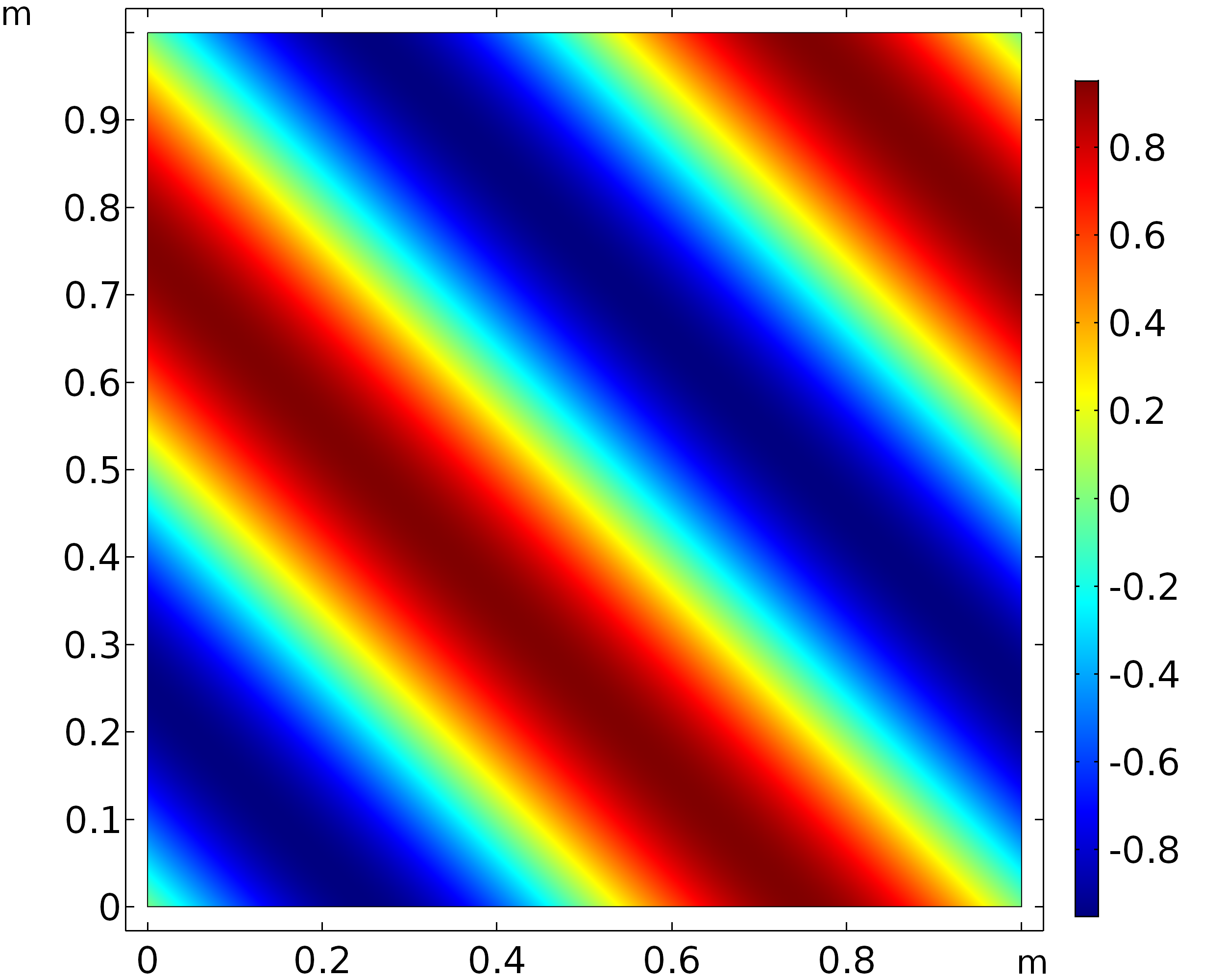}}\\
\subfigure[$ \phi_s $]{\includegraphics[width=0.3\textwidth]{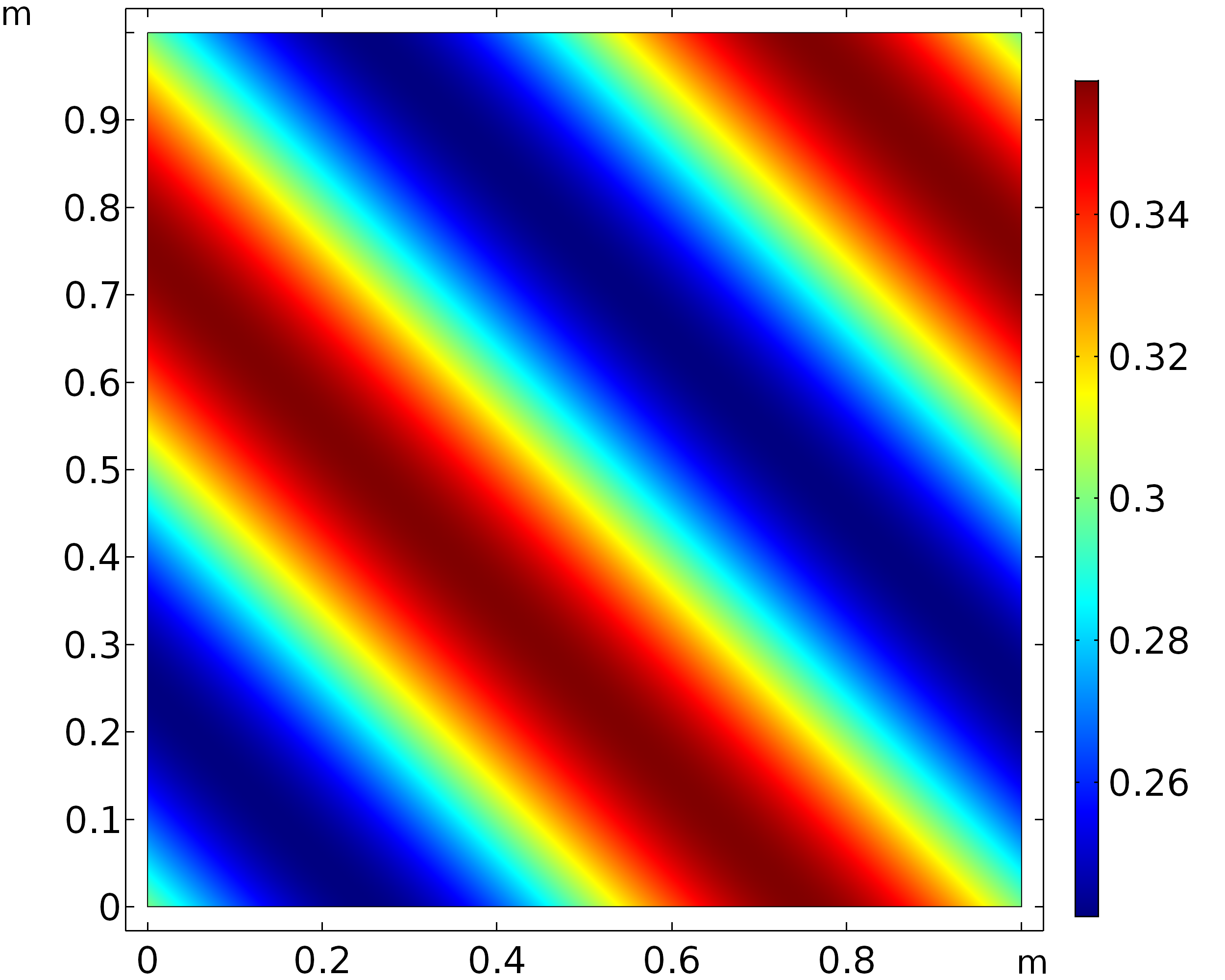}}
\subfigure[$ \phi_2 $]{\includegraphics[width=0.3\textwidth]{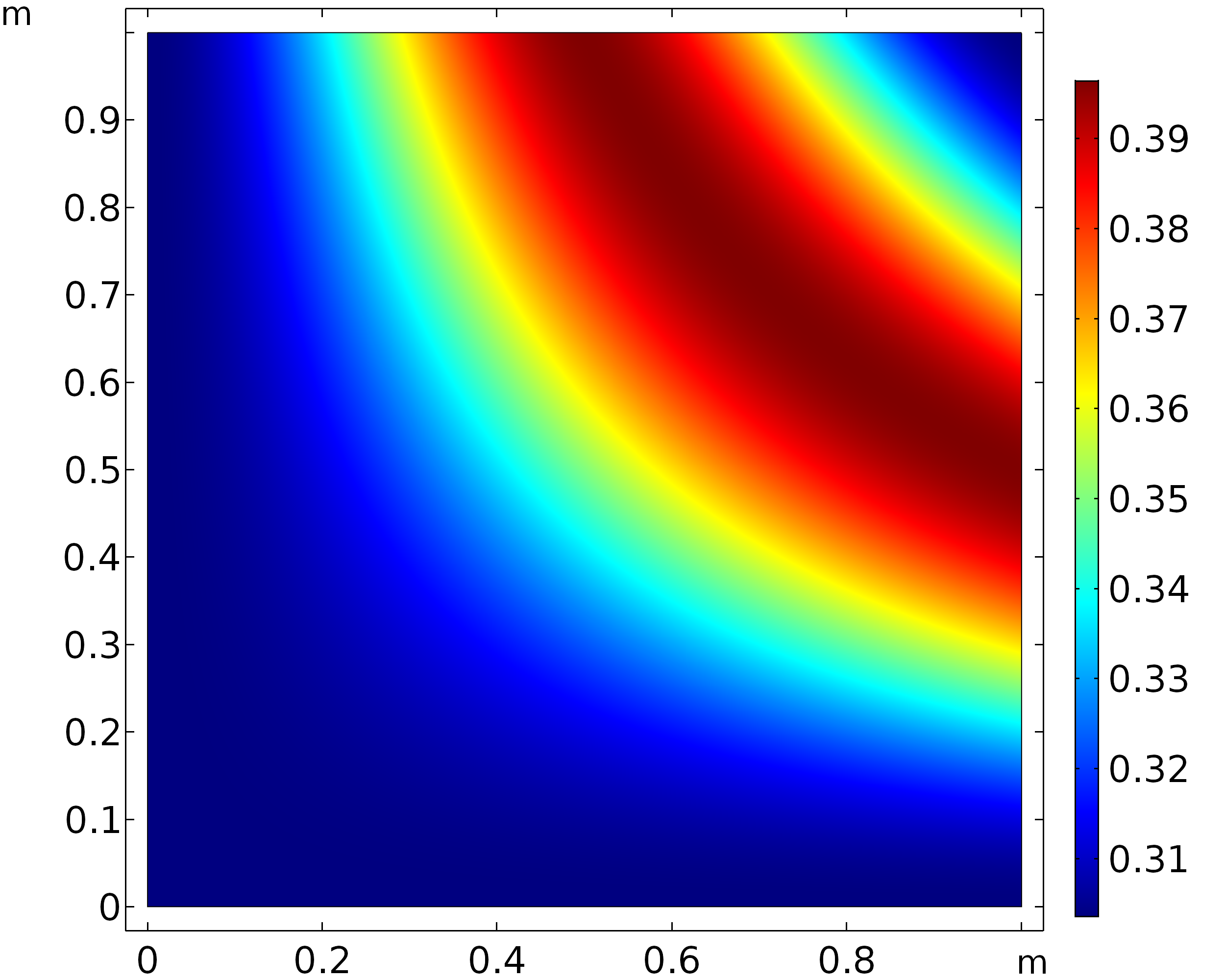}}
\subfigure[$\|\bv{u}_s\|$]{\includegraphics[width=0.3\textwidth]{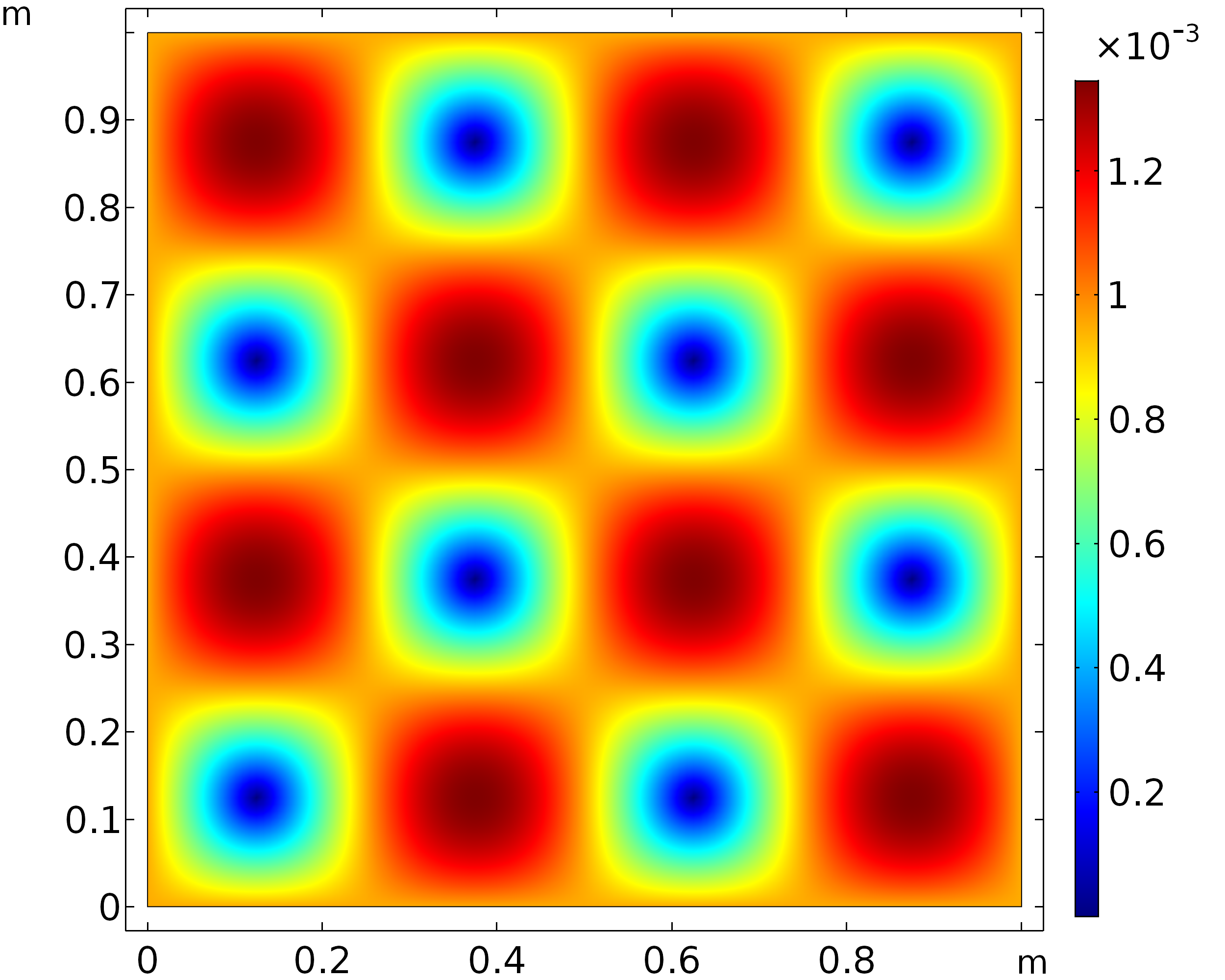}}
\caption{Manufactured solutions for different variables}
\label{fig:Manufactured Solution}
\end{figure}

\paragraph{Discrete Approximation:} 
We partition the domain $\Omega_s$ into a triangulation $\Omega_s^h$ consisting of cells $K$, with diameter $h$. The cells $K$ are chosen to be non-overlapping quadrilaterals in the 2D case and hexahedrons in the 3D case, which span the spatial domain $\Omega_s$. The decomposition of the boundary $\partial\Omega_s$ into Neumann and Dirichlet partitions is also followed by the boundary of $\Omega_s^h$. The interpolation functions over $K$ are chosen to be Lagrange polynomials.

\section{Numerical Results}
\label{sec:Results}
The weak formulation derived in the previous section was implemented in a \ac{FEM} framework, and the resultant numerical model was studied for accuracy. 

\subsection{Model numerical problem}
\label{subsec:Problem description}
For the testing of our numerical scheme, we choose a 2-dimensional square of side length $L = \np[m]{1}$ as the domain $\Omega_s$. The elastic response of the solid is chosen as Neo-Hookean, with the solid energy density given by,
\begin{equation}
W_s(\tensor{C}_s) = \frac{m_s}{2}(\tr(\tensor{C}_s)-3).
\end{equation}
The value of $m_s$ is chosen such that the constitutive response of the solid resembles that of bio-polymers. The viscosities of the fluids were chosen to be of the order of the viscosity of water. The exact parameter values used in our analysis are given in Table \ref{table:parameter values}. The numerical scheme is tested using the Method of Manufactured Solutions \cite{MethodOfManufacturedSolutions}. The functional forms of different variables (manufactured solutions) are chosen as,
\begin{table}[t]
\centering
\begin{minipage}[t]{0.34\textwidth}
\centering
\caption{Parameters}
\begin{tabular}{cc}
	\toprule
	Parameter & Value \\
	\midrule
	$m_s$ & $\np[kg/ms^2]{3e3}$\\
	$\mu_2$ & $\np[kg/sm]{1e-3}$\\
	$\mu_3$ & $\np[kg/sm]{1.5e-3}$\\
	$\kappa_s$ & $\np[m^2]{1e-3}$\\
	\bottomrule
	\label{table:parameter values}
\end{tabular}
\end{minipage}
\begin{minipage}[t]{0.3\textwidth}
\centering
\caption{MS constants}
\begin{tabular}{cc}
	\toprule
	Constant & Value\\
	\midrule
	$\bar{u}_s$ & $\np[m]{0.001}$ \\
	$\bar{v}_{\text{fil}_1}$ & $\np[m/s]{0.1}$ \\
	$\bar{v}_{\text{fil}_2}$ & $\np[m/s]{0.1}$ \\		
	$\bar{p}$ & $\np[Pa]{1}$ \\		
	\bottomrule
	\label{table:formulation constants}
\end{tabular}
\end{minipage}
\begin{minipage}[t]{0.34\textwidth}
\caption{$\phi-$constants}
\begin{tabular}{ccccc}
	\toprule
	Case & $\check{\phi}_2$ & $\bar{\phi}_2$ & $\check{\phi}_s$ & $\bar{\phi}_s$\\
	\midrule
	1 & 0.2 & 0.1 & 0.5 & 0.2 \\
	2 & 0.35 & 0.15 & 0.3 & 0.1 \\
	3 & 0.2 & 0.1 & 0.2 & 0.1 \\
	\bottomrule
	\label{table:volume fraction constants}
\end{tabular}
\end{minipage}
\end{table}
\begin{align}
\bv{u}_s &= \bar{u}_s\sin(\frac{2\pi t}{t_0})\left[\cos(2\pi\frac{x+y}{L}) \hat{\imath}
+  \sin(2\pi\frac{x-y}{L}) \hat{\jmath}\right], \\
\bv{v}_{\text{fil}_2} &= \bar{v}_{\text{fil}_2}\cos(\frac{2\pi t}{t_0})\left[ \sin(2\pi\frac{x^2+y^2}{L^2})\hat{\imath} + \cos(2\pi \frac{x^2-y^2}{L^2})\hat{\jmath} \right], \\
\bv{v}_{\text{fil}_3} &= \bar{v}_{\text{fil}_3}\cos(\frac{2\pi t}{t_0})\left[ \cos(2\pi\frac{x^2}{L^2})\sin(2\pi \frac{y^2}{L^2})\hat{\imath} + \sin(2\pi \frac{x^2-y^2}{L^2})\hat{\jmath} \right], \\
p &= \bar{p}\sin(\frac{2\pi t}{t_0})\sin(2\pi\frac{x+y}{L}), \label{eq:Manufactured solutions}\\
\phi_2 &= \check{\phi}_2 + \bar{\phi}_2\cos( \frac{2\pi t}{t_0})\cos(2\pi\frac{x+y}{L}), \\
\phi_s &= \check{\phi}_s + \bar{\phi}_s\sin( \frac{2\pi t}{t_0})\sin(2\pi\frac{x+y}{L}),
\end{align}
where $\hat{\imath}$ and $\hat{\jmath}$ are the base vectors of the adopted Cartesian coordinate system.
The values chosen for different constants are given in Table \ref{table:formulation constants}. The constants $\check{\phi}_2$, $\bar{\phi}_2$, $\check{\phi}_s$ and $\bar{\phi}_s$ are adjusted so as to cover different phases of degradation. The three combinations of the volume fraction constants used in our analysis are given in Table \ref{table:volume fraction constants}. Fig. \ref{fig:Manufactured Solution} shows representative solution field distributions for all the variables over the domain $\Omega_s$. The solution is obtained using a mesh of size $h = L/128$, for the case 2 mentioned in Table \ref{table:formulation constants}, at time = $\np[s]{0.7}$. The nature of the solution is qualitatively the same for all the cases, at all times, and for all refinements levels.


\subsection{Solver and mesh specifications}
The equations presented in this paper constitute a transient differential-algebraic system, which was solved in COMSOL Multiphysics\textsuperscript{\textregistered}. The weak formulation was studied for stability and convergence for the two sets of boundary conditions discussed in the previous \hyperref[para:boundary]{section}. The primary variables for the numerical study were $\bv{u}_s, \bv{v}_{\text{fil}_2}, \bv{v}_{\text{fil}_3}, \phi_s, \phi_2$ and $p$.
The solid velocity, $\bv{v}_s$ was calculated in a post-processing step, using the kinematic relation: $\bv{v}_s = \partial_t{\bv{u}_s}\big|_{\bv{X}_s}$. Implicit time stepping was implemented by the method of lines \cite{methodoflinesbook, COMSOLmanual}, using the variable-order variable time step \ac{BDF}. The order of BDF was constrained between 2 to 5, whereas a maximum time step size of $\np[s]{1e-3}$ was imposed. The resultant non-linear system of equations for every time step was solved using Newton's method (with no damping). 
PARDISO (\cite{PARDISO-1, PARDISO-2, PARDISO-3}) was chosen as the linear solver for the resultant linear system of equations. The unit square domain was discretized uniformly into square elements of side length $1/h$, for $h = 8, 16, 32, 64, \np[m]{128}$. 
Piece-wise Lagrange interpolation polynomials were used for all fields. No additional stabilization technique has been implemented for the numerical formulation.

\subsubsection{Choice of order of interpolation functions.}
It was observed that the solvability and accuracy of the unstabilized quasi-static formulation is dependent on the choice of \ac{IO} of the interpolating functions. The numerical scheme did not provide satisfactory results when pressure was interpolated using an \ac{IO} equal to or greater than that of the vector variables. This phenomenon was observed in the fully dynamic case (the formulation in which the time derivative and convective terms were included in the momentum balance equations) of non-reactive poro-elastic flow, \cite{costanzo2017arbitrary} as well. However, unlike the quasi-static case presented in their paper, we observe that having equal \ac{IO} for pressure and vector fields does not yield convergent solutions to our numerical problem. This condition likely stems from a Brezzi-Babu\v{s}ka (inf-sup) type condition.

In order to get acceptable numerical results, the \ac{IO} of pressure interpolation functions was set to be one less than that of the vector variables (velocities and solid displacement). 
We choose \acp{IO} of 2 and 3 for the vector variables, and corresponding order of 1 and 2 respectively for pressure. The \ac{IO} of the volume fractions was chosen to be either equal to, or less than that of pressure for each of the above choice of \acp{IO} for the rest of the variables. It was observed that having a higher order for volume fractions did not produce acceptable numerical results. Thus, we will be discussing three combinations of interpolation polynomials:
\begin{itemize}
\item \label{o-1}
Case O-1: Quadratic polynomials for vector variables, linear for pressure and volume fractions,
\item \label{o-2}
Case O-2: Cubic polynomials for vector variables, quadratic for pressure whereas linear polynomials for volume fractions,
\item \label{o-3}
Case O-3: Cubic polynomials for vector variables and quadratic elements for pressure as well as volume fractions.
\end{itemize}
The \ac{IO} of the auxiliary boundary variable $\lambda_b$ for the implementation of the impermeability boundary condition for the filtration velocity was chosen to be equal to, or less than the order of $\bv{v}_{\text{fil}_b}$, for $b = 2, 3$. Thus, Lagrange linear and quadratic \acp{IO}  were chosen for $\lambda_b$ for case O-1 and linear, quadratic and cubic polynomials for cases O-2 and O-3. 
We would also like to mention that acceptable results were not obtained when linear interpolation polynomials were used for $\lambda_b$ while using cubic polynomials for filtration velocities (cases O-2 and O-3). However, exact same convergence rates were obtained using either quadratic or cubic polynomials for $\lambda_b$ for cases O-2 and O-3. Similarly, the convergence rates obtained using either linear or quadratic polynomials for case O-1 were exactly the same. Hence, we arbitrarily choose to report the results for linear $\lambda_b$ interpolation polynomials for case O-1 and quadratic for case O-2 and O-3.

\subsection{Formulation results}

\begin{figure}[t!]
\includegraphics[width=0.45\textwidth]{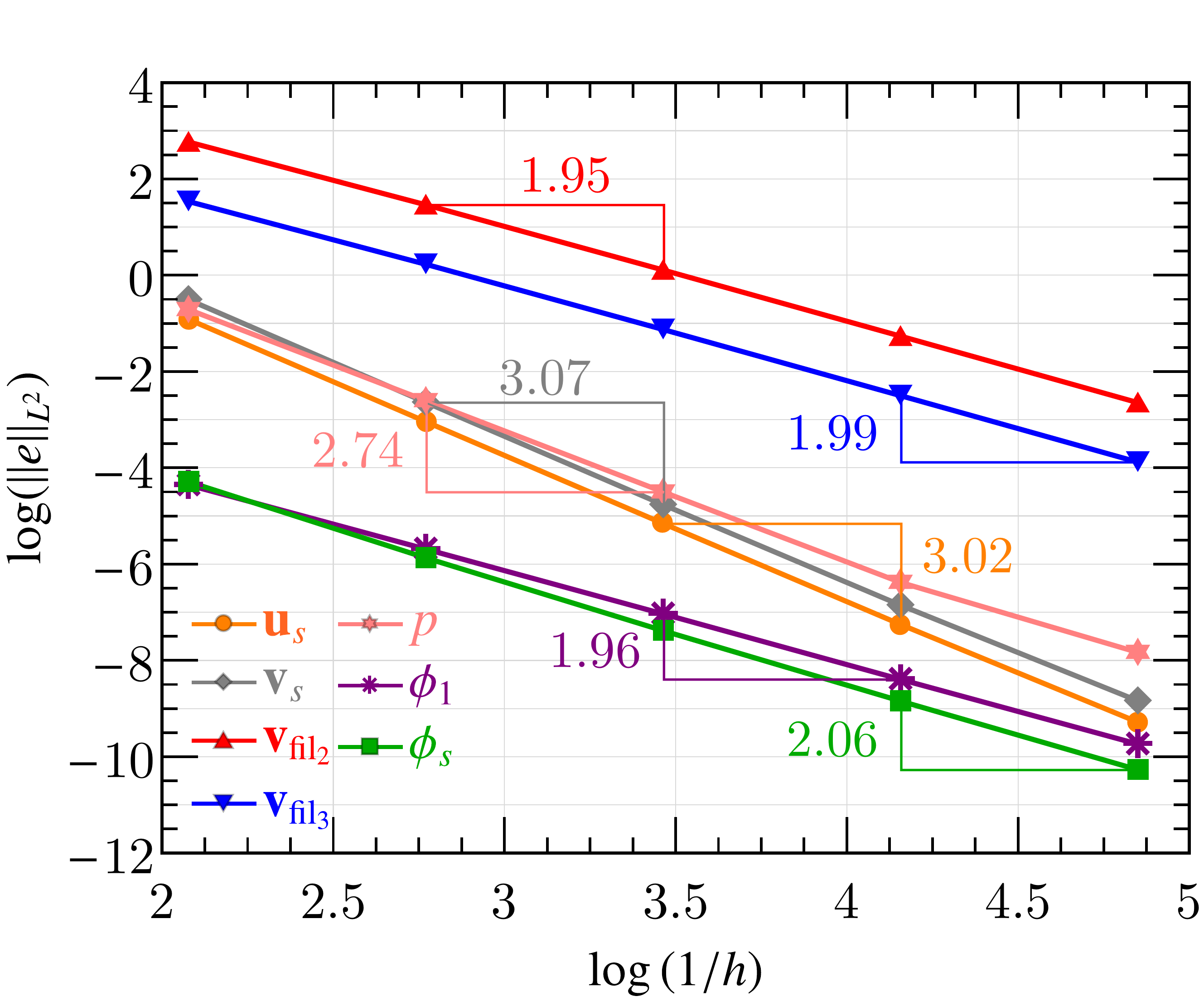}\hspace{0.01\textwidth}\includegraphics[width=0.44\textwidth]{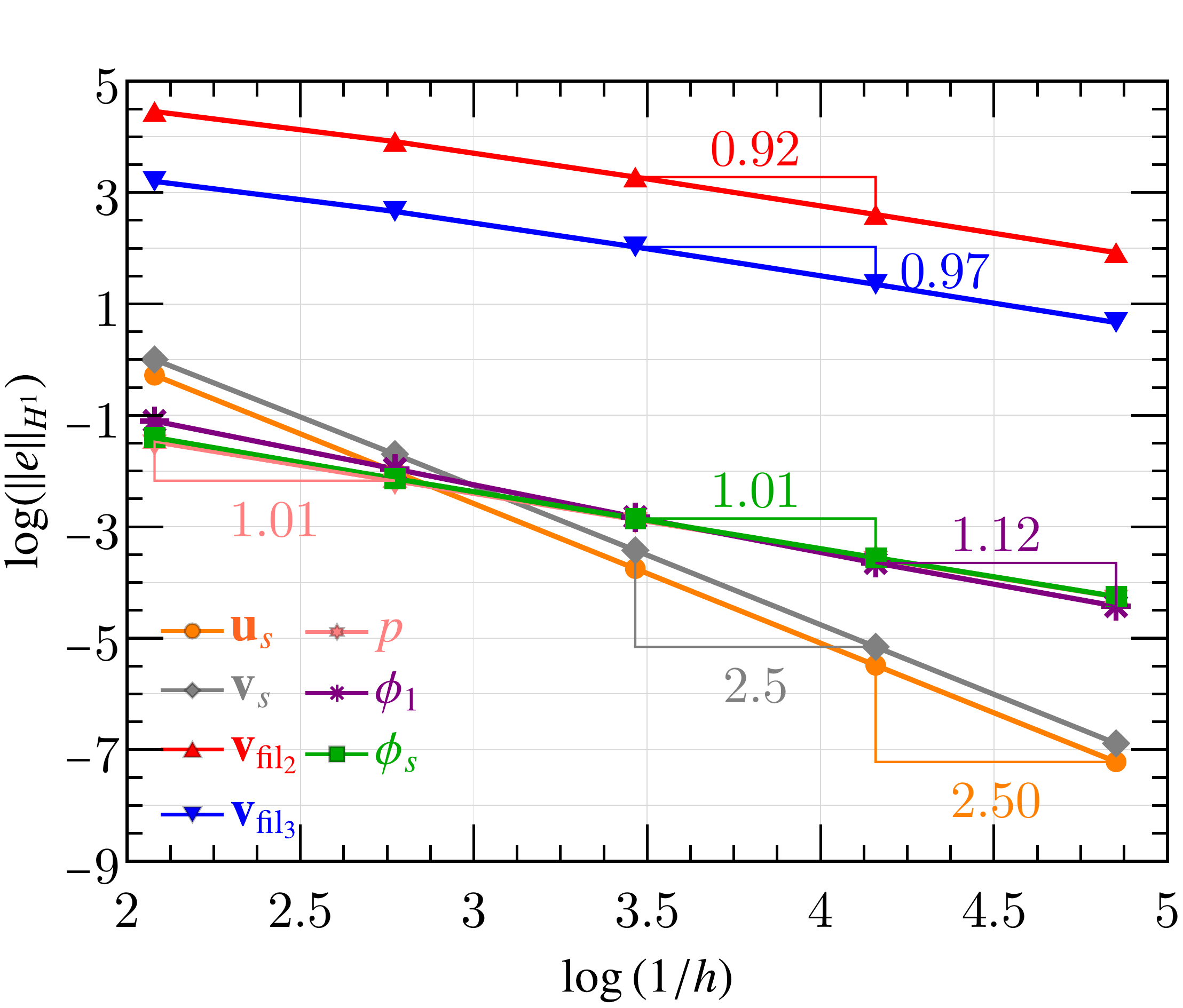}\\ \includegraphics[width=0.45\textwidth]{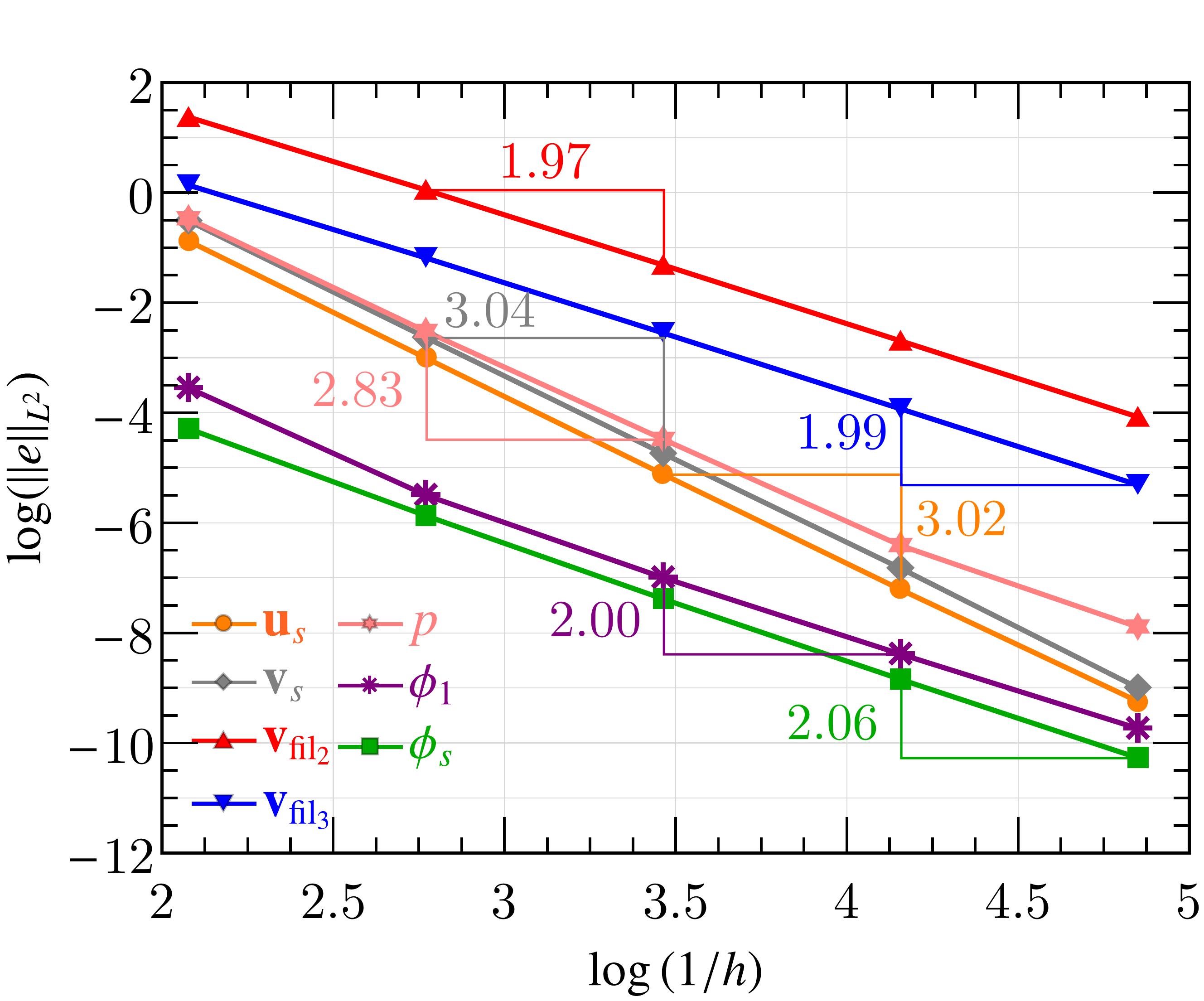}\includegraphics[width=0.45\textwidth]{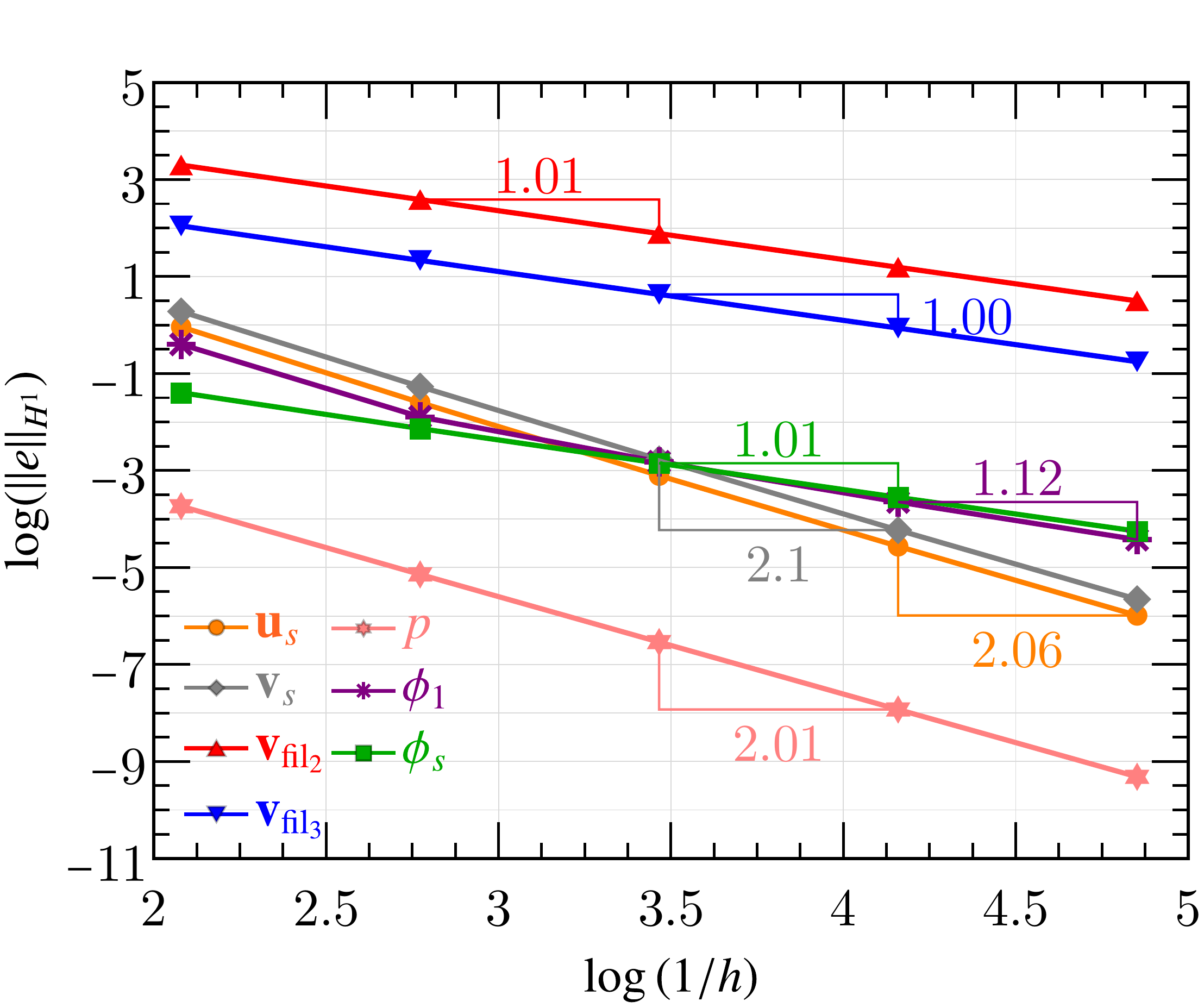}\\ \includegraphics[width=0.45\textwidth]{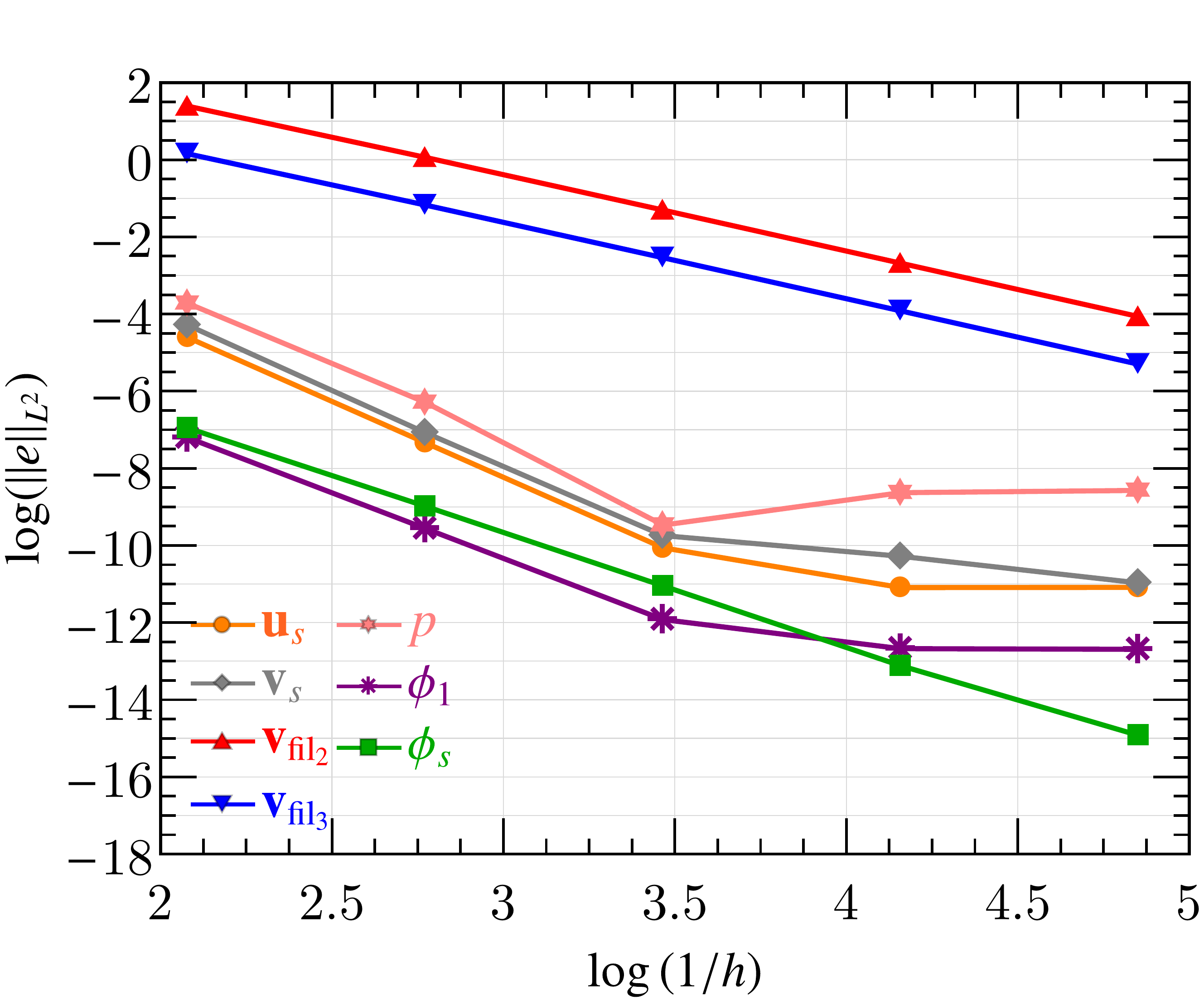}\includegraphics[width=0.45\textwidth]{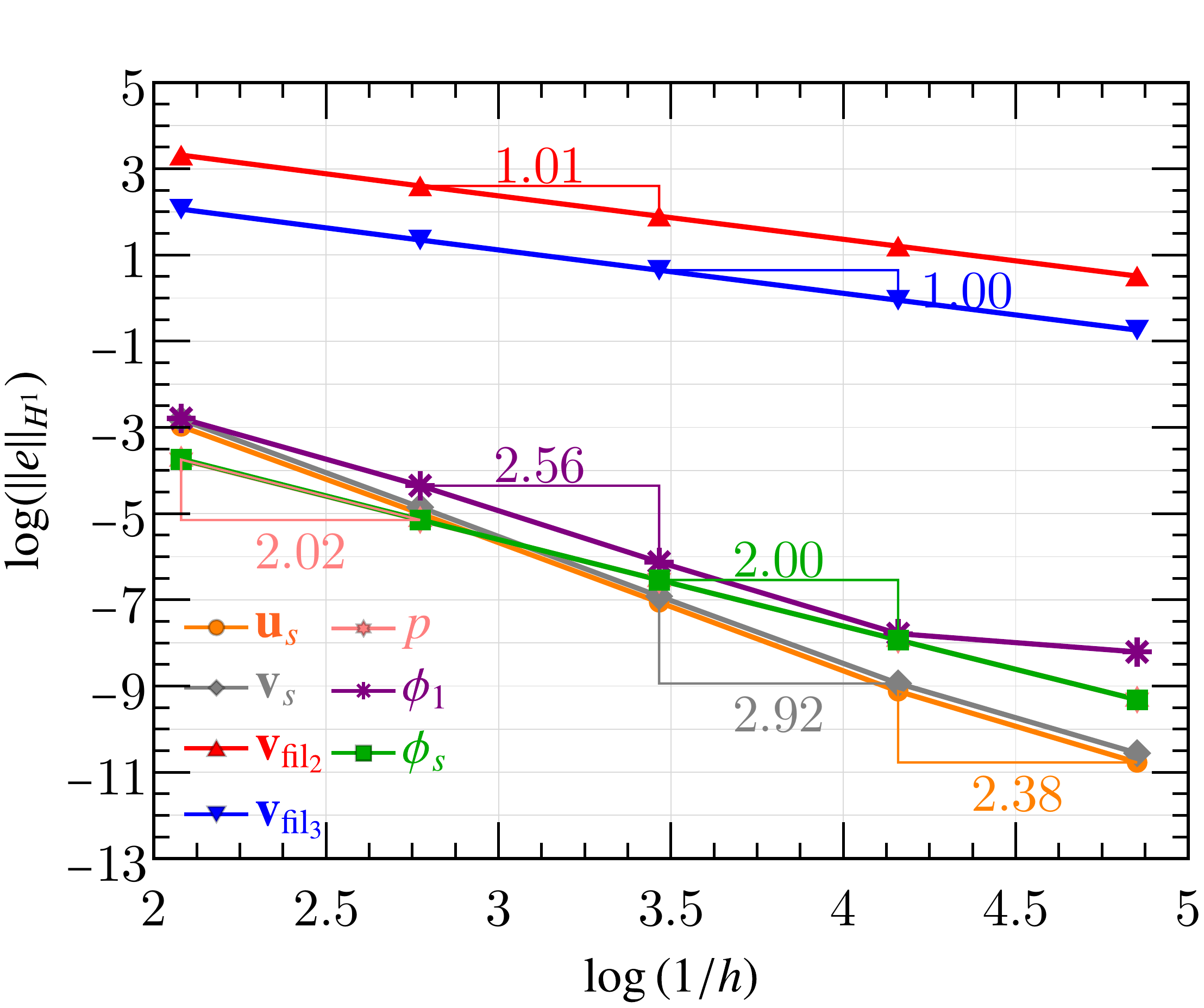}
\caption{$L^2$ and $H^1$ error norms for case 1 in Table \ref{table:volume fraction constants}, for cases O-1, O-2 and O-3 (from top to bottom respectively.), for the weak formulation along with Neumann boundary condition for $\bv{u}_s$}
\label{fig:one}
\end{figure}

\begin{figure}[t!]
\includegraphics[width=0.45\textwidth]{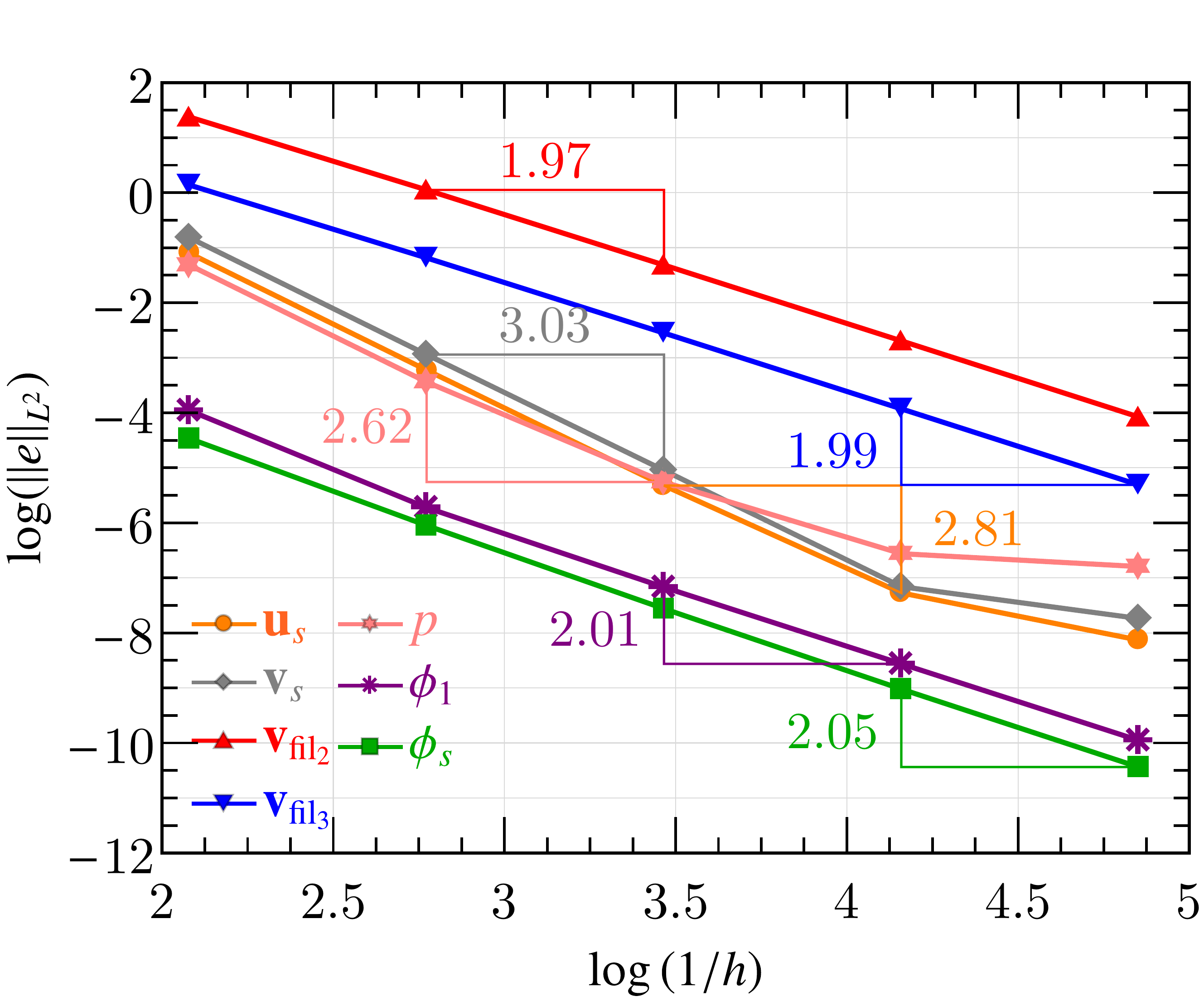}\hspace{0.01\textwidth}\includegraphics[width=0.44\textwidth]{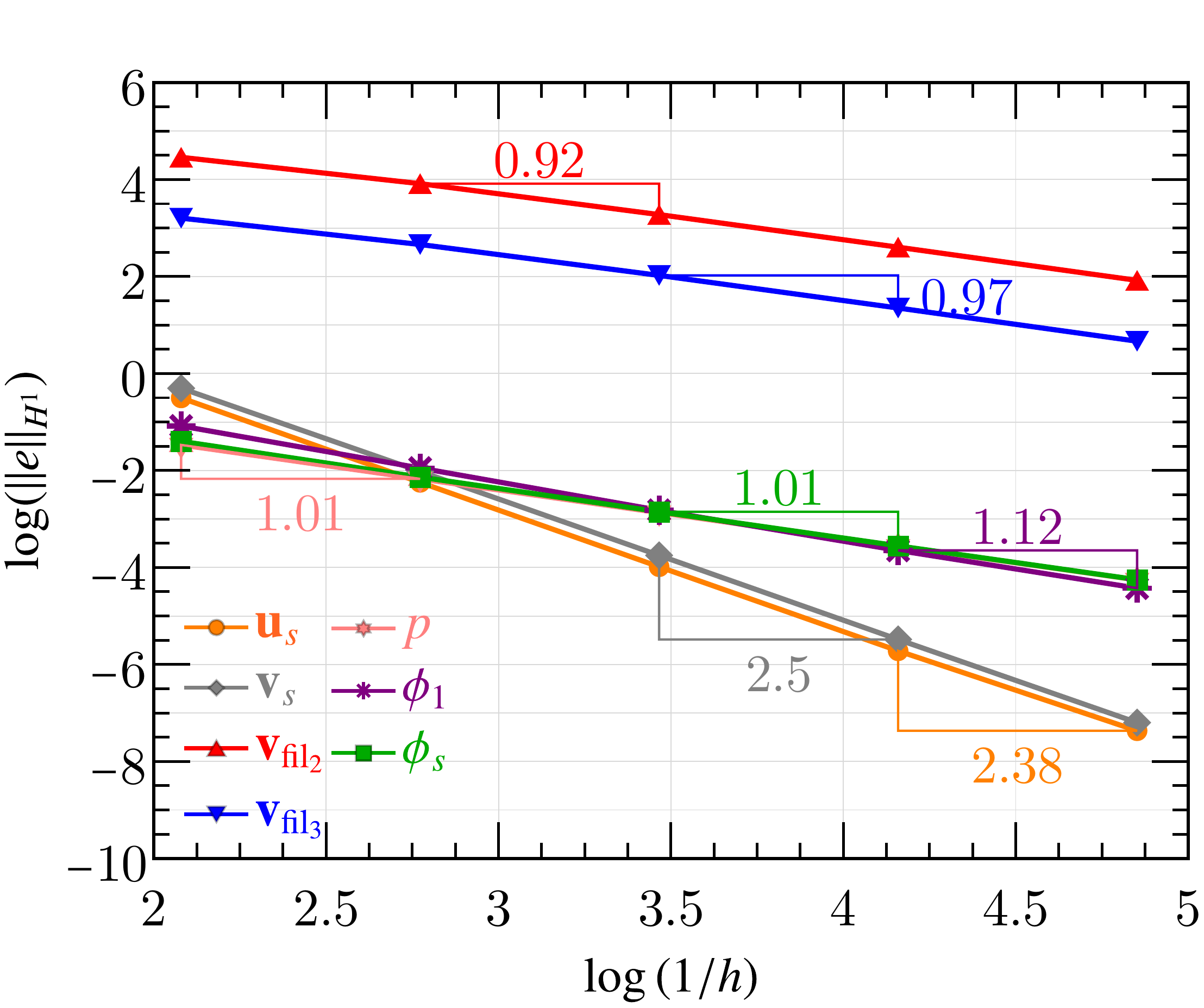}\\ \includegraphics[width=0.45\textwidth]{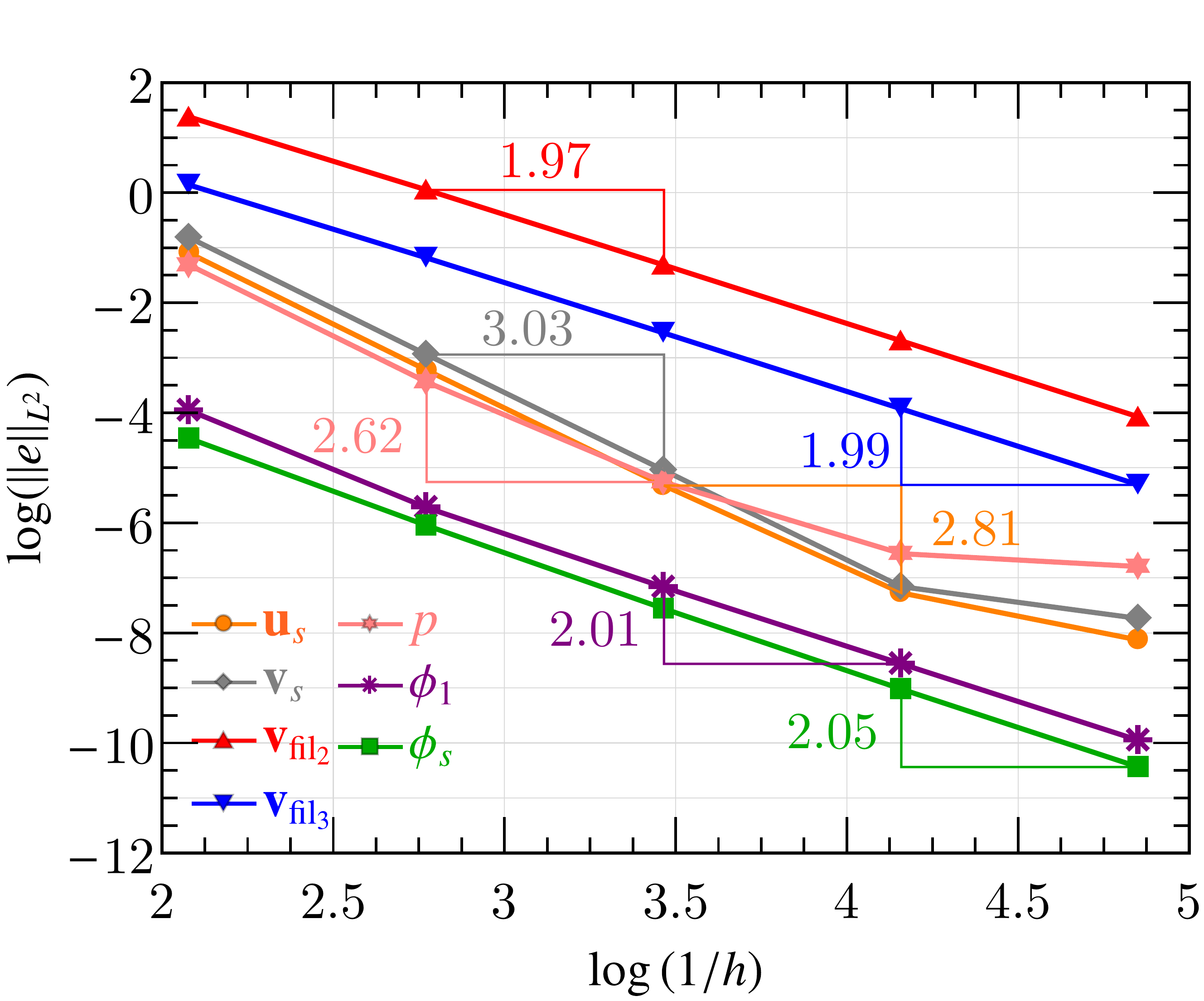}\includegraphics[width=0.45\textwidth]{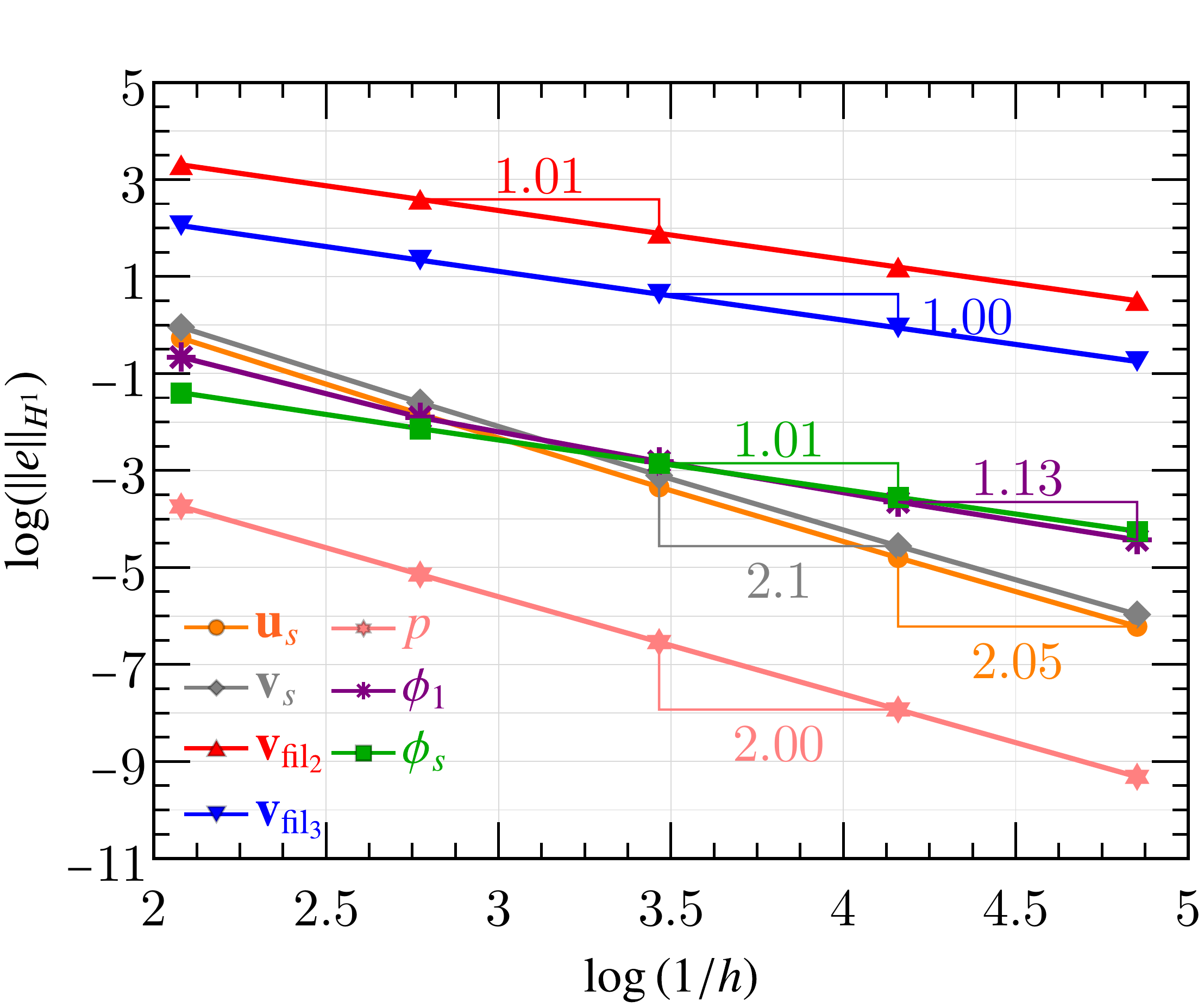}\\
\includegraphics[width=0.45\textwidth]{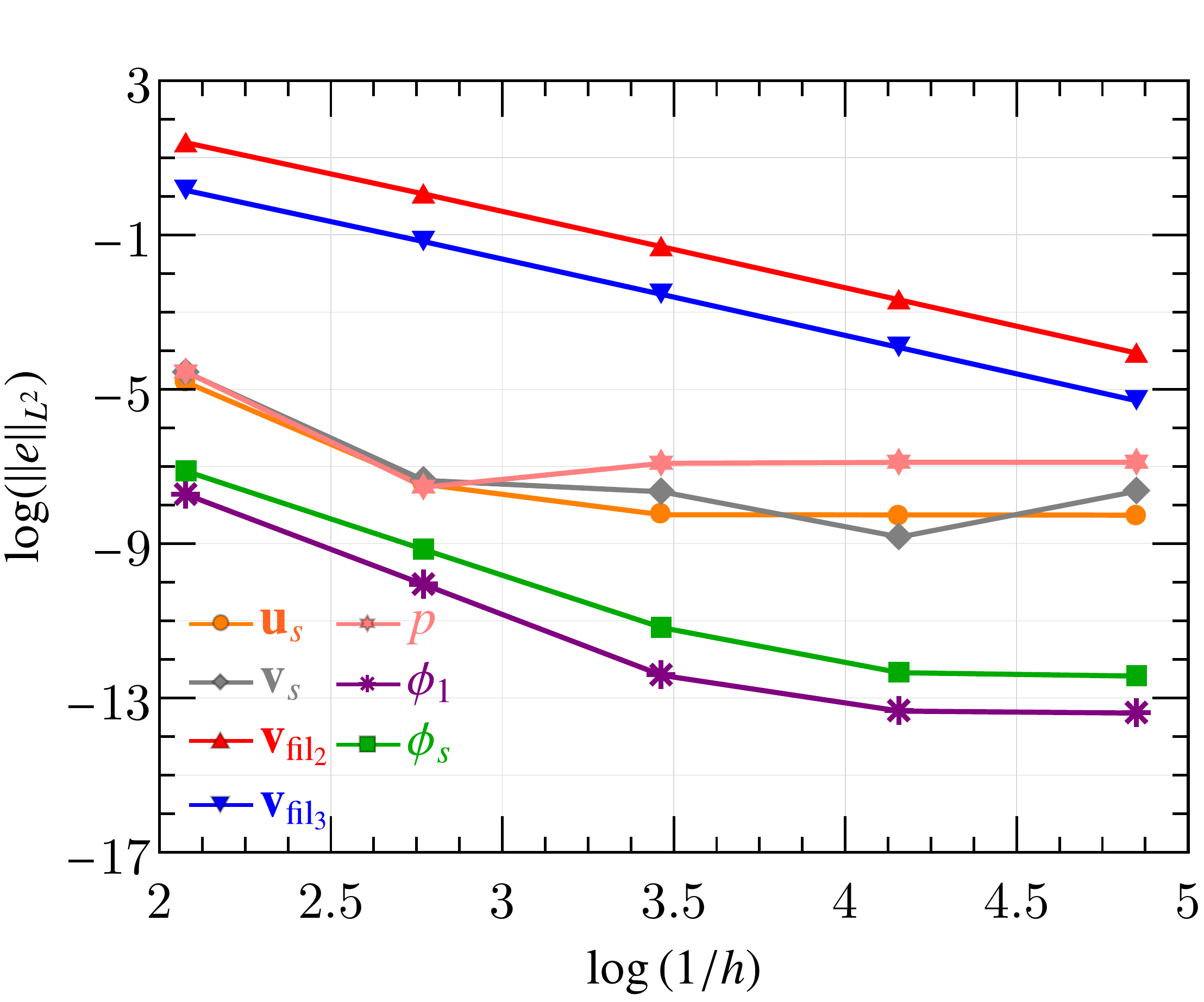}\includegraphics[width=0.45\textwidth]{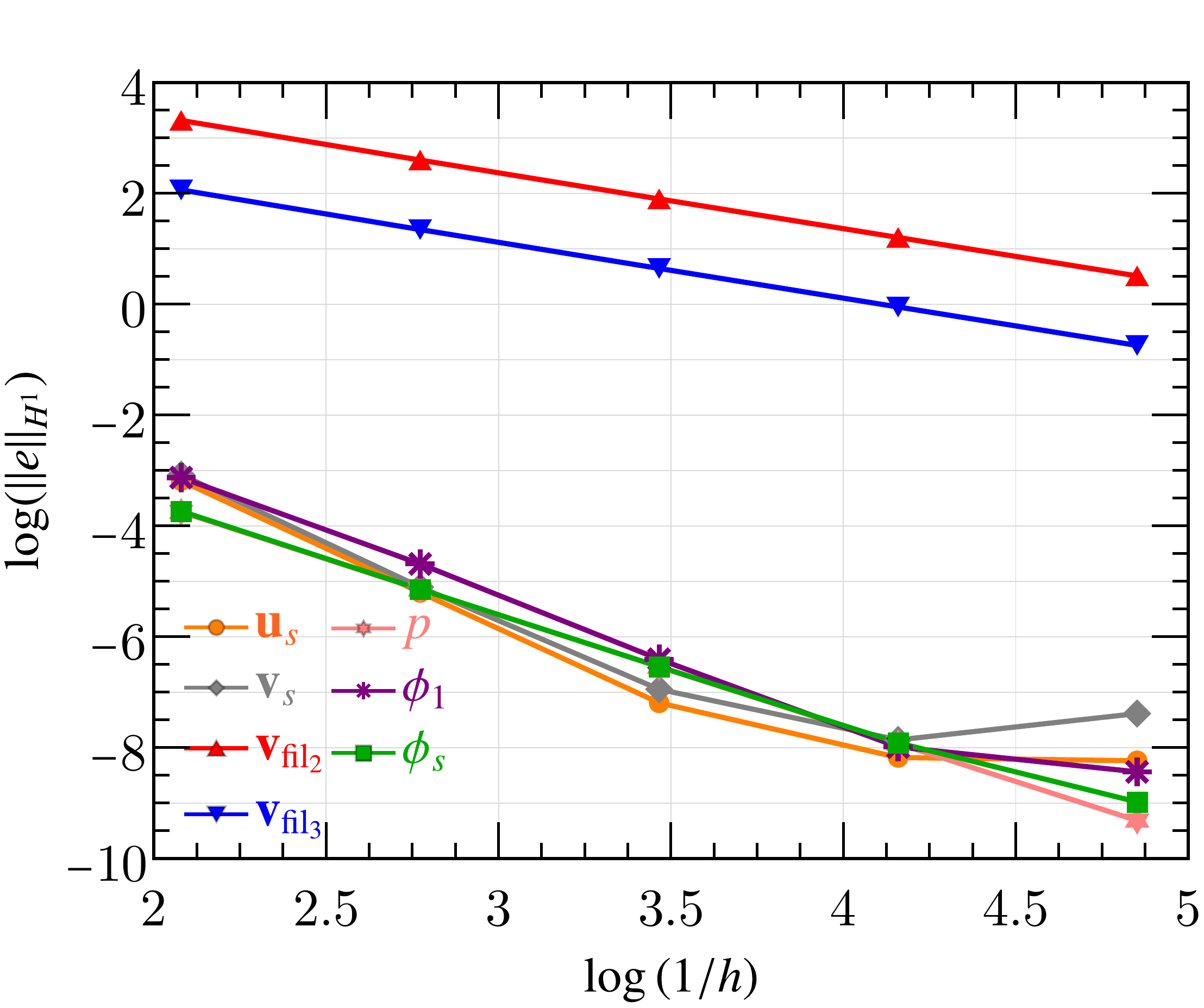}
\caption{$L^2$ and $H^1$ error norms for case 2 in Table \ref{table:volume fraction constants}, for cases O-1, O-2 and O-3 (from top to bottom respectively.), for the weak formulation along with Neumann boundary condition for $\bv{u}_s$}
\label{fig:two}
\end{figure}

\begin{figure}[t!]
\includegraphics[width=0.45\textwidth]{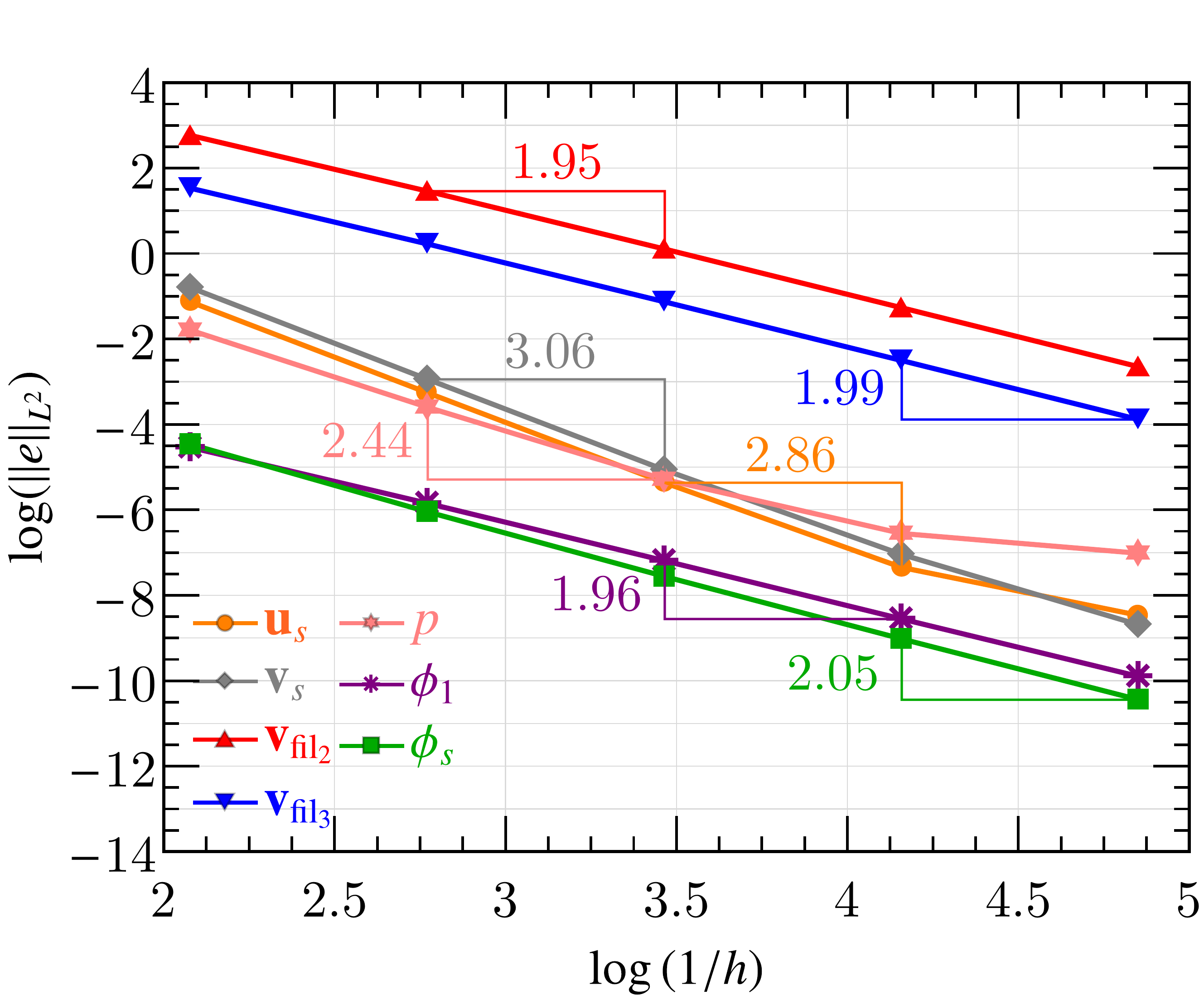}\hspace{0.01\textwidth}\includegraphics[width=0.44\textwidth]{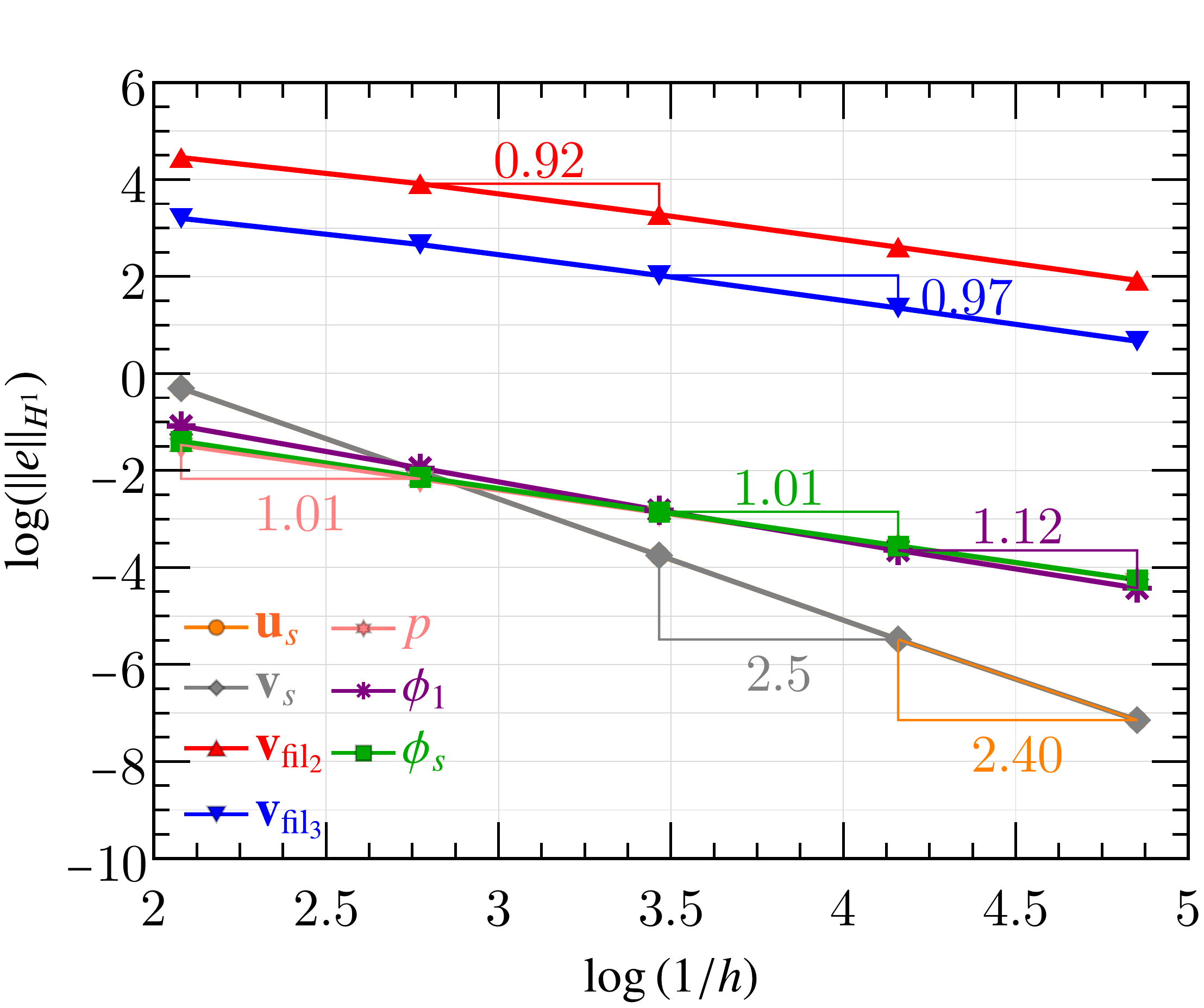}\\ \includegraphics[width=0.45\textwidth]{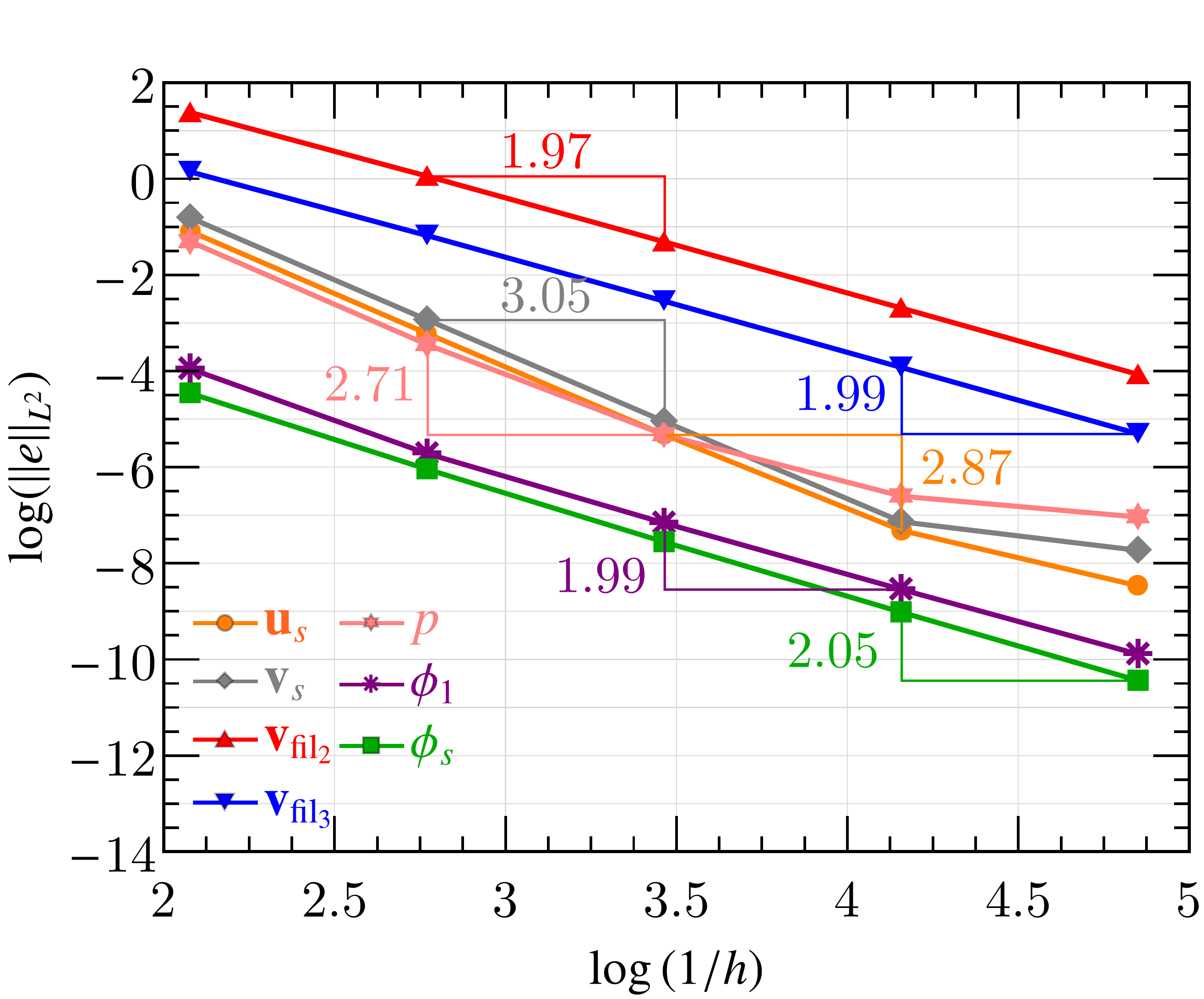}\includegraphics[width=0.45\textwidth]{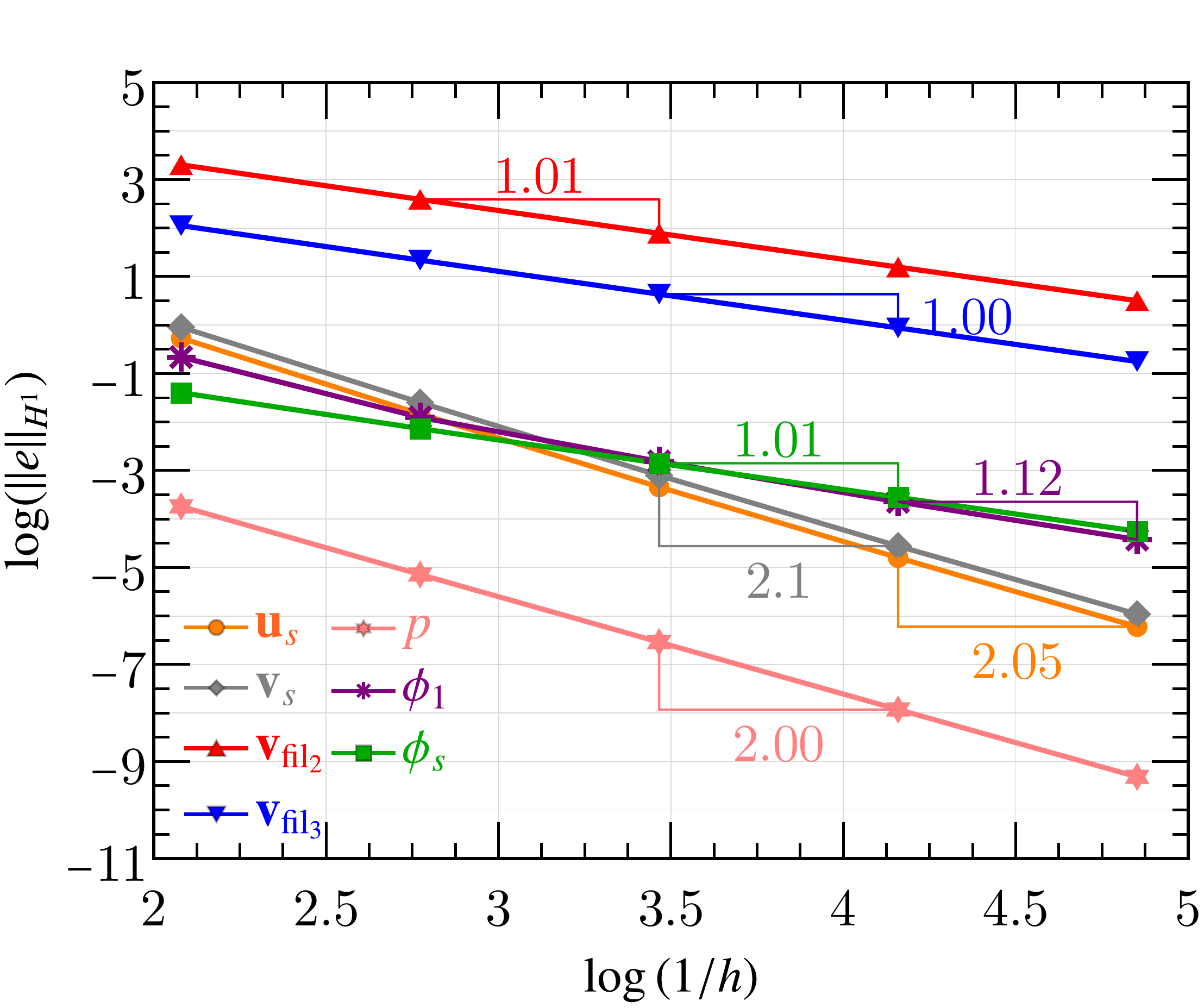}\\ \includegraphics[width=0.45\textwidth]{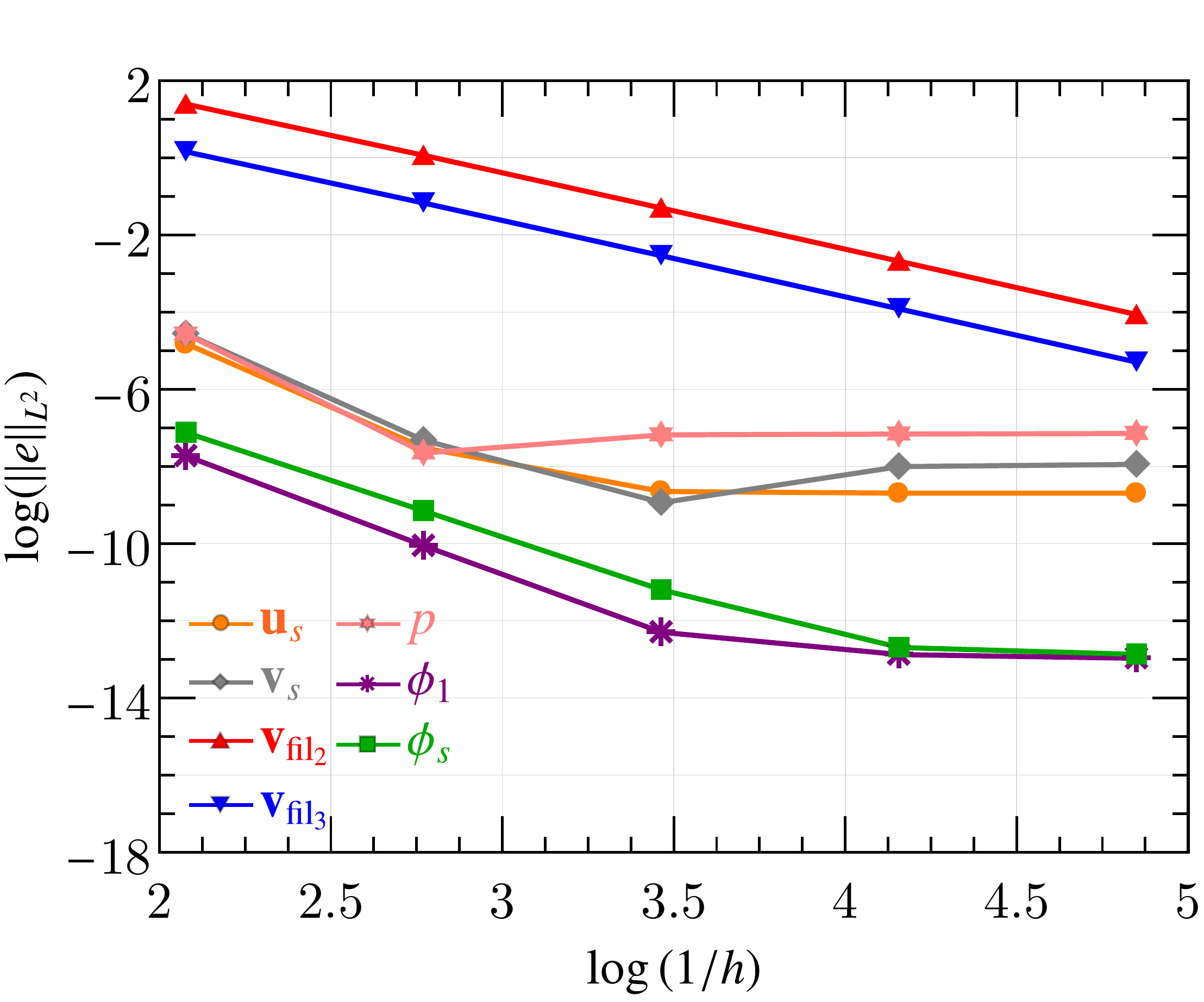}\includegraphics[width=0.45\textwidth]{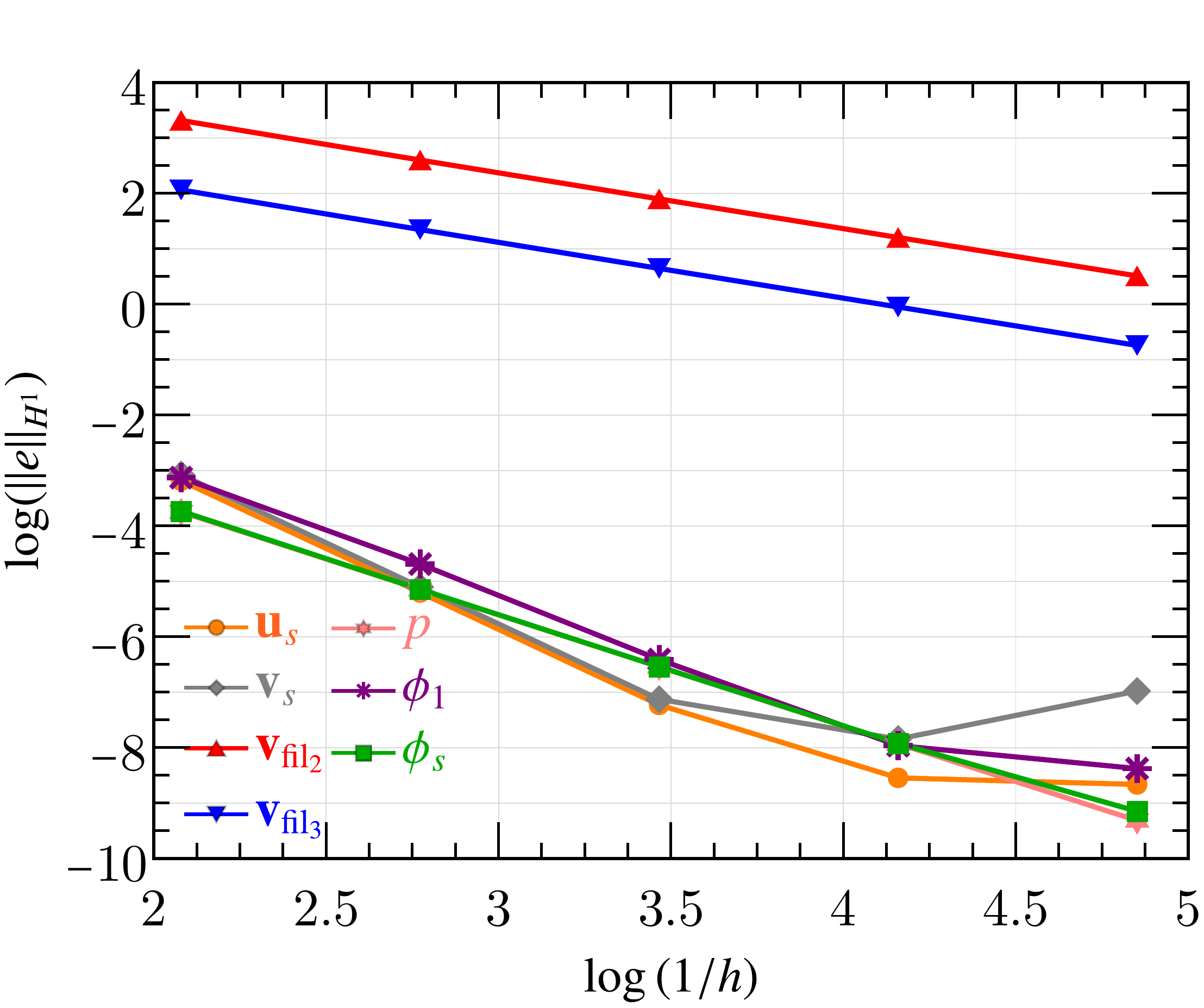}	
\caption{$L^2$ and $H^1$ error norms for case 3 in Table \ref{table:volume fraction constants}, for cases O-1, O-2 and O-3 (from top to bottom respectively.), for the weak formulation along with Neumann boundary condition for $\bv{u}_s$}
\label{fig:three}
\end{figure}

\begin{figure}[t!]
\includegraphics[width=0.45\textwidth]{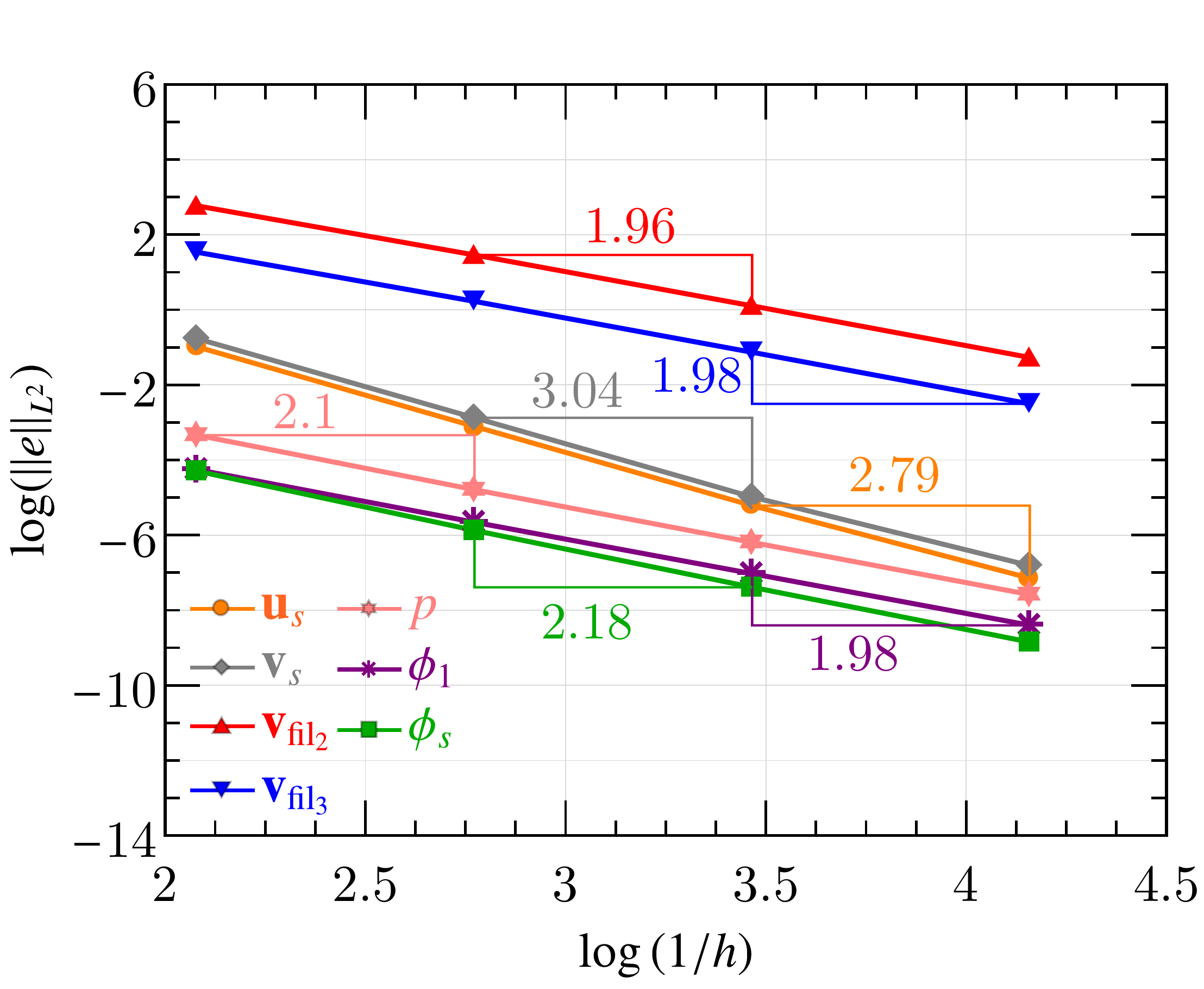}\hspace{0.01\textwidth}\includegraphics[width=0.44\textwidth]{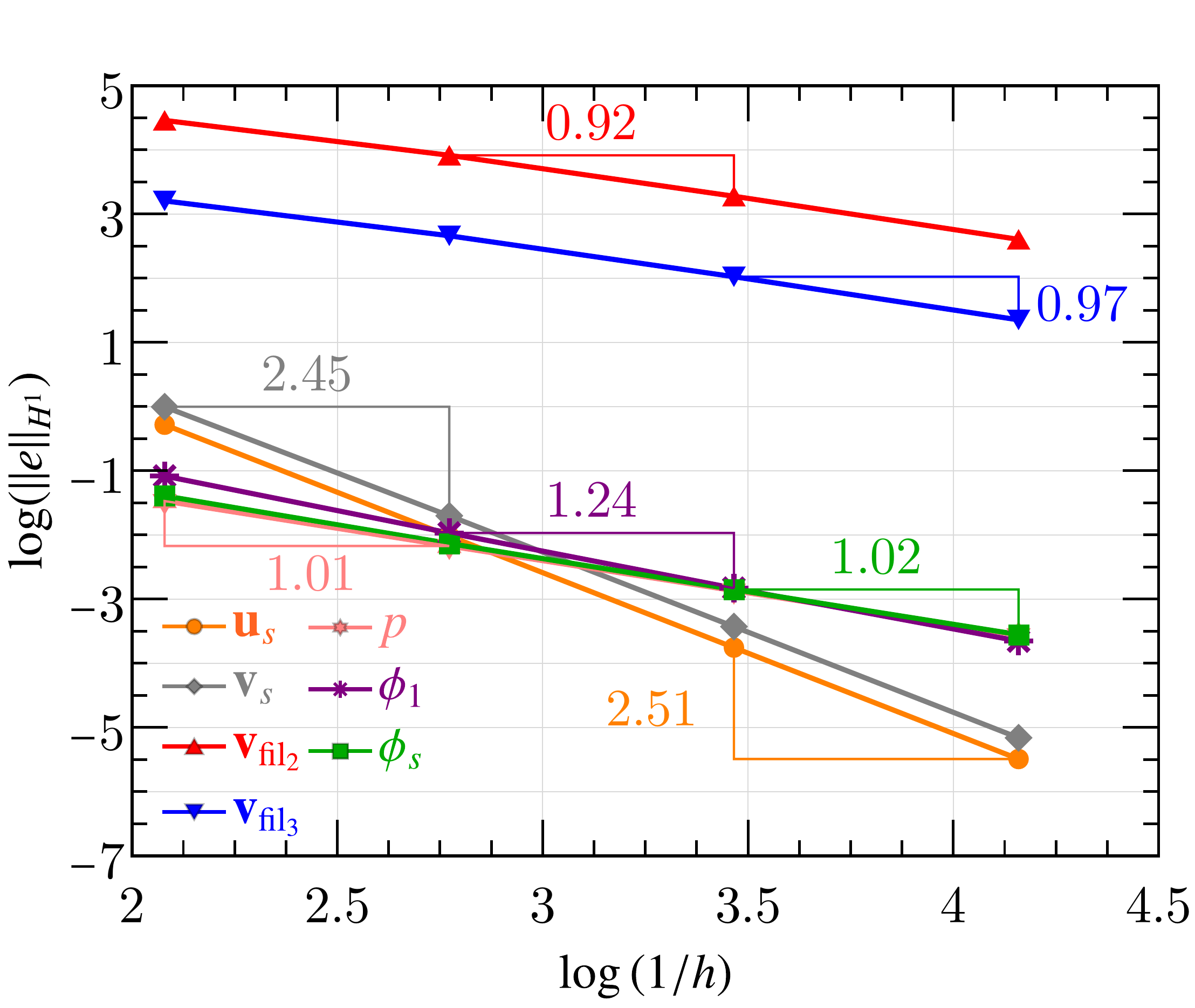}\\ \includegraphics[width=0.45\textwidth]{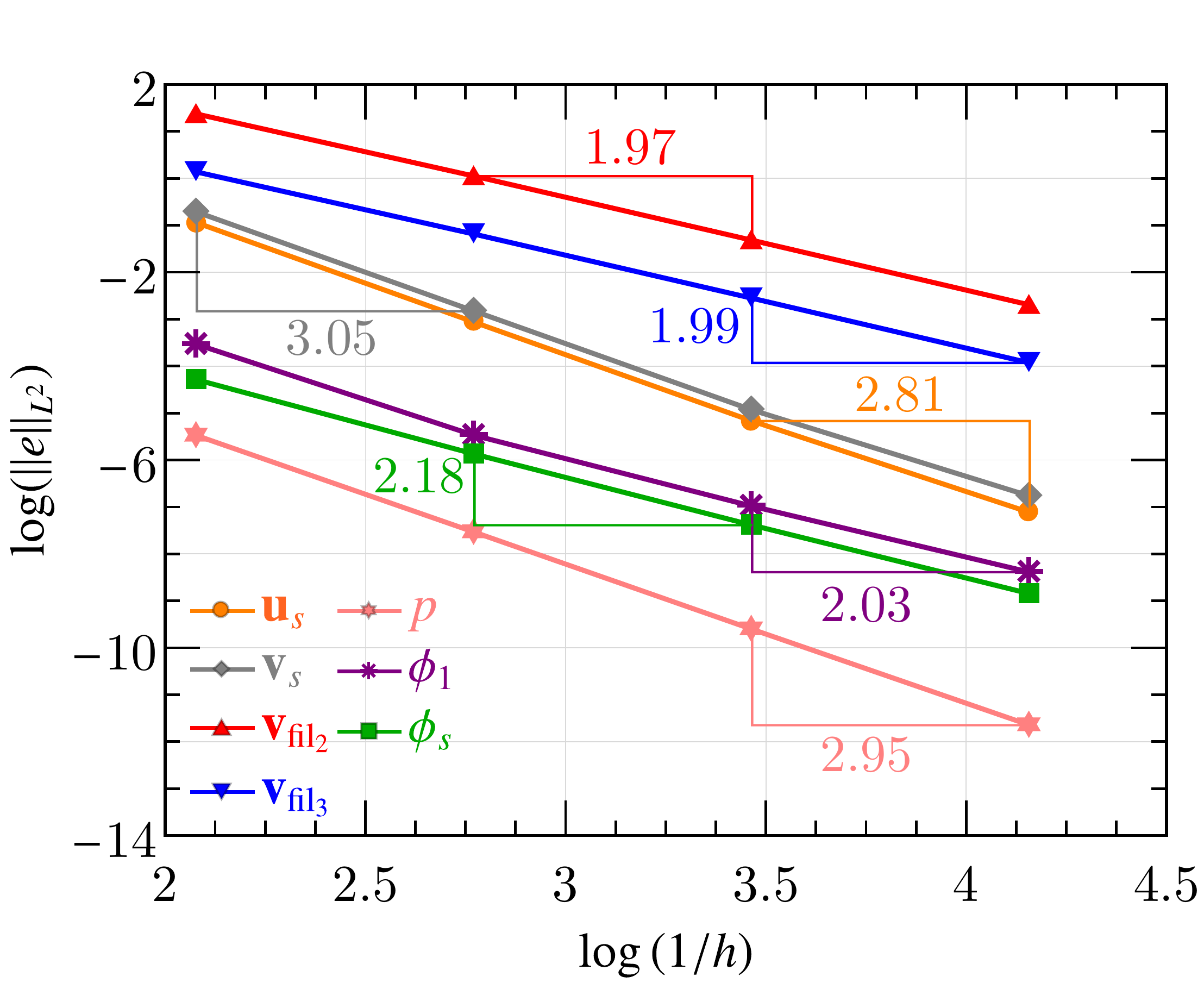}\includegraphics[width=0.45\textwidth]{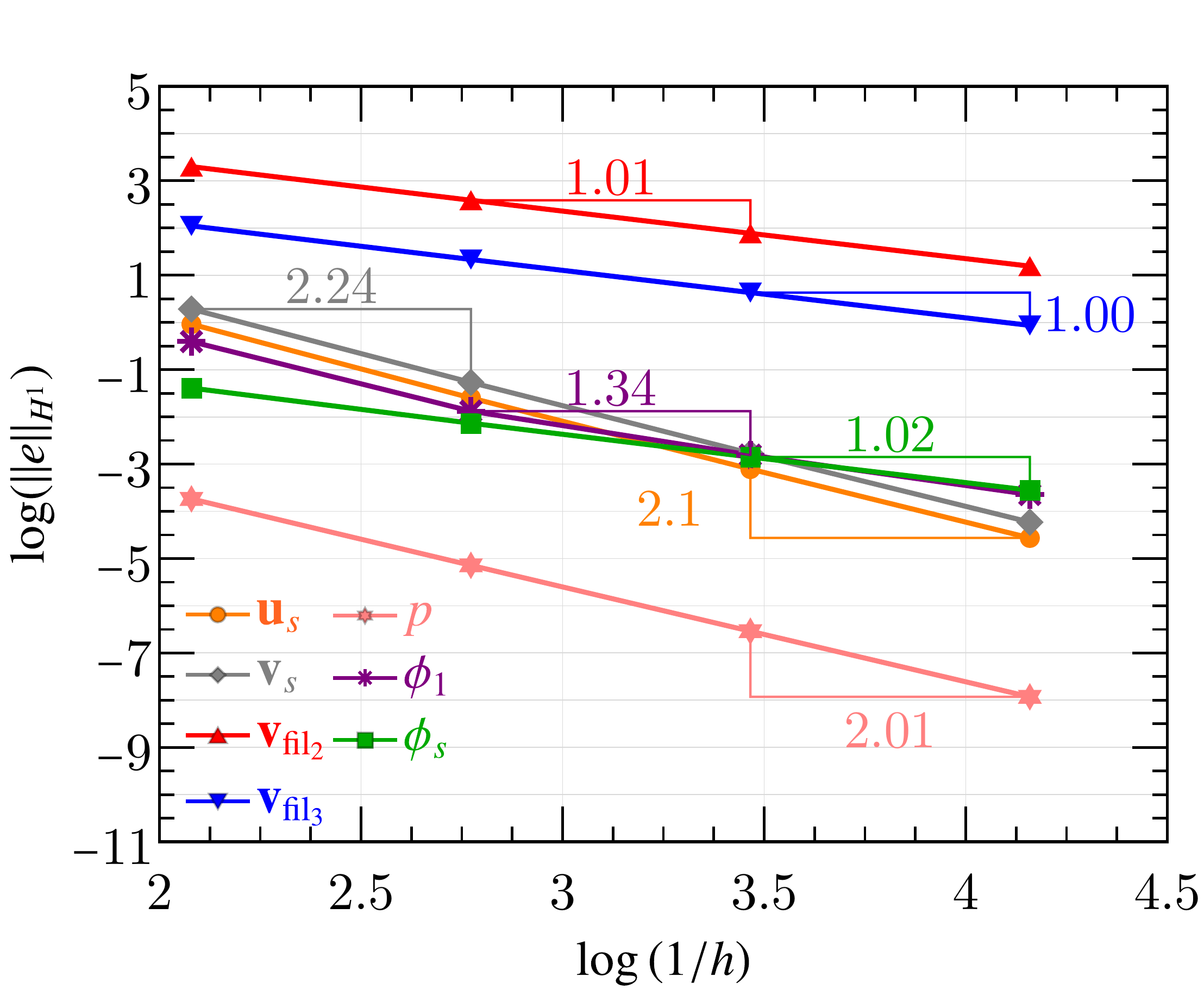}	
\caption{$L^2$ and $H^1$ error norms for case 1 in Table \ref{table:volume fraction constants}, for cases O-1 and O-2 (top and bottom respectively.), for the weak formulation along with pure Dirichlet boundary condition on all boundaries for $\bv{u}_s$ and the global integration constraint on $p$.}
\label{fig:PInt}
\end{figure}

We now present results of our \ac{FEM} implementation of the 2D Manufactured Solution problem discussed earlier. The simulations were run for the parameters and constants discussed in Tables \ref{table:parameter values} and \ref{table:formulation constants} and the three cases described in Table \ref{table:volume fraction constants}. We observed that even without any stabilization, the numerical scheme did not produce any numerical instabilities for these parameter values.

Similar convergence rates were obtained for both the sets of boundary conditions (Neumann boundary condition for $\bv{u}_s$; and global constraint on pressure along with pure Dirichlet boundary conditions for $\bv{u}_s$), with the impermeability condition implemented using the method of Lagrange multipliers. We first present the results for the `Neumann boundary condition for $\bv{u}_s$' case. Results for convergence rates are given in Figs. \ref{fig:one}, \ref{fig:two} and \ref{fig:three} for the three cases mentioned in Table \ref{table:volume fraction constants}.

Convergence results are presented in terms of the logarithms of both the $L^{2}$ norm and the $H^{1}$ semi-norm of all the quantities determined in the solution. Such results are presented as `empirical evidence’ collected in the numerical experiments. This said, by reporting convergence in the $H^{1}$ norm we do not mean to suggest that said norm is controlled by the data of the problem for all of the quantities at hand.  Proper expectations on the various fields comprising the problem have been stated in the choice of their respective functional spaces. Hence, for example, the filtration velocities are only expected to converge in the $L^{2}$ norm and $H(\text{div})$ semi-norm.  Clearly, if the numerically observed data show convergence in the $H^{1}$ semi-norm, $H(\text{div})$ convergence is implied.

The convergence behavior of the numerical scheme is observed to be similar at different times. Hence, without loss of generality, the norms have been reported at $t=\np[s]{0.3}$. For all the three cases, we observe that near-constant convergence rate of about 2 was obtained for the $L^2$ error norms for both the filtration velocities, irrespective of the order of interpolation functions. Similarly, the $L^2$ and $H^1$ error norms for pressure approach optimum values of 3 and 2 respectively for the cases O-1 and O-2. These results are comparable with the `quasi-static' formulation case in the non-reactive flow through poro-elastic solid study presented in \cite{costanzo2017arbitrary}.

The $L^2$ error norm for solid displacement is seen to be above 2.5 in cases O-1 and O-2 for all ranges of volume fractions, whereas the $H^1$ error norm is above 2 in these cases. Solid velocity achieves optimum $L^2$ norm of about 3 in cases O-1 and O-2 as well. In case of O-3, we observe a relatively erratic behavior of the error norms of solid displacement at higher resolutions. 

The two additional scalar fields with respect to the non-degrading poro-elastic case, namely the volume fractions, were seen to have optimal $L^2$ convergence rates of about 2 for the combination of interpolation polynomials given by cases O-1 and O-2. However, in the case of O-3, the behaviour of the $L^2$ error norm is unacceptable for higher resolutions. Thus, linear Lagrange polynomials seem to be the best choice for the volume fraction fields, along with the choice of shape functions discussed earlier for pressure and vector fields. 

A point worth mentioning is that the numerical solver failed to convergence when the solid volume fraction went above a value of about 0.85 or below 0.1
. We would like to note, however, that the case when the solid volume fraction is extremely low (0.1 or less), would correspond to a physical scenario in which most of the solid has already degraded, thus leading to a loss of its initial porous structure. Under these circumstances, we expect that the constitutive equations for $\tensor{P}_e$ and $W_s$ described earlier will cease to describe the elastic behaviour of the solid accurately, since the `amount' of solid will be too low in the mixture. 
The failure of the numerical scheme to converge at very high volume fractions of the solid were also observed in the case of non-degrading mixture of solid and single fluid in \cite{costanzo2017arbitrary}.

As mentioned earlier, similar tests were also conducted using the Method of Manufactured solutions for the second boundary condition, viz. Dirichlet boundary condition for $\bv{u}_s$ and the global constraint on pressure. The convergence rates are similar to those obtained in the case of Neumann boundary condition for $\bv{u}_s$. Hence, to avoid repetition, we choose to report the results for just one representative case (case 1) for the `Dirichlet boundary condition for $\bv{u}_s$' case. The results are shown in Fig. \ref{fig:PInt}.

We also conducted some numerical experiments to study the effect of relative magnitude of the filtration velocities with respect to the solid velocity (or displacement) on the numerical outcome. The solid displacement amplitude, $\bar{u}_s$ was held constant at $\np[m]{0.01}$, while the amplitudes of both the filtration velocities $\bar{v}_{\text{fil}_2}$ and $\bar{v}_{\text{fil}_3}$ were identically varied. No changes were made to the rest of the variable amplitudes.
The numerical scheme failed to converge for $\bar{v}_{\text{fil}_2}$ and $\bar{v}_{\text{fil}_3}$ values of $\np[m/s]{1}$ and higher. For such high values of filtration velocities, it would be more appropriate to use the full dynamic formulation (cf.\ Eq.~\eqref{eq:momentum_original_a}) along with a suitable stabilization scheme. This is beyond the scope of this paper and would be the focus of our future work.

\section{Conclusions}
A mixture theoretic formulation for the study of degradation of a poro-elastic solid immersed in a base fluid is derived
. Our transient 3-component system consists of the solid, the product of degradation known as the monomeric fluid, and the base fluid. The solid is embedded in the base fluid, which also initiates the hydrolytic degradation reaction of the solid. The product of degradation reaction remains in the system, and is expected to affect the rate of degradation as well.
The constitutive equations for the three component system are derived in accordance with the principals of continuum mixture theory, so as to satisfy the second law of thermodynamics for the mixture.
The model is easily extendable to include more number of fluids, or reaction products as well. The equations are applicable for flow of multiple incompressible Newtonian fluids flowing through a degrading incompressible poro-elastic solid. The simplified, filtration velocity-based, quasi-static \ac{ALE} formulation was implemented into an \ac{FEM} based numerical model. The formulation was studied for stability and convergence rates using the method of manufactured solutions. The numerical model is stable over a wide range of practically relevant values of filtration and solid velocities, without any additional stabilization. Future work will focus on implementing the full analytic model, including the convection terms into a numerical model, with the possibility of including stabilization for the fully dynamic case. The eventual goal is to apply the model to studies pertaining to real bio-degradable scaffold materials like \ac{CUPE} when inserted at real target cites.

\section*{Acknowledgments}
This work was supported by the U.S.~National Science Foundation (CMMI~1537008).


	} 
	
	\bibliographystyle{unsrtnat}
	\bibliography{ref_final_for_degradation_paper.bib}
	
\end{document}